\documentclass[12pt,reqno]{article}

\usepackage[dvips]{graphicx}
\usepackage{pict2e}
\usepackage{stfloats}
\usepackage{cite,epic,eepic,euscript,verbatim,afterpage,bm,paralist,stmaryrd,wrapfig}
\usepackage{float}
\usepackage{subfigure}

\usepackage{lineno}
\usepackage{hyperref}
\usepackage{amsmath}
\usepackage{amssymb}
\usepackage{amsthm}
\usepackage{amscd}
\usepackage{amsfonts}
\usepackage{epstopdf}
\usepackage{graphicx}
\usepackage{epsf}
\usepackage{graphics}              
\usepackage{color}
\usepackage{psfrag} 
\usepackage{graphicx,url,newverbs}
\usepackage{caption}
\usepackage{multirow}
\usepackage{nicematrix}
\usepackage{tikz}
\usetikzlibrary{calc,intersections}
\usetikzlibrary{positioning}
\usepackage{cancel}
\usepackage{dcolumn}
\usepackage{fancyhdr}
\usepackage{makeidx}
\usepackage{epsfig}
\usepackage{pgfplots}
\usepackage[ruled,vlined]{algorithm2e}
\usepackage{algorithmic}

\newcommand{\reff}[1]{{\rm (\ref{#1})}}

\numberwithin{equation}{section}

\newcommand{\R}{{\mathbb{R}}}
\newcommand{\br}{{\mathbf{r}}}
\newcommand{\bR}{{\mathbf{R}}}
\newcommand{\bx}{{\mathbf{x}}}

\newcommand{\ba}{\begin{array}}
\newcommand{\ea}{\end{array}}
\newcommand{\be}{\begin{equation}}
\newcommand{\ee}{\end{equation}}
\newcommand{\bd}{\begin{displaymath}}
	\newcommand{\ed}{\end{displaymath}}

\begin{document}

\allowdisplaybreaks

\title{
Explicit-Solute Implicit-Solvent Molecular Simulation
with Binary Level-Set, Adaptive-Mobility, and GPU}

\author{
Shuang Liu\footnote{All authors have contributed equally.}
\thanks{Department of Mathematics, University of California, San Diego, 
9500 Gilman Drive, La Jolla, California 92093-0112, United States. 
Email:shl083@ucsd.edu.
}
\and
Zirui Zhang$^*$\thanks{
Department of Mathematics, University of California, San Diego, 
9500 Gilman Drive, La Jolla, California 92093-0112, United States.
Email:zzirui@ucsd.edu.
}
\and
Li-Tien Cheng$^*$\thanks{
Department of Mathematics, University of California, San Diego, 
9500 Gilman Drive, La Jolla, California 92093-0112, United States. Corresponding author. 
Email:l3cheng@ucsd.edu.
}
\and
Bo Li$^*$\thanks{
Department of Mathematics, University of California, San Diego, 
9500 Gilman Drive, La Jolla, California 92093-0112, United States.
Email:bli@ucsd.edu.
}
}


\maketitle

\begin{abstract}

\noindent
Coarse-grained modeling and efficient computer simulations are critical to the 
study of complex molecular processes with many degrees of freedom and multiple spatiotemporal scales. 
Variational implicit-solvent model (VISM) for biomolecular solvation
is such a modeling framework, and its initial success has been demonstrated consistently. 
In VISM, an effective free-energy functional of solute-solvent interfaces  
is minimized, and the surface energy is a key component of the free energy. 
In this work, we extend VISM to include the solute mechanical interactions, 
and develop fast algorithms and GPU implementation for the extended
variational explicit-solute implicit-solvent (VESIS) 
molecular simulations to determine the underlying molecular equilibrium conformations. 
We employ a fast binary level-set method for minimizing the solvation free energy of solute-solvent interfaces and construct an adaptive-mobility gradient descent method for solute atomic optimization. 
We also implement our methods in GPU. Numerical tests and applications 
to several molecular systems verify the accuracy, stability, and efficiency of
our methods and algorithms. 
It is found that our new methods and GPU implementation improve the efficiency 
of the molecular simulation significantly over the CPU implementation. Our fast computational techniques may enable us to simulate very large systems such as protein-protein interactions and membrane dynamics for which explicit-solvent all-atom molecular dynamics simulations can be very expensive. 


\bigskip


\end{abstract}

\section{Introduction}
\label{s:Introduction}

Computer simulations are basic tools 
in the study of complex biomolecular processes with multiple temporal and 
spatial scales and many-body interactions. 
Efficiency and computational costs, however, are bottlenecks in such simulations for large systems 
with long time scales of biological interest. 
Examples of such systems include protein-protein interactions, 
membrane dynamics, and aggregation of biopolymer networks.  
The development of coarse-grained biophysical and mathematical modeling, together with 
fast numerical algorithms and computer implementation, is therefore 
critical to the success of computational studies of complex biomolecular systems. 

Implicit-solvent models are a class of coarse-grained models in which solvent is
efficiently treated in comparison with explicit-solvent all-atom molecular dynamics simulations.
In recent years, variational implicit-solvent model (VISM) 
has shown its initial success in efficient modeling of 
biomolecular conformations and recognition.  VISM is a mesoscale description of 
the solvation of charged molecules, particularly biomolecules such as proteins, 
in an aqueous environment  \cite{DSM06a, DSM06b}. 
The central quantity of such a model is a macroscopic free-energy functional of all 
possible solute-solvent interfaces each of which separates the solute molecules
from the aqueous solvent (i.e., water or salted water). 
Minimizing such a functional leads to an equilibrium molecular conformation that is 
often metastable, and the corresponding minimum free energy.  
The free energy consists mainly of the solute-solvent interfacial energy, 
solute-solvent van der Waals (vdW) interaction energy, and 
the electrostatic interaction energy that can be described by a continuum electrostatics model. 
Implemented by the level-set method, a numerical method for interface motion, 
VISM is capable of capturing qualitatively or semi-quantitatively many 
key features of charged molecular processes, such as 
the dry and wet solvation states and the effect of electrostatic interactions, 
and providing reasonably good estimates of the solvation free energy
\cite{CDML_JCP07,WangEtal_VISMCFA_JCTC12,Zhou_VISMPB_JCTC2014,GNF_Che,GNF_Che2014,
zhou_ls-vism_2015,Ricci_Rev2018}.  
We note that several other solvation models
have been developed \cite{LCW99,Borgis_JPCB2005,BatesEtal_JMB09}.

In this work, we extend VISM to include the flexibility of solute atoms, and 
develop fast algorithms and GPU implementation for 
the extended variational explicit-solute implicit solvent (VESIS) simulations of molecular 
conformational change and binding process. 
This study is motivated by our recent work that couples VISM with Monte Carlos (MC) method
to simulate the binding of proteins p53 and MDM2 that are treated as rigid bodies, 
where each MC move is followed by a solvation free-energy calculation \cite{zhang2021coupling}.   
A new and fast binary level-set algorithm that we have developed enables us to carry out such
intensive MC-VISM simulations with hundreds of thousands MC moves. 
Clearly, the rigid-body approximation can hardly make our MC-VISM simulations
reach the final p53-MDM2 bound complex.   
However, explicit-solvent all-atom molecular dynamics (MD) simulations starting from
our MC-VISM conformations reach quickly to the final complex.  It is therefore naturally 
for us to further develop our mesoscale molecular simulation approach to 
allow the solute atoms to move around as in the real system. This is what we do in our current study. 

Our main results include the following: 
\begin{compactenum}
\item[(1)]
We extend VISM to include the solute-solute atomic interactions with a usual force field
to construct our VESIS model. 
Such interactions include the mechanical bonding, bending, and torsion,
 vdW interactions modeled by Lennard-Jones (LJ) potentials, and the electrostatic interactions
by Coulomb's law. The coupling between these solute interactions and 
the implicit solvent is through the solute-solvent interactions described by
a sum of integrals over the solvent region, summing over all the solute atoms. 
\item[(2)]
We design an adaptive-mobility gradient descent  optimization method 
to relax all the solute atoms, and couple it with our fast binary level-set method
to minimize the VESIS free-energy functional. 
\item[(3)]
We implement our methods and algorithms in GPU, and test our code to verify its accuracy, 
stability, and efficiency.  
\item[(4)]
We apply our VESIS model and GPU implementation to simulate several molecular systems,
including the protein BphC and the protein complex p53-MDM2, 
to demonstrate the significant improvement of efficiency of our new algorithms and implementation
over  the CPU implementation. 
\end{compactenum}

First introduced in \cite{zhang2021coupling}, 
the binary level-set method is based on the approximation of surface area 
of an interface  separating two regions by 
the convolution of the characteristic functions of these regions with a compactly supported kernel. 
This combines two steps,  diffusion and threshold, 
in the method of threshold dynamics \cite{MBO_1992} (cf.\ also 
\cite{Ruuth_DiffuGen_JCP1998, Ruuth_Convol_JCP1999, RuuthMerriman_Rev_JCP2001, Esdoglu_JCP2017, wang_efficient_2017})
into one step.  An energy functional of the interface that includes
the surface area and other related quantities can then 
be expressed as the sum over finite-difference gird cells. 
Cells in the two regions separated by the interface
are marked by $-1$ and and $+1$. Equivalently, the interface is determined by 
a binary level-set function taking the value $-1$ or $+1$ on all the grid cells.   
The approximated total free-energy value can then be expressed as the 
sum of those values over all the grid cells. 
When a given interface is spatially perturbed, the energy change only occurs from those
cells around the interface. The method then proceeds with flipping the cells (i.e., changing the sign
of the binary level-set function on the cells) near the interface and only accept the change
of sign when the energy is decreased. 
The algorithm is seemingly simple yet is significantly more efficient than the classical 
continuous level-set method \cite{zhang2021coupling}.  
A key factor contributing to such efficiency is that the flipping is done only locally around
the interface instead of globally in the computational box 
\cite{Tai_IEEE2006,Tai_MathComp2006,gibou2005fast}.

Our new, adaptive-mobility gradient  descent optimization  method is designed to efficiently optimize 
a multi-variable objective function that may have many local minima and saddle points
and that the gradient may vary significantly. The method is of the type of the gradient descent. But the descent is not uniform for all the iteration steps. Instead, mobility constants are adaptively changed during the iteration steps. This way, one may speed up the convergence. 

In section~\ref{s:Model}, we describe our VESIS modeling framework. 
In section~\ref{s:NumericalMethods}, 
we present our fast binary level-set method for interface motion and adaptive-mobility
optimization method for relaxing atomic positions, as well as the simulation algorithm. 
Section~\ref{s:GPU} is devoted to the description of our GPU implementation. 
In section~\ref{sec:numericalexperiments}, 
we present the numerical tests and applications to several molecular systems, 
and demonstrate the efficiency of our methods and implementation.  
Finally, in section~\ref{s:Conclusions}, 
we draw conclusions and discuss our future work.   
Appendix collects some calculations and formulas that are used in our modeling and 
numerical methods. 

\section{A Variational Explicit-Solute Implicit-Solvent Model}
\label{s:Model}

\begin{figure}[ht]
 	\centering 
	\includegraphics[width=0.41\linewidth]{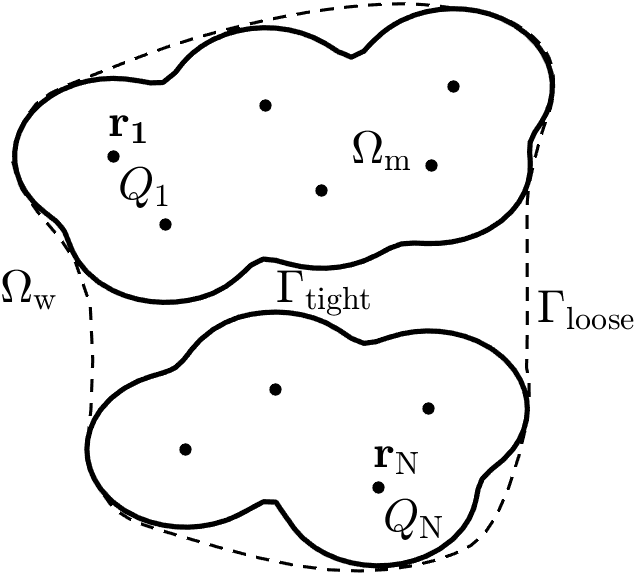}
\vspace{-2.5 mm}
	\caption{ A schematic view  of a solvation system with explicit solute and implicit solvent.
The entire system region $\Omega$ is divided by a solute-solvent interface into 
the solvent region $\Omega_{\rm w}$ 
and the solute region $\Omega_{\rm m}$ containing all the solute atoms at $\br_i$ $(1 \le i \le N).$ 
Two different solute-solvent interfaces are shown. 
One is a tight interface $\Gamma_{\rm tight}$ (solid line) and the other 
a loose interface $\Gamma_{\rm loose}$ (dashed line).  
}
	\label{fig:solute-solvent-system}
\end{figure}

We consider a few molecules immersed in an aqueous solvent (i.e., water or salted water).
This system of molecular solvation is confined spatially in a bounded region $\Omega\subset \R^3;$
cf.\ Figure~\ref{fig:solute-solvent-system}. 
We assume that there are $N$ atoms of these solute molecules, 
located at $\br_1, \dots, \br_N \in \Omega $ 
and carrying partial charges $Q_1, \dots, Q_N$, respectively. 
A closed surface $\Gamma$ inside $\Omega$ and enclosing all the solute atoms $\br_i $ $(1 \le i \le N)$
is called a solute-solvent interface or dielectric boundary. Such an interface, which 
may have several disjoint connected components, divides the entire solvation region $\Omega$ into 
two parts. One is the solute region, denoted $\Omega_{\rm m}$ 
(m stands for molecule),  which is the interior of the surface $\Gamma$, and the other
is the solvent region, denoted $\Omega_{\rm w}$ (w stands for water) and 
defined by $\Omega_{\rm w} = \Omega \setminus \overline{\Omega}_{\rm m}$   
(a bar denotes the closure of a set).

\vspace{-2 mm}

\vspace{-2 mm}
Our basic assumption is that an experimentally observed
equilibrium solvation system is determined by its solute-solvent
interface and solute atomic positions that together 
minimize an effective free-energy functional \cite{CXDMCL_JCTC09}
\begin{equation}
\label{eq:freeenergyfn}
G[\Gamma,\mathbf{R}]=G_{\rm VISM}[\Gamma, \bR]+G_{\rm ss} [\bR],
\end{equation}
over all possible solute-solvent interfaces $\Gamma$ and solute atomic positions 
$\bR = (\br_1, \dots, \br_N).$ Here, the first part is the solvation free energy
approximated by the VISM free energy
and the second part is the solute-solute interaction potential or force field. 

The VISM free energy is given by 
\cite{DSM06a, DSM06b, WangEtal_VISMCFA_JCTC12,Zhou_VISMPB_JCTC2014}
\begin{align}
\label{G_VISM}
G_{\rm VISM}[\Gamma, \bR] &= 
\gamma_0 {\rm Area}\, (\Gamma) +  \rho_{\rm w}\sum_{i=1}^{N}\int_{\R^3 \setminus \Omega_{\rm m}} U_{\rm LJ}^{(i)} 
(|\mathbf{r}-\mathbf{r}_i|)dV_\mathbf{r}
\nonumber \\
&\quad  + \frac{1}{32\pi^2\varepsilon_0} 
\left (\frac{1}{\varepsilon_{\rm w}}-\frac{1}{\varepsilon_{\rm m}}\right) 
\int_{\R^3 \setminus \Omega_{\rm m}} 
\left | \sum_{i=1}^{N}\frac{Q_i(\mathbf{r}-\mathbf{r}_i)}{|\mathbf{r}-\mathbf{r}_i|^3} \right |^2 
dV_{\mathbf{r}}.
\end{align}
The first term here is the solute-solvent interfacial energy, 
where $\gamma_0$ is the surface tension constant. 
The second term describes the van der Waals (vdW) type 
interactions between the solute atoms located at $\br_i$ $(1 \le i \le N)$ 
 and solvent molecules that are treated as 
a continuum, where $\rho_{\rm w}$ is the bulk solvent density and 
each $U_{\rm LJ}^{(i)}$ is a Lennard-Jones (LJ) potential of the form
\begin{equation}
\label{eq:LJpotential}
	U_{LJ}(r)=4\varepsilon \left [\left (\frac{\sigma}{r} \right)^{12}
- \left (\frac{\sigma}{r} \right )^{6} \right],
\end{equation}
where the length parameter $\sigma$ and energy parameter $\varepsilon$ can depend on 
individual solute atoms.  The last term in \reff{G_VISM} is the Coulomb-field 
approximation (CFA) of the electrostatic
interaction energy, where $\varepsilon_0$ is the vacuum permittivity, and 
$\varepsilon_{\rm w}$ and $\varepsilon_{\rm m}$ are the 
relative permittivities of the solvent and solute, respectively.
Note that the integrals in \reff{G_VISM}
 are over the region $\R^3\setminus \Omega_{\rm m}$, instead
of $\Omega_{\rm w}$ which is bounded. This is to account for the long-range effect 
of the vdW and Coulomb interactions.  

We remark that one can include more terms in the VISM free energy. However, to keep our numerical 
implementation robust, we shall focus on this version of VISM free-energy functional. 

The solute-solute interaction potential in \reff{eq:freeenergyfn} is given by 
\begin{align}
\label{Gss}
G_{\rm ss} [\mathbf{R}] 
 & = \sum_{(i,j)} \frac{1}{2} A_{ij}(r_{ij}-r_{0ij})^2 
+ \sum_{(i,j,k)}  \frac{1}{2}B_{ijk}(\theta_{ijk}-\theta_{0ijk})^2 
\nonumber \\
& \quad 
+  \sum_{(i,j,k,l)} \sum_{n=0}^{6}C_n[1+\cos(n\tau_{i,j,k,l}-\psi_n)] 
+ \sum_{(i,j)'}U_{\rm LJ}^{(i,j)} (r_{ij})
+ \sum_{(i,j)'} \frac{Q_i Q_j }{4\pi\varepsilon_0\varepsilon_{\rm w} r_{ij}  }. 
\end{align}
Here, the first three terms account for the mechanical interaction energy from bonded solute atoms. 
The first term is the bonding energy of solute atoms, where the sum is taken over all 
pairs $(i,j)$ of bounded solute atoms, $r_{ij} = | \br_i - \br_j|$, 
and $r_{0ij}$ and $A_{ij}$ are the corresponding equilibrium distance and spring constant, respectively. 
The second term in \reff{Gss} 
is the bending energy of solute atoms, where the sum is taken over all triplets  
$(i, j, k)$ such that both pairs of solute atoms  
$(\mathbf{r}_i,\mathbf{r}_j)$ and $(\mathbf{r}_j,\mathbf{r}_k)$ are bonded. 
For such a triplet, $\theta_{ijk}$ is the angle between the vectors $\mathbf{r}_i-\mathbf{r}_j$  
and $\mathbf{r}_k-\mathbf{r}_j$, $\theta_{0ijk} \in [0, \pi]$ 
is the corresponding equilibrium angle,  
and $B_{ijk}$ is a constant parameter. 
The third term in \reff{Gss} accounts for the torsion energy of solute atoms
\cite{hopfinger1985computer}.  The sum 
is taken over all quadruples $(i,j,k,l)$ 
such that $(\mathbf{r}_i,\mathbf{r}_j)$, $(\mathbf{r}_j,\mathbf{r}_k)$,  
and $(\mathbf{r}_k,\mathbf{r}_l)$ are all bonded.
For such a quadruple $(i,j,k,l)$,
$\tau_{ijkl}$ is the torsion angle that 
is the angle between the plane determined by 
$(\mathbf{r}_i,\mathbf{r}_j, \mathbf{r}_k)$ and 
that determined by $(\mathbf{r}_j, \mathbf{r}_k, \mathbf{r}_l)$, 
$n$ is the multiplicity, $\psi_n$ is the  phase factor, and all $C_n$ are constants. 

The last two terms in \reff{Gss} account for the interaction energies from non-bonded solute atoms 
indicated by $(i,j)'$ in the summation.  The fourth term is the solute-solute vdW interaction energy, 
where each $U_{\rm LJ}^{(i,j)}$ is an LJ potential of the form \reff{eq:LJpotential}. 
The last term is the solute-solute Coulomb interaction energy. 

\section{Numerical Methods}
\label{s:NumericalMethods}

We minimize the free-energy functional $G[\Gamma, \bR]$ defined in
(\ref{eq:freeenergyfn}) numerically by an iteration scheme. 
Each iteration step consists of two parts. In the first part, we 
fix the solute atomic positions and minimize numerically the VISM solvation free-energy
functional \reff{G_VISM} by a binary level-set method  
to obtain an optimal solute-solvent interface. 
In the second part, we fix the interface obtained in the first part, and minimize
the energy functional $G[\Gamma, \bR]$ that is a multi-variable function of 
the solute atomic positions $\bR = (\br_1, \dots, \br_N)$  
by an adaptive-mobility gradient descent method. 
The binary level-set method was introduced and used in our rigid-body MC-VISM simulations of 
protein binding \cite{zhang2021coupling}. 
Here, we briefly recall the method, referring to \cite{zhang2021coupling} for more details. 
We also describe in details our new, adaptive-mobility  gradient descent 
optimization method for minimizing the function $G[\Gamma, \bR]$ with $\Gamma$ fixed. 
We present our step-by-step algorithm at the end of this section. 

\subsection{A Binary Level-Set Method}
\label{ss:BLSM}

We set the solvation system region to be $\Omega = (-L, L)^3$ for some $L > 0$. 
The side length $L$ is chosen to be large enough so that the region $\Omega$
includes all the solute atoms $\br_i$ $(1 \le i \le N)$ 
whose geometrical center can be shifted to the origin, if necessary; 
cf.\ Figure~\ref{fig:solute-solvent-system}.  This region $\Omega$ is also our computational box.  
We cover it by a uniform finite-difference grid of size $h.$  
A solute-solvent interface $\Gamma = \partial \Omega_{\rm m}$ is approximated
by a binary level-set function $\phi$ that is defined on all the grid cells with
$\phi = -1  $ and $\phi = +1  $ on cells interior and exterior  to $\Gamma$, respectively.

We discretize the VISM solvation free-energy functional \reff{G_VISM} 
with all the solute atoms fixed at $\br_i $ $(1 \le i \le N).$ 
Let us first rewrite this functional as 
\begin{equation}
\label{GGR2}
G_{\rm VISM}[\Gamma, \bR] = \gamma_0 \mbox{Area}\, (\Gamma) +  
\int_{\Omega \setminus \Omega_{\rm m}} U (\br) \, dV_{\br} 
+ \int_{\R^3 \setminus \Omega} U (\br) \, dV_{\br}, 
\end{equation}
where 
\begin{equation}
\label{Ur}
U(\br) =  \rho_{\rm w}\sum_{i=1}^{N} U_{\rm LJ}^{(i)} (|\mathbf{r}-\mathbf{r}_i|)
 + \frac{1}{32\pi^2\varepsilon_0} 
\left (\frac{1}{\varepsilon_{\rm w}}-\frac{1}{\varepsilon_{\rm m}}\right) 
\left | \sum_{i=1}^{N}\frac{Q_i(\mathbf{r}-\mathbf{r}_i)}{|\mathbf{r}-\mathbf{r}_i|^3} \right |^2. 
\end{equation}

{\bf Approximation of the surface energy.}
The surface area of the solute-solvent interface $\Gamma$ can be expressed as \cite{zhang2021coupling} 
\begin{equation}
\label{Area}
{\rm Area}(\Gamma)= \frac{C_0}{\delta^4} \int_{\mathbf{x}\in \Omega_{\rm m}} 
\int_{\mathbf{y}\in \Omega_{\rm w}} K\left(\frac{\mathbf{x}-\mathbf{y}}{\delta}\right) 
d\mathbf{y} d\mathbf{x} +O(\delta^2) \quad for\quad0< \delta \ll 1,  
\end{equation}
where
	\[C_0  = \left( \int_0^1  \int_{B(\mathbf{0},1) \cap \{y_3>s\}} K(\mathbf{y}) \, d\mathbf{y}
	\, ds \right)^{-1} 
\] 
is a constant, $B(\mathbf{0}, A) $ for any $A > 0$ is the ball centered at the origin 
$\mathbf{0}$ with radius $A$, and $y_3$ is the third component of the position vector $\mathbf{y}$. 
The kernel function $K=K(\mathbf{x})$ ($\mathbf{x} \in \mathbb{R}^3$)
is chosen to be non-negative, compactly supported in the closure of 
the unit ball $B_1(\mathbf{0}) $ of 
$\R^3$, and spherically symmetric (i.e., it is a function of $|\bx|$). In our implementation,
we set $K(\mathbf{x})= \sin^2(\pi |\mathbf{x} |)$ if $|\mathbf{x} | \leq 1$ and $0$ elsewhere. 
The small parameter $\delta > 0$ is the rescaled kernel radius, defined as the radius
of the ball of the support of $K(|\bx|/\delta).$ 

{\bf Discretization of the VISM free energy in the solvation region.}
By employing the center-point numerical integration rule, one can discretize 
the double-integral in \eqref{Area} with an optimal choice $\delta = \lambda \sqrt{h},$ 
where $\lambda > 0$ is a constant. 
Consequently, we obtain the following expression of an approximation of the surface energy,  
the first term in \reff{GGR2} \cite{zhang2021coupling}:

\begin{align*}
	\gamma_0 \mbox{Area}\, (\Gamma) = 
	\frac{\gamma_0 C_0h^{4}}{ \lambda^{4} }  \sum_{\bx_j \in \Omega_{\rm m}} 
	\sum_{\substack{\bx_k\in \Omega_{\rm w}\\
			|\bx_k-\bx_j|\leq \lambda \sqrt{h} }} K(\bx_j-\bx_k)   + O(h),
\end{align*}
where $\bx_j \in \Omega_{\rm m}$ and $\bx_k \in \Omega_{\rm w}$ 
are the centers of grid cells in $\Omega_{\rm m}$ and $\Omega_{\rm w}$, respectively. 
The second term in \reff{GGR2} can be approximated by the center-point integration rule: 
\[
\int_{\Omega \setminus \Omega_{\rm m}} U (\br) \, dV_{\br} 
= h^3 \sum_{{\mathbf{x}_j}\in \Omega_{\rm w}} U(\bx_j) + O(h).  
\]
Therefore, these approximations and \reff{GGR2} lead to 
the approximation 
\begin{align}
	\label{4flip} 
	G_{\rm VISM}[\Gamma, \bR] &\approx  
	\frac{\gamma_0 C_0h^{4}}{ \lambda^{4} }  \sum_{\bx_j \in \Omega_{\rm m}} 
	\sum_{\substack{\bx_k\in \Omega_{\rm w}\\
			|\bx_k-\bx_j|\leq \lambda \sqrt{h} }} K(\bx_j-\bx_k) 
	\nonumber \\
	& \quad 
	+ h^3 \sum_{{\mathbf{x}_j}\in \Omega_{\rm w}} U(\bx_j) 
	+ \int_{\R^3 \setminus \Omega} U (\br) \, dV_{\br}.  
\end{align}
The last integral can be written analytically as iterated integrals using the
spherical coordinates and evaluated by one-dimensional numerical quadrature; 
 cf.\ \cite{cheng2010level}.

{\bf Flipping grid cells to decrease the energy.}
Given a solute-solvent interface defined by a binary level-set function, 
we relax its VISM free energy \reff{4flip}
by flipping the grid cells, i.e., by changing the sign of the binary level-set function on the grid
cells.  The flipping is only done for grid cells that around the interface. 
This is because that any grid cells centered at $\bx_j \in \Omega_{\rm m}$ 
and $\bx_k \in \Omega_{\rm w}$ with $|\bx_j - \bx_k| > \lambda \sqrt{h}$ do not contribute to 
 the first term in \reff{4flip}. 

We pick up a grid cell that is immediate next to be the interface, 
flip its sign, and calculate the change of the approximate energy based on \reff{4flip}. 
Note that the last term in \reff{4flip} does not change if we flip a grid cell. 
If the cell is centered at $\bx_j \in \Omega_{\rm m}$, so the sign of the cell is $-1$,
then the flip of the cell leads to the change of energy
\begin{equation}
\label{dGj1}
\Delta (G_{\rm solv})_j = \frac{\gamma_0 C_0h^{4}}{ \lambda^{4} }
	\sum_{\substack{\bx_k\in \Omega_{\rm m}\\
			|\bx_k-\bx_j|\leq \lambda \sqrt{h} }} K(\bx_j-\bx_k) -\frac{\gamma_0 C_0h^{4}}{ \lambda^{4} }
		\sum_{\substack{\bx_k\in \Omega_{\rm w}\\
		|\bx_k-\bx_j|\leq \lambda \sqrt{h} }} K(\bx_j-\bx_k) +U(\mathbf{x_j}).
\end{equation}
Otherwise, if the  cell is centered at $\bx_j \in \Omega_{\rm w}$, 
so the sign of the cell is $+1$, then the flip of the cell leads to the change of energy 
\begin{equation}
\label{dGj2}
\Delta (G_{\rm solv})_j = 	\frac{\gamma_0 C_0h^{4}}{ \lambda^{4} }
		\sum_{\substack{\bx_k\in \Omega_{\rm w}\\
				|\bx_k-\bx_j|\leq \lambda \sqrt{h} }} K(\bx_j-\bx_k) -\frac{\gamma_0 C_0h^{4}}{ \lambda^{4} }
		\sum_{\substack{\bx_k\in \Omega_{\rm m}\\
				|\bx_k-\bx_j|\leq \lambda \sqrt{h} }} K(\bx_j-\bx_k) -U(\mathbf{x_j}).
\end{equation}
After calculating $\Delta (G_{\rm solv})_j$ by flipping grid cells near the interface, we put $\Delta (G_{\rm solv})_j$ in a Min-Heap. We flip the grid cell with the smallest $\Delta (G_{\rm solv})_j$ in the heap if $ \Delta (G_{\rm solv})_j < 0$.  With the new interface, we add energy changes for new grid cells near the new interface to the heap, delete old grid cells in the heap which are not near the new interface, update $\Delta (G_{\rm solv})_j$  for old grid cells near the new interface in the heap, then we sort the Min-Heap. This flipping process stops until all  $\Delta (G_{\rm solv})_j  \ge 0$, indicating the energy reaches a minimum. 

\begin{figure}[H]
\centering 
\includegraphics[width=3.2 in,height=3in]{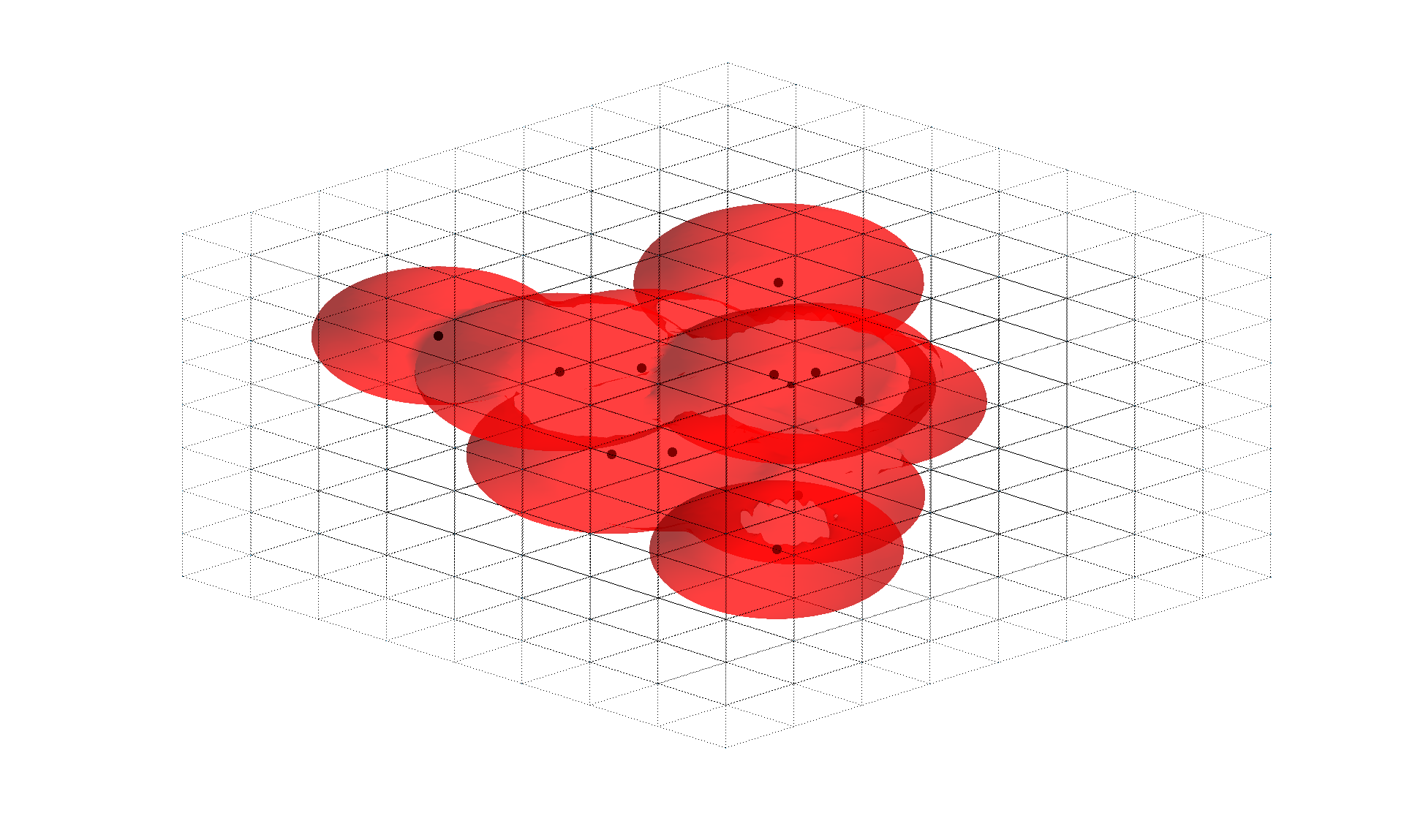}
\hspace{-10.2 mm}
\includegraphics[width=3.2 in, height=3 in]{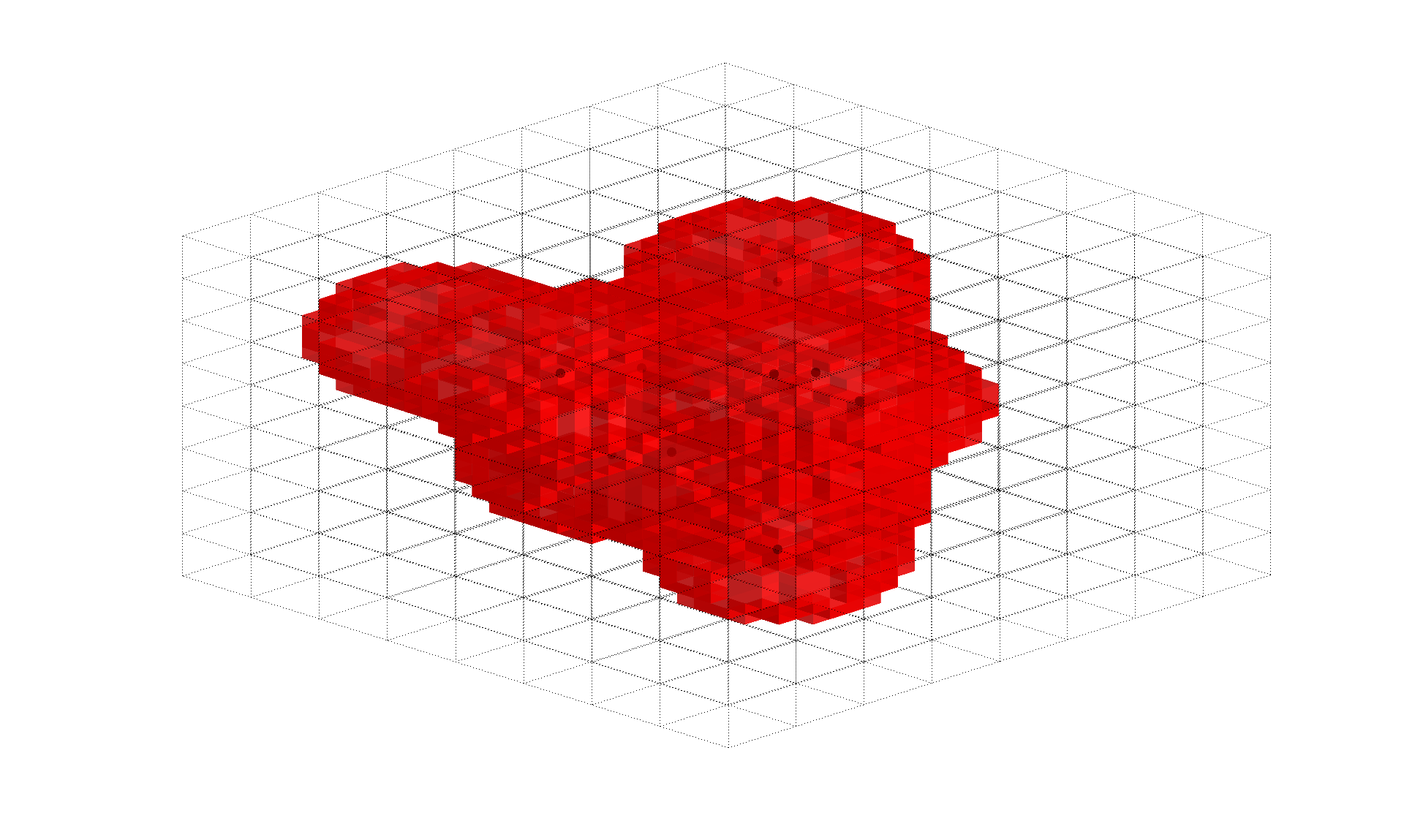}
\vspace{-8 mm}
\vspace{-3 mm}
\caption{A tight initial solute-solvent interface constructed as  the union of vdW spheres 
centered at solute atoms (black dots) by a continuous level-set function (Left) and a binary 
level-set function (Right), respectively.}
\label{fig:simulationbox}
\end{figure}

{\bf Initial surfaces.} 
Our method for minimizing the VISM free-energy functional is of steepest descent type. It starts with an 
initial surface and iteratively moves it with the free energy decreased in each step of the iteration.  
The final free-energy minimizing surface is a local minimizer of the functional and depends
on the initial surface. Different initial surfaces 
can lead to different (meta)stable equilibrium conformations that are of interest;  
see Figure~\ref{fig:solute-solvent-system}. 
In order to capture multiple local minimizers, we often use two types of initial surfaces. 
One is a loose initial surface that can be a large sphere enclosing all the solute atoms.  
The other is a tight initial surface that wraps up all the solute atoms tightly with vdW radii. 
Such a surface is the zero level-set of the continuous function
\begin{equation*}
	\varphi(\mathbf{r})=\min_{1\leq i\leq N}(|\mathbf{r}-\mathbf{r}_i|-d_i), 
\end{equation*}
where $d_i>0$ is the vdW radius of the $i$th solute atom 
located at $\mathbf{r}_i$ $(i=1,\cdots,N).$
The binary level-set function 
for the surface can then be obtained by
setting  its value at the center of a grid cell to be the sign of $\varphi$-value at that center. 
Figure~\ref{fig:simulationbox} shows a tight surface constructed by both a continuous and the 
corresponding binary  level-set function. 

\vspace{-4 mm}

\subsection{Adaptive-Mobility Gradient Descent Method for the Relaxation of Solute Atoms}
\label{ss:RelaxSoluteAtoms}

With a fixed solute-solvent interface $\Gamma$, we minimize the free energy $G[\Gamma, \bR]$
defined in (\ref{eq:freeenergyfn}), \eqref{G_VISM}, and \eqref{Gss} 
as a function of $\bR = (\br_1, \dots, \br_N)$ by solving for a steady-state
solution to the system of the gradient descent equations
\begin{equation}
\label{eq:moveatoms}
\frac{d(\mathbf{r}_n(t))_l}{dt}
=-M_{nl}(\nabla_{\mathbf{r}_n}G[\Gamma,\mathbf{R}])_l, \quad n=1,\cdots, N, \quad l=1,2,3, 
\end{equation}
where $({\bf a})_l$ denotes the $l$th component of a vector $\bf a \in \R^3$ $(1 \le l \le 3)$ 
and all $M_{nl}>0$ are constants (called mobility constants).  
The formula of the gradient 
$\nabla_{\mathbf{r}_n}G[\Gamma,\mathbf{R}]$ is given in \reff{GradientG} in Appendix. 
We use the forward Euler method to solve these equations  iteratively with a fixed time step.

Due to the complex molecular interactions of an underlying system, the gradient
of $G[\Gamma, \bR]$ can vary significantly with solute atoms and with different components.  
If the mobility constant $M_{nl}$ is too large, then the motion of the particle $\mathbf{r}_n$ 
in its gradient descent direction may possibly overshoot and increase the free energy. 
If $M_{nl}$ is too small, the free energy may decrease very slowly. 
To improve the stability and efficiency, we therefore adaptively change the mobility constants $M_{nl}$ 
in each step of iteration  based on the magnitude of gradient and the total free-energy value. 

In our implementation, the mobility constants are chosen to be $\omega M$, 
where the ``base" mobility $\omega$ is updated in each interaction step
of the forward Euler method and $M$ is also adjustable. 
The adjustment of $M$ is based on the decrease or increase of the energy 
with a high energy threshold.  It is controlled by a relative energy value 
$\delta G^* > 0$ depending on iteration steps, and a shrinking parameter  $\alpha \in (0, 1)$ 
that shrinks $M$ if the energy increases too much and too often 
tracked by a counting number $N_{\rm cnt}$ which has a threshold value $N_{\rm cnt}^*$.  
Initially, we set $M = 1$ and $N_{\rm cnt} = 0.$ 

Given all the atomic positions after a forward Euler iteration step,
with some $M > 0$ and $ 0\le N_{\rm cnt} < N_{\rm cnt}^*$,   
and the corresponding energy $G[\Gamma, \bR]$ calculated and denoted $G_{\rm old}.$ 
We  calculate all the gradient of $G[\Gamma, \bR]$ at those positions and set
\begin{equation}
\label{eq:adjustmml}
\omega= \frac{ 1}{\max_{1 \le m \le N, 1 \le k \le 3}
| (\nabla_{\br_m}(G[\Gamma, \bR])_k |}.
\end{equation}
We then move the atomic positions in one time step by the forward Euler iteration for 
solving \reff{eq:moveatoms} with all $M_{nl} = \omega M$. 
We also calculate the energy for the new atomic positions and denote it by $G_{\rm new}.$ 
If $G_{\rm new} < G_{\rm old}$, then we accept the moved solute atomic positions, 
do not change the value of $M$, increase $N_{\rm cnt }$ by $1$, and continue the iteration. 
If $G_{\rm new} >  G_{\rm old}$, then we choose the threshold $\delta G^*$ to be some
fraction of $G_{\rm old}$ and consider two cases. 
If  $G_{\rm new} \ge  G_{\rm old} + \delta G^*$, then  
we do not update the atomic positions but shrink $M$ to  $M := \alpha M$ 
reset $N_{\rm cnt} = 0$ and start over with the next iteration step. 
If $G_{\rm new} < G_{\rm old} + \delta G^*$, then  
we check the counting number $N_{\rm cnt} $ and consider two cases: 
\begin{compactenum}
\item[(1)]
If $N_{\rm cnt}  {\geq}  N_{\rm cnt}^*$, which means that the same $M$ has been used for $N_{\rm cnt}^*$
times, then we do not accept the new positions, shrink $M $ to $M:=\alpha M$, and 
reset $N_{\rm cnt} = 0.$ 
\item[(2)]
If $N_{\rm cnt} <  N_{\rm cnt}^*$, we accept the new atomic positions, keep the same $M$, 
and increase $N_{\rm cnt} $ by $1$. 
\end{compactenum}
Note that $N_{\rm cnt}$ is the number of steps where the same $M $ is used consecutively.  
In our implementation, we choose $\alpha = 1/\sqrt{2}$, $N_{\rm cnt}^*=5$, and $\delta  G^*$ to be $ 5\% $ of $ G_{\rm old}$.

\subsection{Numerical  Algorithm}
\label{ss:Algorithm}

\begin{compactenum}
\item[Step 1.]
Initialization.
Input all the model parameters from \reff{G_VISM}--\reff{Gss}. 
In particular, position $N$ solute atoms with the center of geometry 
at the origin by a coordinate translation if necessary. Set the computational box $\Omega=(-L, L)^3$
and discretize the box with a uniform finite-difference grid. Set an initial binary level-set function $\phi^{(0)}$ to define the initial solute-solvent interface $\Gamma^0$, solute region $\Omega_{\rm m}^0$, and solvent region $\Omega_{\rm w}^0$. 

Set counter $N_{\rm cnt} =0$, count tolerance $N_{\rm cnt}^* =5$,  an initial uniform mobility constant $M=1$, and the time step $dt=1$. Set the initial iteration number $k = 0$ and $k_{\rm max}$$=3000$. Set the shrinking parameter $\alpha=1/\sqrt{2}$. Set the error tolerance 
${\rm Tol}_1=1e$-$4$  for the gradient  and ${\rm Tol}_2=1e$-$5$ for atomic positions update.

{\item[Step 2.] Get the optimal  solute-solvent interface  $\Gamma^{k+1}$ with atomic positions $\bR^k$ by the binary level-set method. }

\begin{compactenum}
\item[Step 2.1.]
Calculate by \reff{dGj1} and \reff{dGj2} the change of solvation free energy $\Delta (G_{\rm solv})_j$ 
on all grid cells next to the interface $\Gamma^k$.  Sort solvation energy change $\Delta (G_{\rm solv})_j$  in a Min-Heap. 

\item[Step 2.2.]
Flipping Process: \\
$\mathbf{while}$ (Smallest $\Delta (G_{\rm solv})_j <0$) $\mathbf{do}$\\
 $~~~~\mathbf{Flip:}$ flip the corresponding grid cell with smallest $\Delta (G_{\rm solv})_j$.\\
$~~~~\mathbf{Update:}$ check for new grid cells next to the new interface, calculate $\Delta (G_{\rm solv})_j$ $~~~~~~~~~$for new grid cells, and update $\Delta (G_{\rm solv})_j$  for old grid cells in the heap. \\
$~~~~\mathbf{Sort:}$ sort solvation energy change $\Delta (G_{\rm solv})_j$ in a Min-Heap.\\
$\mathbf{endwhile}$

\item[Step 2.3.]
Define $\Gamma^{k+1}$ to be the optimal solute-solvent interface from the flipping process. 

\end{compactenum}

\item[Step 3.] Update the solute atomic positions by solving the system 
of equations \reff{eq:moveatoms}. 

\begin{compactenum}
\item[Step 3.1.]
Calculate by \reff{GradientG} the gradient $\nabla_{\mathbf{r}_n}G[\Gamma^{k+1},\mathbf{R}^k]$ 
for all $n =1, \cdots, N.$ 

\item[Step 3.2.]
Test the convergence. If absolute values of $\nabla_{\mathbf{r}_n}G[\Gamma^{k+1},\mathbf{R}^k]$ for each solute atom in each coordinate $<\rm{Tol}_1$, then stop the algorithm.

\item[Step 3.3.]
Update positions of all moving solute atoms according to equation (\ref{eq:moveatoms}) and (\ref{eq:adjustmml}). Calculate the total free energy change $\delta G[\Gamma^{k+1},\mathbf{R}^{k+1}]=G[\Gamma^{k+1},\mathbf{R}^{k+1}]-G[\Gamma^{k+1},\mathbf{R}^{k}]$. 
Set $\delta G^{*}=5\%G[\Gamma^{k+1},\mathbf{R}^{k}]$.
\item[Step 3.4.] Check the total free energy change:

$\mathbf{if}$  ($\delta G[\Gamma^{k+1},\mathbf{R}^{k+1}] > \delta G^{*})$~or~$(~N_{\rm cnt} \geq N_{\rm cnt}^* $ and $0<\delta G[\Gamma^{k+1}, \mathbf{R}^{k+1}] \leq \delta G^{*}$)

Put all moving solute atoms back to $\mathbf{R}^k$.

$M=\alpha M$, $N_{\rm cnt} =0$, go to Step 3.3. 

$\mathbf{else}$ \\
$N_{\rm cnt} =N_{\rm cnt} +1;$\\
$\mathbf{endif}$
\end{compactenum}
\item[Step 4.]
Calculate the absolute error  $\Delta G[\Gamma^{k+1},\mathbf{R}^{k+1}]_{\rm abs}$ and 
 relative error  $\Delta G[\Gamma^{k+1},\mathbf{R}^{k+1}]_{\rm rel}$.  

\item[Step 5.]

Test the convergence. If either $\Delta G[\Gamma^{k+1},\mathbf{R}^{k+1}]_{\rm rel}<\rm {Tol}_2$ or $\Delta G[\Gamma^{k+1},\mathbf{R}^{k+1}]_{\rm abs}<\rm {Tol}_2$  stays continuously for 100 steps, or if the number of iterations reaches to $k_{\rm max}$, stop the algorithm. Otherwise, go to Step 2. 

\end{compactenum}
 
We remark that there are two error tolerances and stopping criteria:  
 One is a small tolerance $1$e-$4$ for the gradient descent for updating
 the solute atomic positions.  The other stop criterion is a small tolerance of  relative difference or absolute difference of the total free energy. In our experiments, 
we set the relative difference stop criterion to be $1$e-$5$, and absolute difference stop criteria to be $1$e-$5$. To avoid the situation that  the free energy functional decreases slowly because of small mobility factor $M_{nl}$, we determine that the system stops only when the  relative difference stop criterion or the absolute  difference stop criterion is satisfied continuously for 100 steps. 
\section{GPU Implementation}
\label{s:GPU}

In this section, we discuss the parallel implementation of aspects of
our free-energy functional minimization algorithm for the fast execution
of our programs.

Parallel computing concerns strategies for performing simultaneous 
computations, usually through the use of multiple processors.  This approach
has become more and more important as the abilities of individual processors 
reach their limits under Moore's law.
The recent advent of the use of the graphics processing unit (GPU) for general
purpose parallel computing, instead of traditionally multiple central 
processing units (CPU's), has allowed for algorithms that can take advantage 
of its high throughput and hundreds or thousands of cores to achieve new 
heights in speed.  This has, for example, revolutionized the subject of deep 
learning in artificial intelligence.

We introduce parallel programming, using OpenCL, employing the GPU for 
operations in our free-energy functional minimization algorithm.  The 
operations that are particularly amenable to this kind of parallelization
are usually made up of a large number of smaller, simpler ones, to 
take advantage of the large number of cores in the GPU that, alternatively, 
must work in lockstep.  We find such operations in our computations of the 
LJ and CFA of the electrostatics in the VISM free energy 
\reff{G_VISM}, the solute-solute interaction energy \reff{Gss}, and all of 
their derivatives, combined in \reff{GradientG}. We separate the 
parallelization into two cases that are treated differently, one handling 
VISM LJ and electrostatic terms, and their derivatives, and the other 
handling solute-solute interaction terms and their derivatives.

For the first case, we begin by describing the procedure introduced in 
\cite{zhang2021coupling} for the LJ and CFA contributions, though 
in more general terms.  These contributions notably both contain integrals 
of the form
$$
\int_{\R^3\backslash\Omega_{\rm m}} f({\bf r})~dV_{\bf r},
$$
where $f$ has some complexity in the computation of its values; in the case of LJ, 
$
f({\bf r}) = \sum_{i = 1}^N U_{LJ}(|{\bf r} -{\bf r}_i|),
$
which requires some computation when the number of solute atoms is large.  
With far-field approximations handling the integral outside the computational 
box $\Omega$, numerical quadrature for the rest takes the general form
$$
\sum_{k\in I} \alpha_k f({\bf r}_k)\Delta {\bf x}_k,
$$
where ${\bf x}_k$ are grid points of a grid in $\Omega$, and for some 
$\alpha_k\in\R$.  This summation is computationally intensive as $f$ needs
to be evaluated over the grid; in fact, in our problem this needs to be 
performed each time step, when the atoms move.  A GPU parallel implementation 
is introduced in \cite{zhang2021coupling} to handle the evaluation of $f$ over the 
grid which parallelizes over the grid points, passing out the computation
of $f$ at each to the cores.  This works especially well because there of
the large number of grid points, typically hundreds of thousands or millions, 
the GPU cores can work on.  Note, in the parlance of OpenCL, the grid points 
form the work-items, which are instead known as threads under CUDA.  

We apply this idea here not only to LJ and electrostatic terms, 
but also to their derivative terms found in the gradient of the VISM free-energy 
\reff{GradientG}.  These terms also have integrands that grow in complexity 
with the number of solute atoms, thus slowing down computations in the case 
of large numbers of moving atoms. Thus, the same parallelization techniques 
can be adopted to improve runtimes.

For the solute-solute interaction terms and their derivatives, no integration
is present and no grid points are involved.  Instead, all terms involve a
summation of some interaction between solute atoms.  Consider, for example, 
a system's solute-solute LJ interactions: 
$$
\sum_{1\leq i\leq N}\sum_{j\neq i} U_{LJ}(|{\bf r}_i-{\bf r}_j|).
$$
For our parallel implementation, we consider separately the terms
$$
\sum_{j\neq i} U_{LJ}(|{\bf r}_i-{\bf r}_j|),
$$
and parallelize by passing out these computations for each $1\leq i\leq N$ 
to the cores.  Note, however, there are far fewer solute atoms, typically
in the hundreds or thousands, compared to the hundreds of thousands or
millions of grid points.  Thus, the parallelization may not be as efficient 
in comparison with that of our first case.  

One additional note is that while double-precision machine numbers and their
arithmetic are commonly available in CPU architectures, they are not
universally supported on GPU's, where, for traditional graphical purposes,
single-precision has been adequate.  And though more and more GPU architectures
now do support double-precision, due to the expansion of GPU's for general
purpose computing, single-precision and even half-precision arithmetic
operations are still used for faster calculations.  The drawback in the use
of single-precision instead of double-precision is in increased round-off
error.  This especially is of concern when performing a large number of
operations, where round-off errors can accumulate to intolerable levels.
In our application, we find such large numbers of operations in our sums,
with sums over solute atoms, which can be in the thousands, and over grid
points, which can be in the millions.  For our sums, we adopt the strategy
of summing-by-pairs \cite{watkins2004fundamentals}, a binary tree-based approach to order the
operations in such a way as to reduce round-off error.  In our applications,
we find this to result in a nearly negligible amount of error compared to
double-precision results, allowing us to take advantage of the speed afforded
by single-precision computations.  In future work, we may consider computing
with mixed-precision machine numbers to better balance round-off error and
speed.

As we shall show below (cf.\ Tables ~\ref{table:two_plates}, \ref{table:BphC}, 
and \ref{table:p53_MDM2_rigid_body}),  
the combined results of our choices in parallelization and operation orders 
for sums for single-precision arithmetic 
significantly improve in runtimes even in comparison to 
a ported CPU parallelization, where the program for parallelization using the 
GPU instead uses available CPU cores.  In addition, the table reveals that
there are few negative effects in our use of single-precision machine numbers
rather than double-precision.  Our resulting parallel GPU implementation serves 
as the linchpin of our computations, as without it, we would not be able to 
obtain results in any reasonable amount of time due to the requirements of 
the moving atoms.


\section{Numerical Experiments and Applications}
\label{sec:numericalexperiments}

We first apply the binary level-set method and its GPU implementation to 
two molecular systems with 
fixed solute particles, a system of two parallel charged plates, and the protein BphC, 
and show that the binary level-set method is accurate in qualitatively reproducing the known
results of those two systems. 
We then consider the full application of our model and numerical methods to 
two small molecular systems,  a two-particle system and the 
ethane molecule, to show the convergence of our algorithm. 
Finally, we study the p53-MDM2 binding process with solute mechanical interactions
to demonstrate the efficiency of our methods and GPU implementation. 
Table~\ref{t:VISMparameters} summarizes the continuum
model parameters used in all these numerical computations. 

\vspace{-2 mm}

\begin{table}[htbp]
\begin{center}
\caption{Model parameters.}

\vspace{-4 mm}

\label{t:VISMparameters}
\medskip
    \begin{tabular}{| c | c | c | c | }
    \hline
    \hline
  Parameter  & Symbol & Value & Unit \\
    \hline
    \hline
temperature            &  $T$             & $298$  & K \\
solvent number density &  $\rho_{\rm w} $        & $0.0333$     & \AA$^{-3}$   \\
surface tension        &  $\gamma_0 $     & $0.174$  & ${\rm k}_{\rm B} {\rm T} /\mbox{{\AA}}^2$ \\
solute dielectric constant  &   $\varepsilon_{\rm m}$ & $1$ & \\
solvent dielectric constant &  $\varepsilon_{\rm w}$ & $80$ & \\
    \hline
    \hline
    \end{tabular}
\end{center}
\end{table}


\subsection{Free-energy minimization with fixed solute atoms}
\label{ss:TwoPlatesBphC}

We consider two molecular systems each with fixed solute atomic positions, and apply the
binary level-set method with GPU implementation to minimize the solvation free-energy
functional \reff{G_VISM} of solute-solvent interfaces $\Gamma$  
with all atomic positions $\br_i$ $(1 \le i \le N)$ fixed. 
Both systems have been studied extensively with continuous level-set method and CPU computations 
\cite{WangEtal_VISMCFA_JCTC12,Zhou_VISMPB_JCTC2014,Zhou_StochLSM_JCP2015}. 
Here we show the qualitative accuracy, and efficiency, of our new algorithm and implementation. 

\medskip

\noindent
{\bf Two parallel charged plates.} 
Each of these two plates consists of $ 6 \times 6$ $CH_2$ atoms with
a square length of about 3 nm.   The two plates are placed at a center-to-center distance $d$.  
In the following, we investigate how (a) the capillary evaporation, (b) the hydrophobic attraction, and (c) a possible hysteresis in the free energy are affected by charging up the plates. To this end, we assign central charges $q_1$ and $q_2$ to the first and second plates, respectively, with $|q_1|=|q_2|$. The total charges of these two plates are $36q_1$ and $36q_2$, respectively. We study like-charged and oppositely charged plates by choosing the values of $(q_1,q_2)$ to $(+0.2e, +0.2e)$, and $(+0.2e, -0.2e)$. 
The atom-water LJ parameters are 
$\varepsilon=0.262 \, k_BT$, $\sigma=3.15365 \, \mathring{A}$, and the atom-atom  
LJ  are $\varepsilon=0.265 \, k_BT$ and  $\sigma=3.54\,  \mathring{A}$.

\begin{figure}[H]
	\centering 
\includegraphics[width=1.5 in, height=1.1 in]{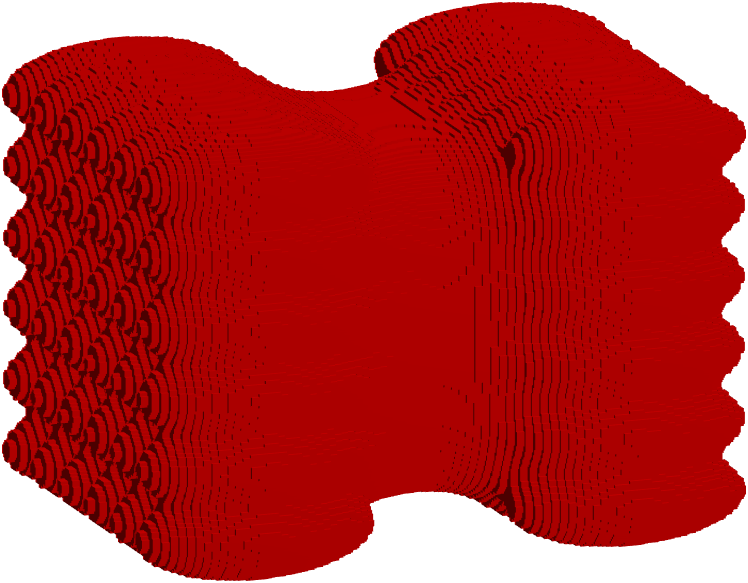}
\hspace{10 mm}
\includegraphics[width=1.5 in, height=1.1 in]{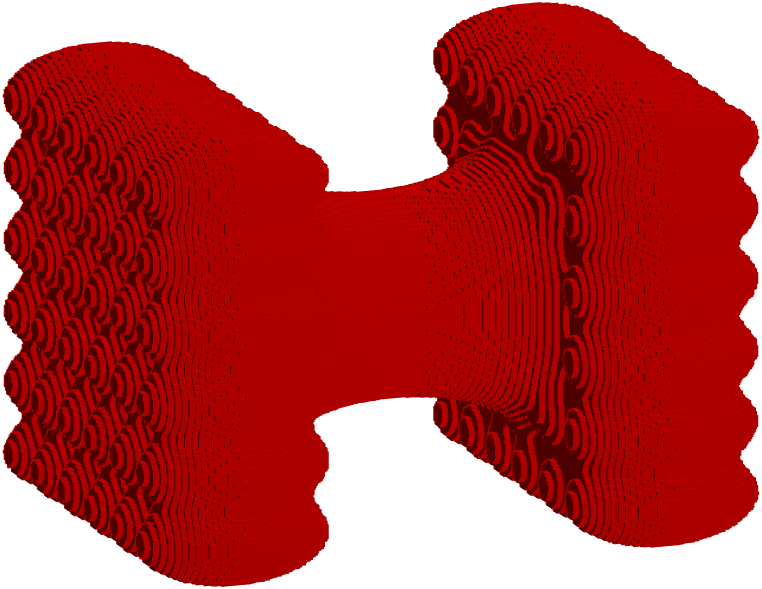}
\hspace{10 mm}
\includegraphics[width=1.5 in, height=1.1 in]{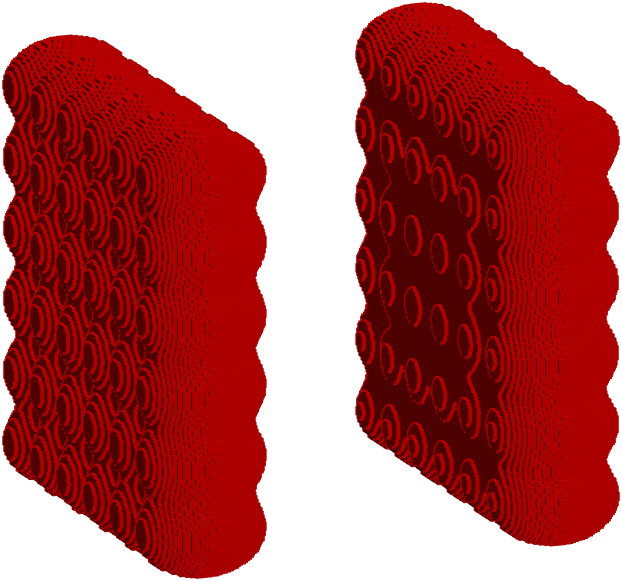}

\hspace{4 mm}
$d = 9$ \AA \hspace{2 mm} loose 
\hspace{20 mm}
$d = 13$ \AA \hspace{2 mm} loose 
\hspace{20 mm}
$d = 16$ \AA \hspace{2 mm} loose 

\vspace{2 mm}

\includegraphics[width=1.5 in, height=1.1 in]{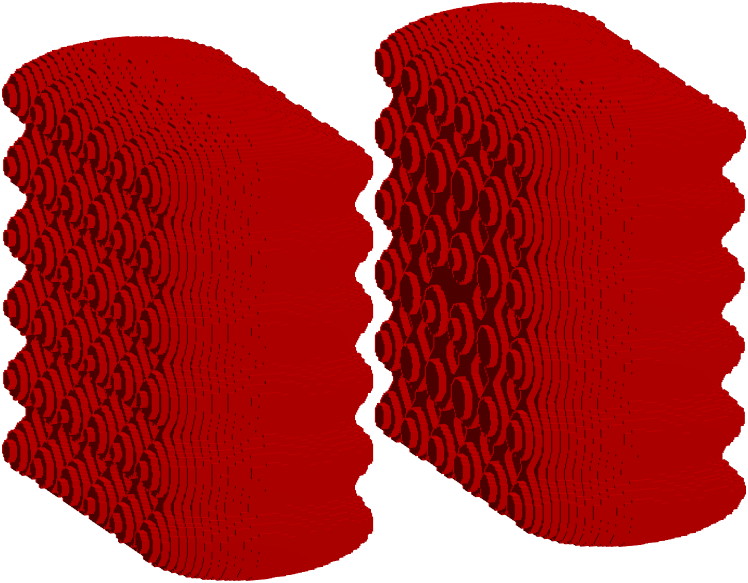}
\hspace{10 mm}
\includegraphics[width=1.5 in, height=1.1 in]{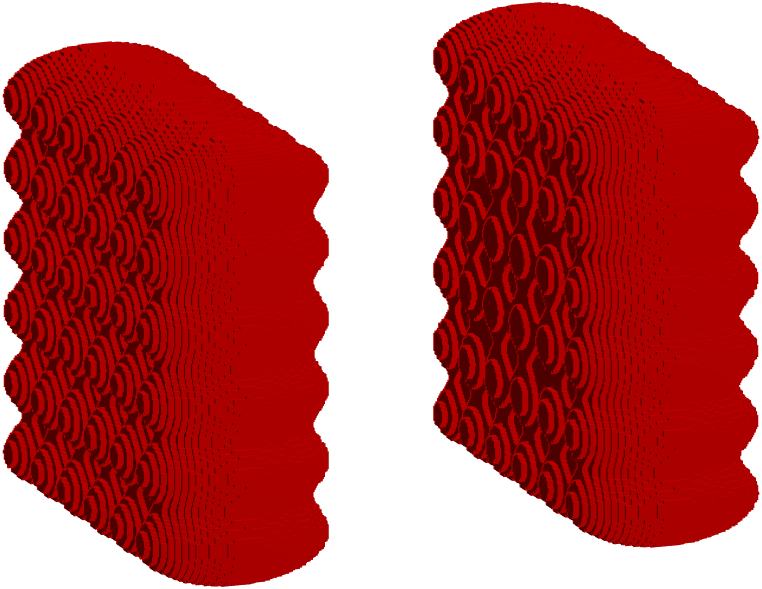}
\hspace{10 mm}
\includegraphics[width=1.5 in, height=1.1 in]{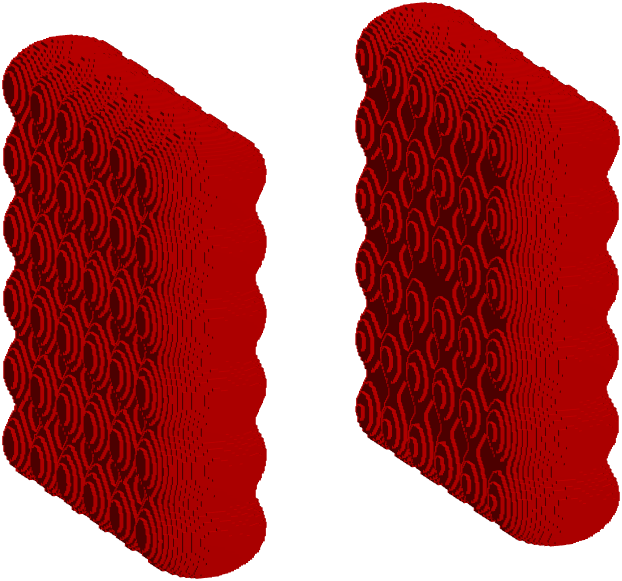}

\hspace{4 mm}
$d = 9$ \AA \hspace{2 mm} tight
\hspace{20 mm}
$d = 13$ \AA \hspace{2 mm} tight 
\hspace{20 mm}
$d = 16$ \AA \hspace{2 mm} tight 
	\caption{Stable 3D equilibrium solute-solvent surfaces of the two parallel charged plates 
obtained by the binary level set VISM calculations with loose (top row) or tight (bottom row) initial interface at $d=9\, \mathring{A}$, $d=13\, \mathring{A} $, and $d=16\, \mathring{A}$. Atomic 
charges are (+0.2e,+0,2e).
}
	\label{fig:two_plates_wet_dry}
\end{figure}

We first investigate  the VISM surfaces of  the two plates at different distances with the like-charge (+0.2e, +0.2e) with different initial configurations.  Figure~\ref{fig:two_plates_wet_dry}
shows a few snapshots of stable 3D equilibrium solute-solvent surfaces of the two parallel charged plates system obtained by the binary level set VISM calculations with loose or tight initial interface at $d=9 \, \mathring{A}$, $d=13 \, \mathring{A}$, and $d=16\, \mathring{A}$. 
In the top row of Figure \ref{fig:two_plates_wet_dry}, with the loose initial interface, a stable capillary bubble remains between the two charged parallel plates at $d=9\, \mathring{A}$ and $d=13\, \mathring{A}$, and the bubble becomes tighter along the enlarging distance. At $d=16 \, \mathring{A} $, the bubble disappears. Comparatively, with the tight initial interface, the equilibrium state is wet at $d=9\, \mathring{A}$, $d=13\, \mathring{A}$, and $d=16\, \mathring{A}$.

We now examine the potential of mean force (PMF) of the two-plate system 
with respect to the plate-plate separation distance $d$. 
This is the VISM free-energy value as a function of $d$, with an additive constant
such that the free energy is $0$ at the infinite plate-plate separation.  
For a given $d$, we may have two VISM free-energy minimizing solute-solvent interfaces
corresponding to a tight and a loose initial surface, respectively. 
We denote by $G_{\rm VISM}^{\rm pmf}(d) $ 
the corresponding minimum free energy of one of the two branches, 
and denote by $G_{\rm geom}^{\rm pmf}(d), $ $G_{\rm vdW}^{\rm pmf}(d), $ 
and $G_{\rm elec}^{\rm pmf}(d)$ the components of the PMF, corresponding to 
the first, second, and third terms in \reff{G_VISM}, respectively.
Precise definition is given in Appendix. 

Figure \ref{fig:two_plates_energy} displays  
the bimodal behavior and hysteresis of the 
two different PMF branches stemming from the equilibria of wet and dry states, 
i.e., the VISM free-energy minimizing surfaces corresponding to initial 
tight and loose surfaces. 
Atomic charges considered here are (+0.2e,-0,2e) and (+0,2e, +0,2e), respectively. 
We can see that like-charged and oppositely charged plates give different free-energy branches and hysteresis. For the like-charged cases in Figure \ref{fig:two_plates_energy}, a strong hysteresis is presented for $8 \lesssim d \lesssim 15 \mathring{A}$. For the oppositely charged plates, strong hysteresis is presented for $6 \lesssim d \lesssim 8 \mathring{A}$.

\begin{figure}[H]
	\centering 
	\includegraphics[width=3.3 in, height=2 in]{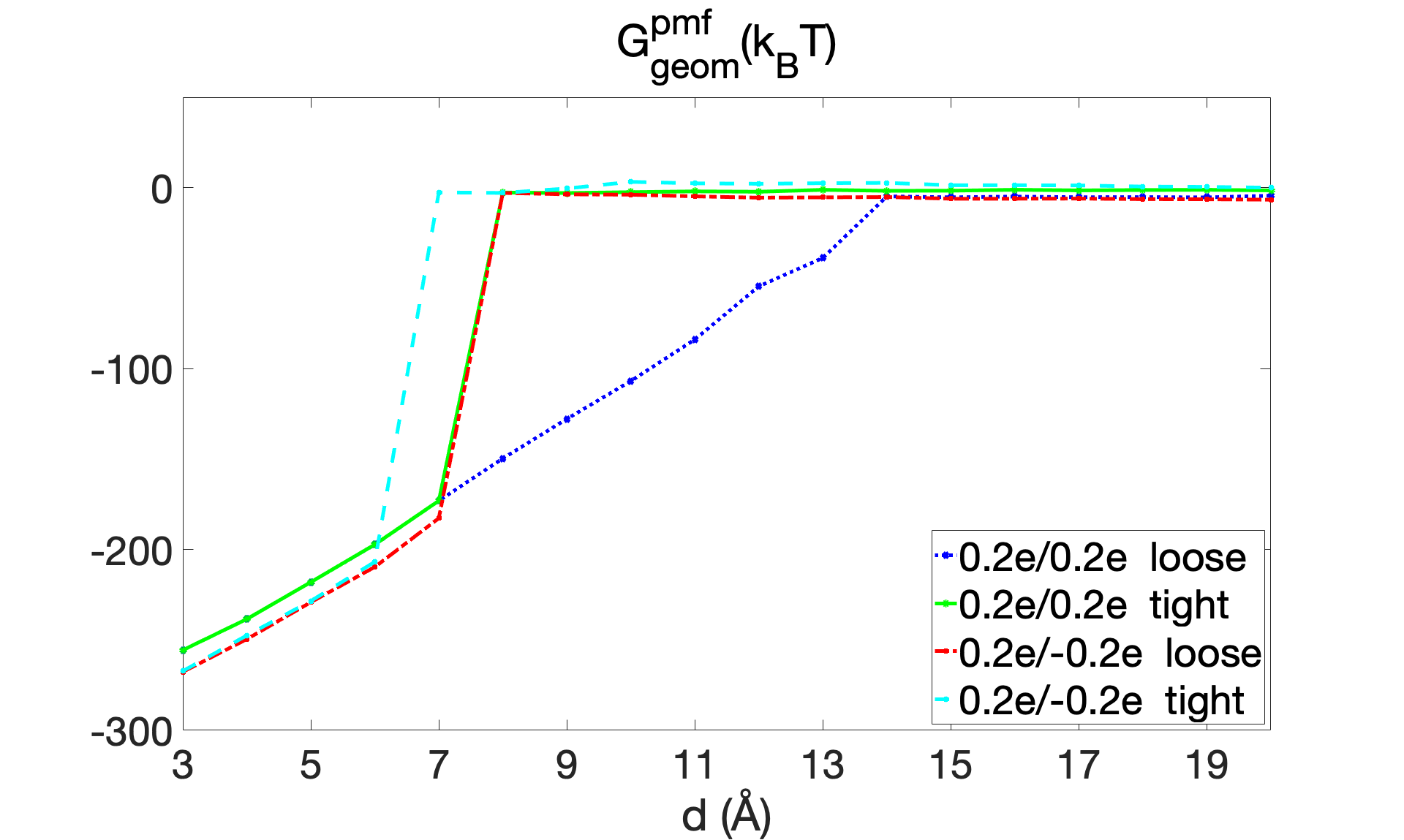}
\hspace{-8 mm}
	\includegraphics[width=3.3 in, height = 2 in]{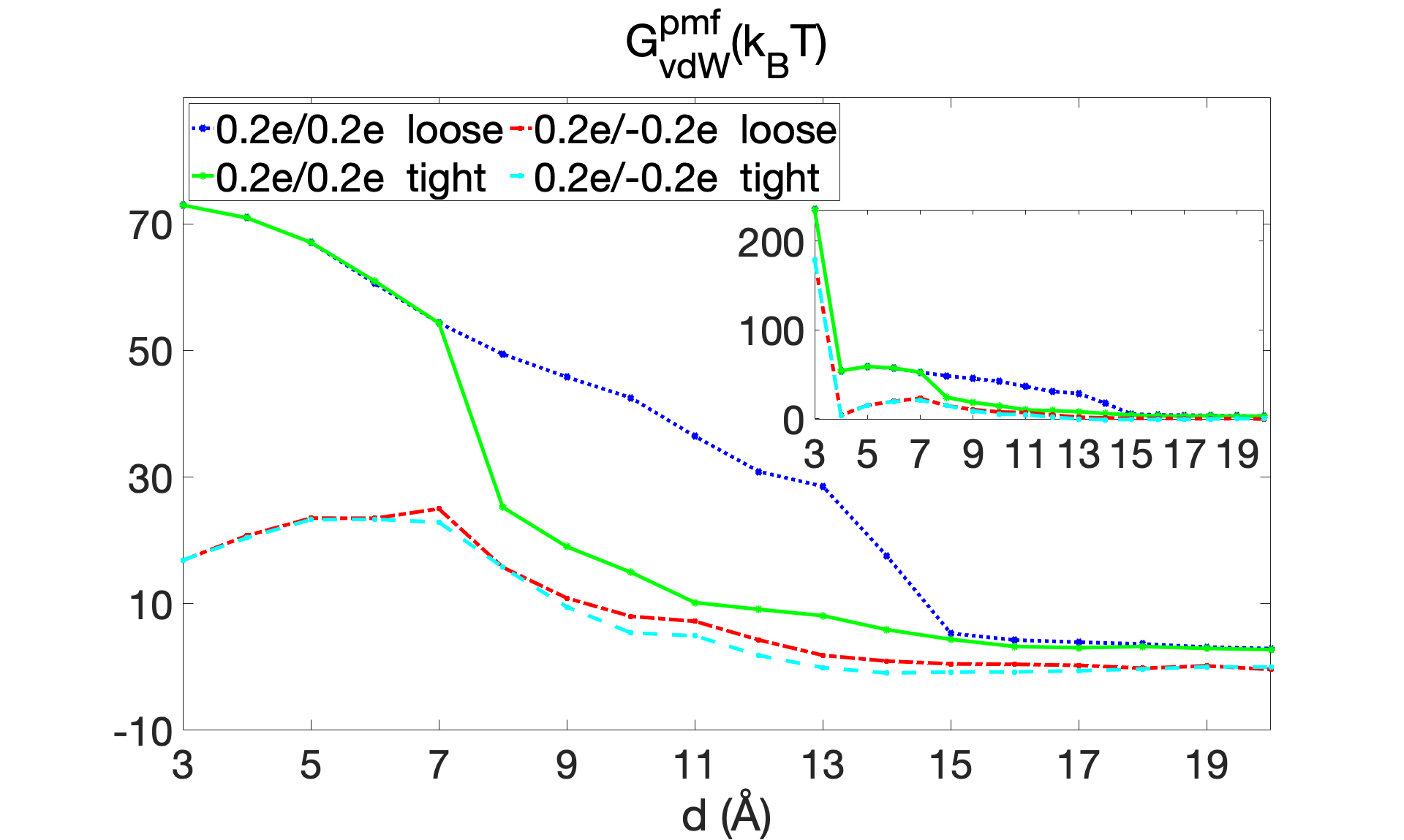}
	\includegraphics[width=3.3 in, height=2  in]{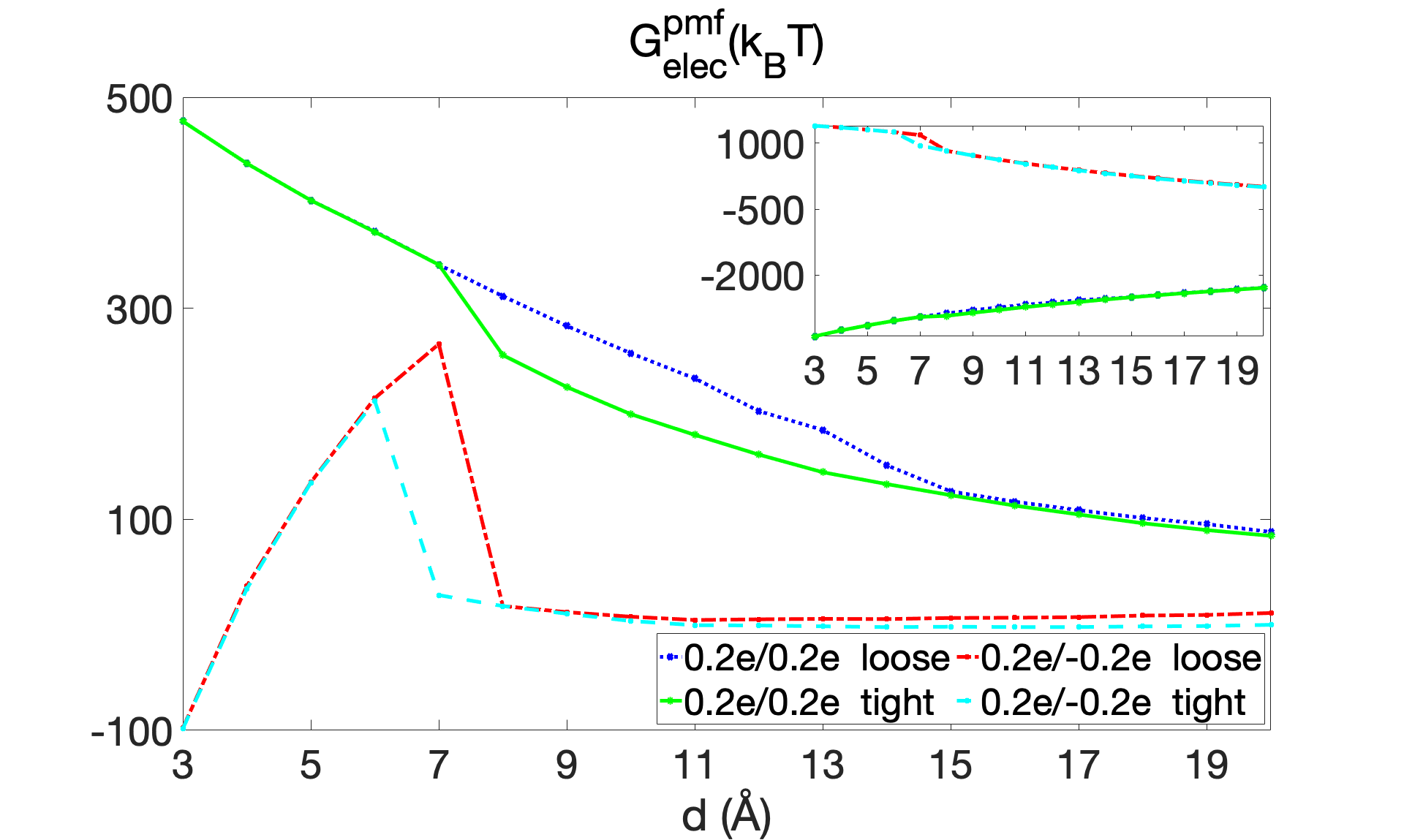}
\hspace{-8 mm}
	\includegraphics[width=3.3 in, height=2 in]{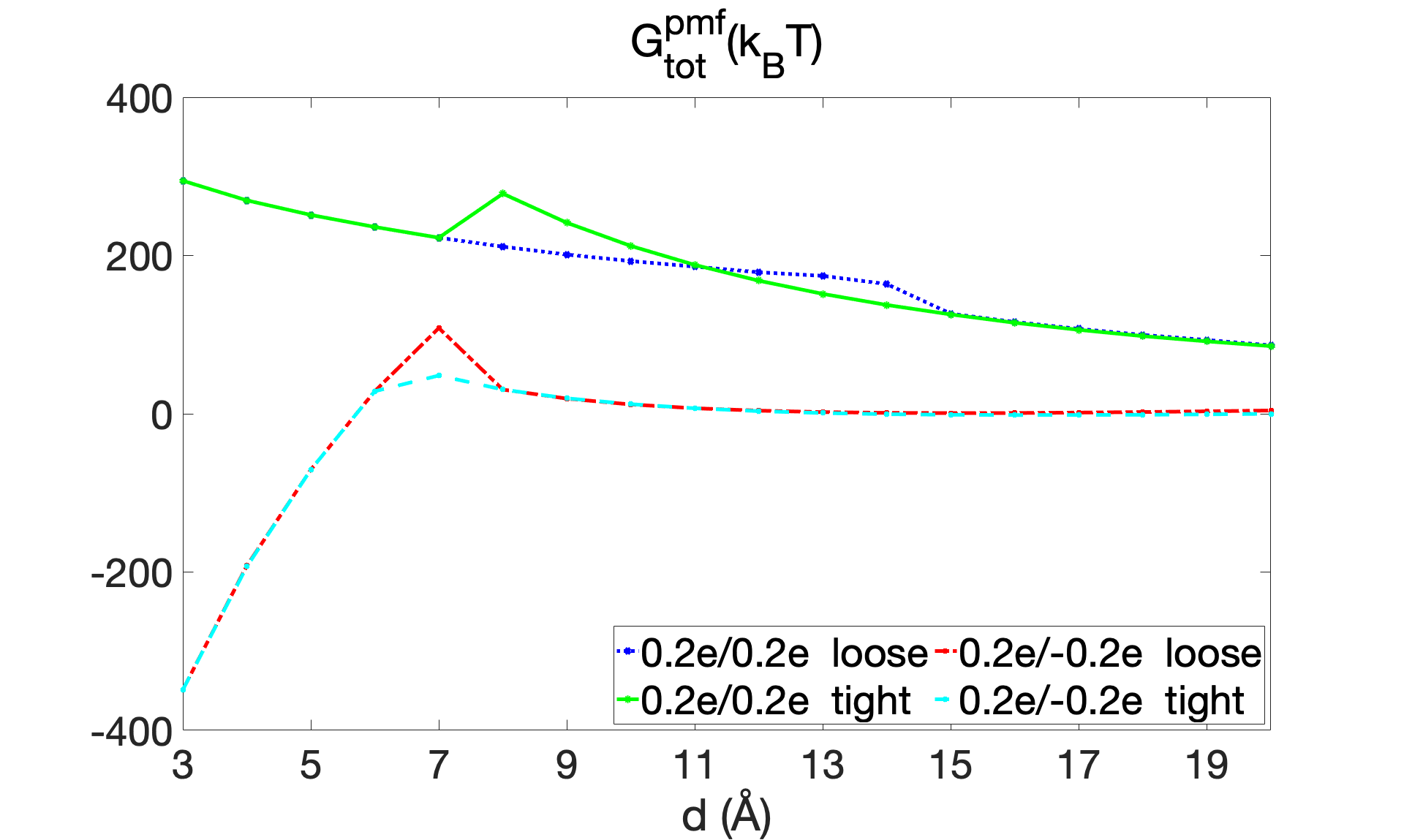}
	\caption{
Different parts of the PMF of the two parallel charged plates with respect to the 
separation distance $d$, with loose and tight initial surfaces. 
(a) The geometrical part $G_{\rm geom}^{\rm pmf}$. (b) The vdW part $G_{\rm vdW}^{\rm pmf}$. 
The solute-solute vdW interactions are excluded in the curves in the main frame but 
included in those in the inset. (c) The electrostatic part $G_{\rm elec}^{\rm pmf}$. 
The solute charge-charge interactions are excluded in the curves in the main frame but 
included in those in the inset. (d) The total PMF $G_{\rm tot}^{\rm pmf}$. The values of $(+0.2e,-0,2e)$ with tight initial interface are used as reference values to show the difference.}
	\label{fig:two_plates_energy}
\end{figure}

In Table \ref{table:two_plates}, we show a comparison of the calculation speed and different parts of the free energy   using the binary level-set method  between GPU single precision code and CPU double precision code of the two charged parallel plates system. Three different grid sizes are shown. We can observe that the  results for free-energy estimates from CPU double precision code and GPU single precision code are nearly the same. The improvement of speed using GPU can be obtained by comparing the time.  The cost time of CPU code is around 5 times of the GPU code with three different grid sizes.

\begin{table}[h]
\caption{Comparison of GPU (single precision) and CPU (double precision) for free energy ($k_BT$) and its components of two parallel charged plates with different charges $(+0.2e,-0,2e)$  at distance $d=10\mathring{A}$. The unit of time is second.}
	\label{table:two_plates}
	\centering
{\small 
\begin{tabular}{|c|c|c|c|c|c|c|c|c|c|c| }
\hline
Grid 
&  \multicolumn{2}{c|}{Total Energy} &\multicolumn{2}{c|}{Surface Energy} &\multicolumn{2}{c|}{vdW Energy} & \multicolumn{2}{c|}{CFA} & \multicolumn{2}{c|}{Total Time}\\
 \cline{2-11} 
Points			& GPU& CPU & GPU& CPU& GPU& CPU & GPU& CPU & GPU& CPU  \\
\hline
			$72^3$&-2099.5&-2099.5&631.8 &631.8 &-98.3&-98.3&-2632.9& -2633.0 &0.6 &3.0 \\
\hline
			$144^3$&-2090.8&-2090.8&640.3&640.3&-97.9&-97.9&-2633.1&-2633.2 &4.0&21.2\\
\hline
			$288^3$&-2082.0&-2082.0	&648.3&648.3&-110.4&-110.4&-2619.9&-2619.9&29.9&163.9\\
\hline
\end{tabular}
}

\end{table}

\medskip

\noindent
{\bf The protein BphC.}
In this example, we consider  biphenyl-2, 3-diol-1, 2-dioxygenase (BphC), an enzyme protein
(PDB code: 1dhy). 

The functional unit of this protein is a homo-octamer, and each subunit consists of two domains. We set up a series of configurations where the two domains are increasingly separated from $d=0$ to $d=20\, \mathring{A}$ apart, perpendicular to their interface. The domain separation $d$ is chosen here to be the reaction coordinate. Note that the zero domain separation corresponds to the native configuration in the crystal structure.

\begin{figure}[!bp]
\centering 
\includegraphics[width=1.2 in, height=1 in]{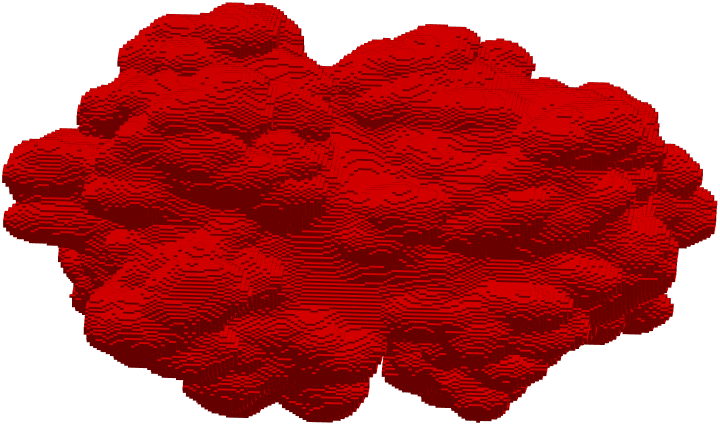}
\hspace{10 mm}
\includegraphics[width=1.2 in, height=1 in]{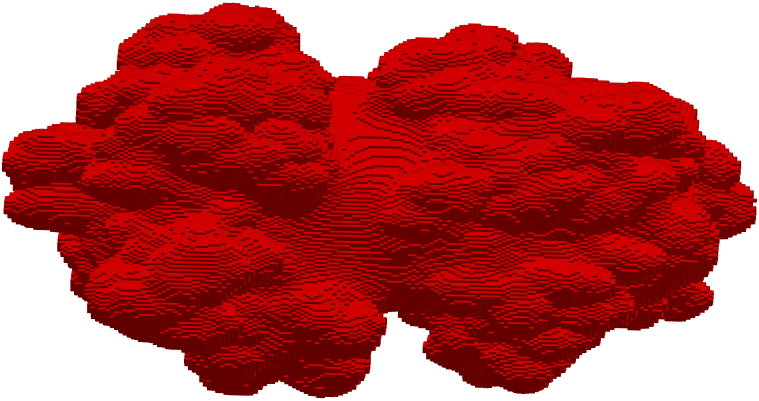}
\hspace{10 mm}
\includegraphics[width=1.2 in, height=1.0 in]{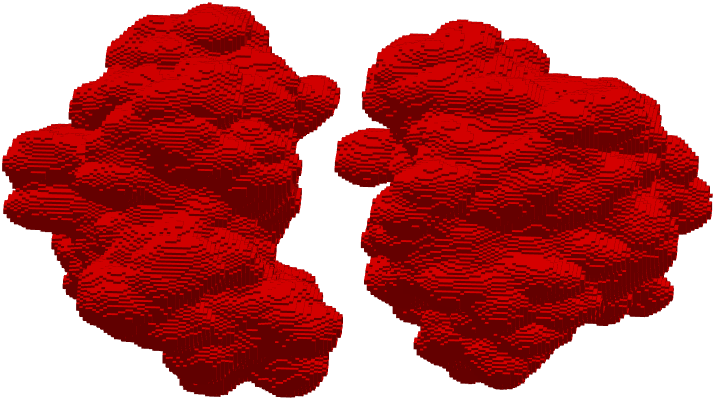}
		
\hspace{4 mm} {$d=8 \mathring{A}$~loose}
\hspace{16 mm} {$d=12 \mathring{A}$~loose} 
\hspace{16 mm} {$d=16 \mathring{A}$~loose}

\vspace{2 mm}

	\includegraphics[width=1.2 in, height=1 in]{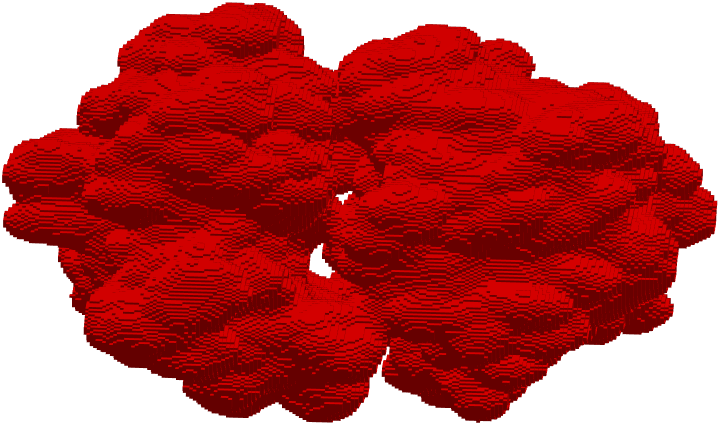}
\hspace{10 mm}
	\includegraphics[width=1.2 in, height=1 in]{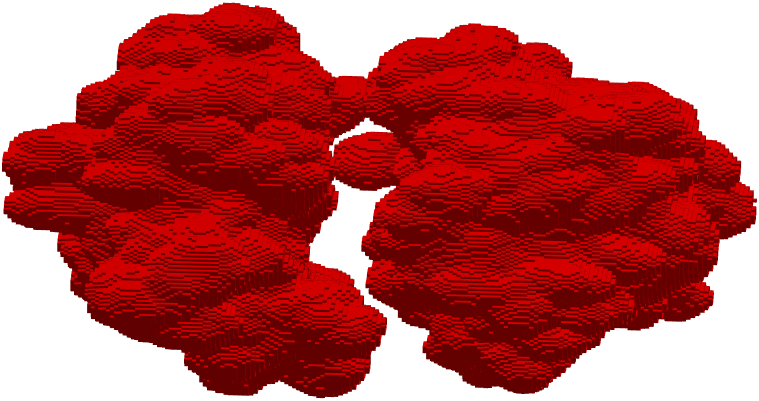}
\hspace{10 mm}
	\includegraphics[width=1.2 in, height=1 in] {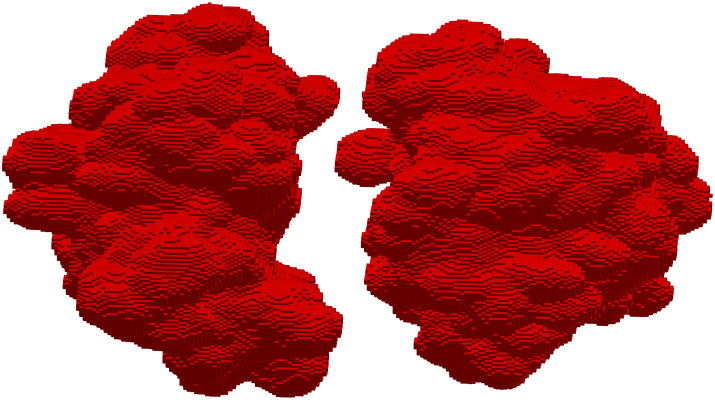}
			
\hspace{4 mm} {$d=8 \mathring{A}$~tight}
\hspace{16 mm} {$d=12 \mathring{A}$~tight}
\hspace{16 mm} {$d=16 \mathring{A}$~tight}
	\caption{Stable 3D equilibrium solute-solvent surfaces of the BphC system obtained by the binary level set VISM calculations with loose (top) or tight (bottome) 
initial interface at $d=8\, \mathring{A}$, $d=12\, \mathring{A}$, and $d=16\, \mathring{A}$. 
}
	\label{fig:BphC}
\end{figure}

Three pairs of stable equilibrium solute-solvent interfaces of BphC at $d=8\, \mathring{A}$, $d=12\, \mathring{A}$, and $d=16\, \mathring{A}$  with tight or loose initial interfaces are presented in Figure \ref{fig:BphC}. The top row is with the loose initial interfaces, and the bottom row is with the tight initial interfaces.  We observe that the equilibria of the loose initial interface wrap the two domains of BphC at $d=8\, \mathring{A}$ and $d=12\, \mathring{A}$, and the equilibria surface becomes tighter along the increasing distance.  At $d=16\, \mathring{A}$, the interfaces of two domains separate, changing to the wet state.   In contract, with the tight initial interface, all equilibria states are wet. 

\begin{figure}[H]
	\centering 
	\includegraphics[width=3.3 in, height=2 in]{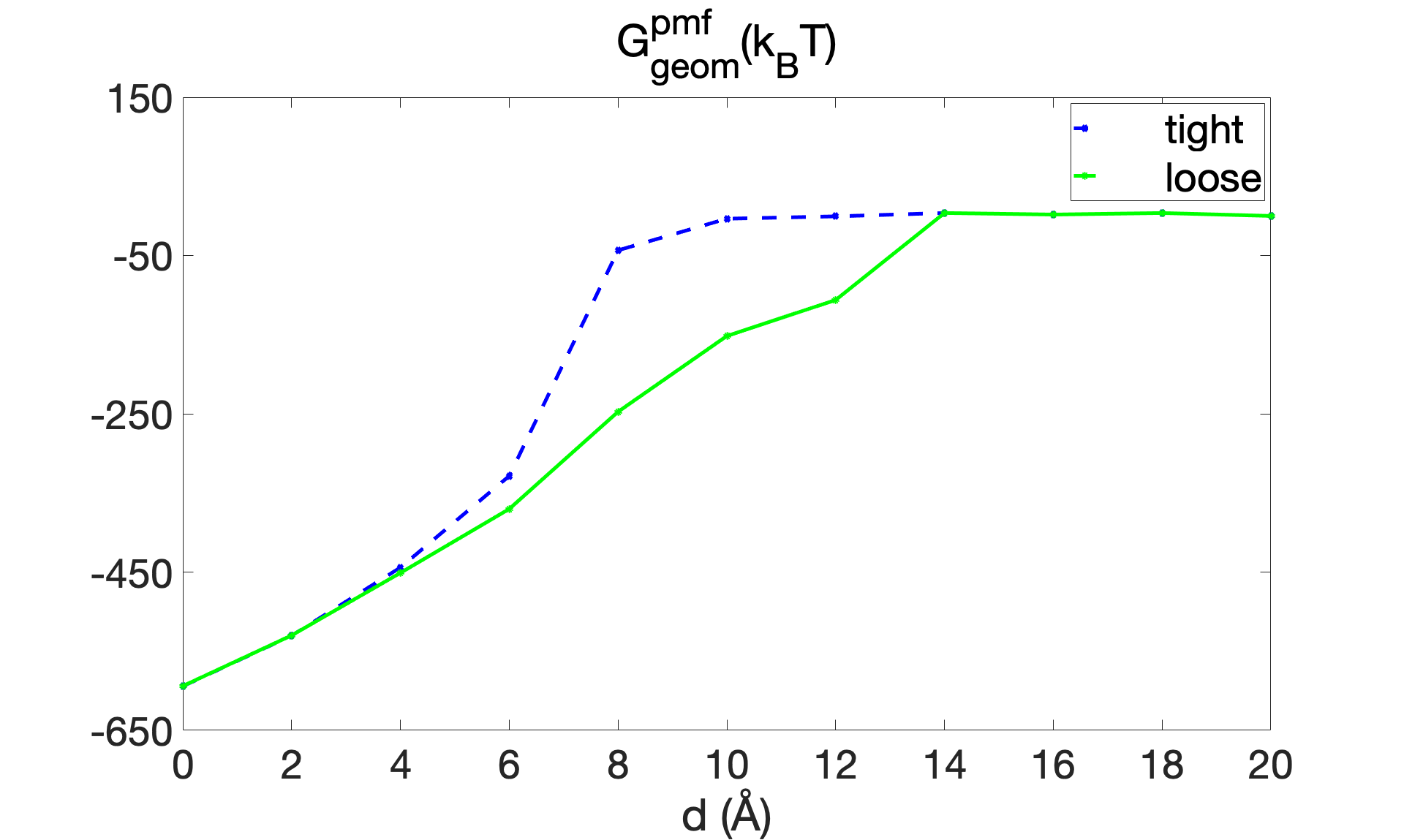}
	\hspace{-8 mm}
	\includegraphics[width=3.3 in, height = 2 in]{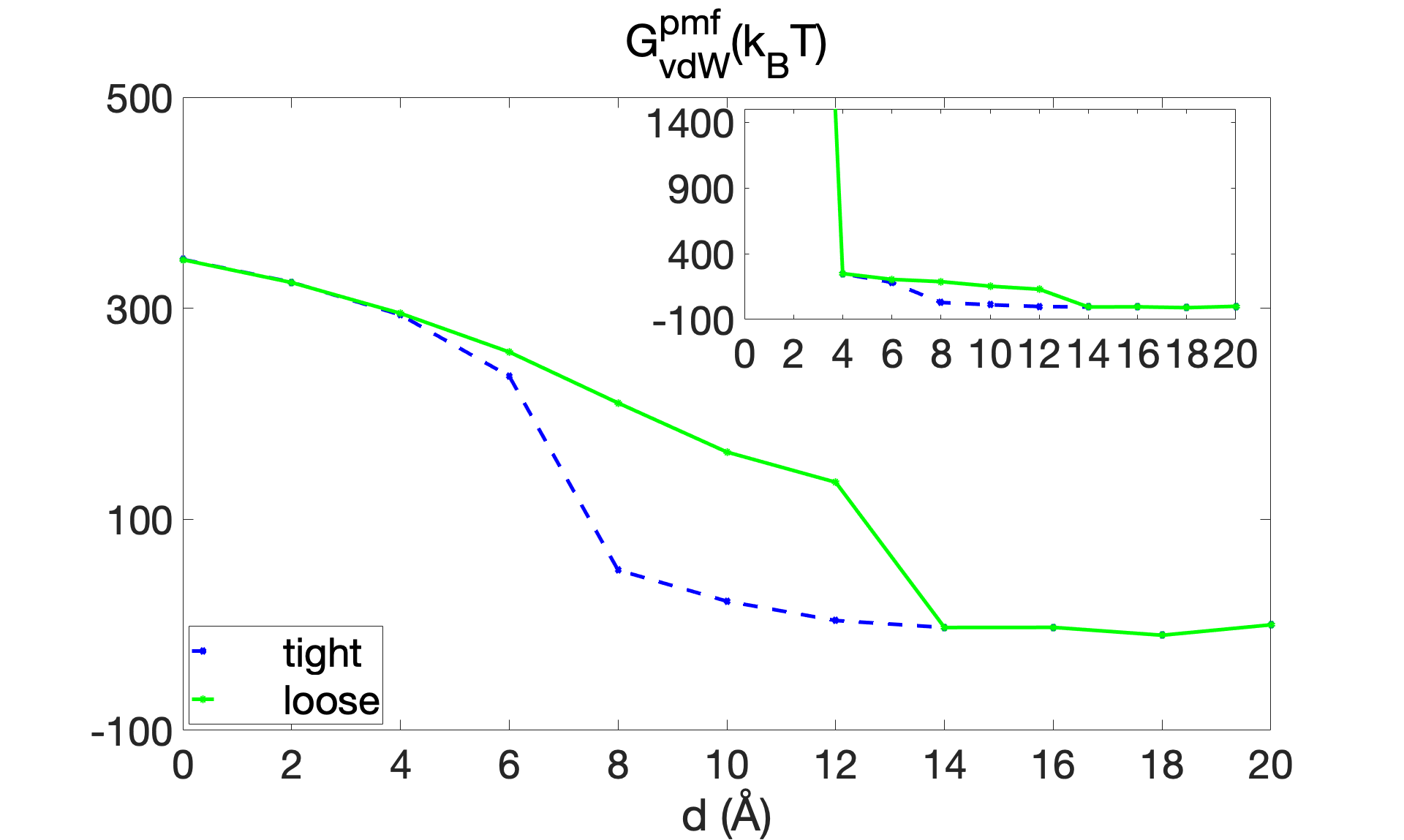}
	\includegraphics[width=3.3 in, height=2  in]{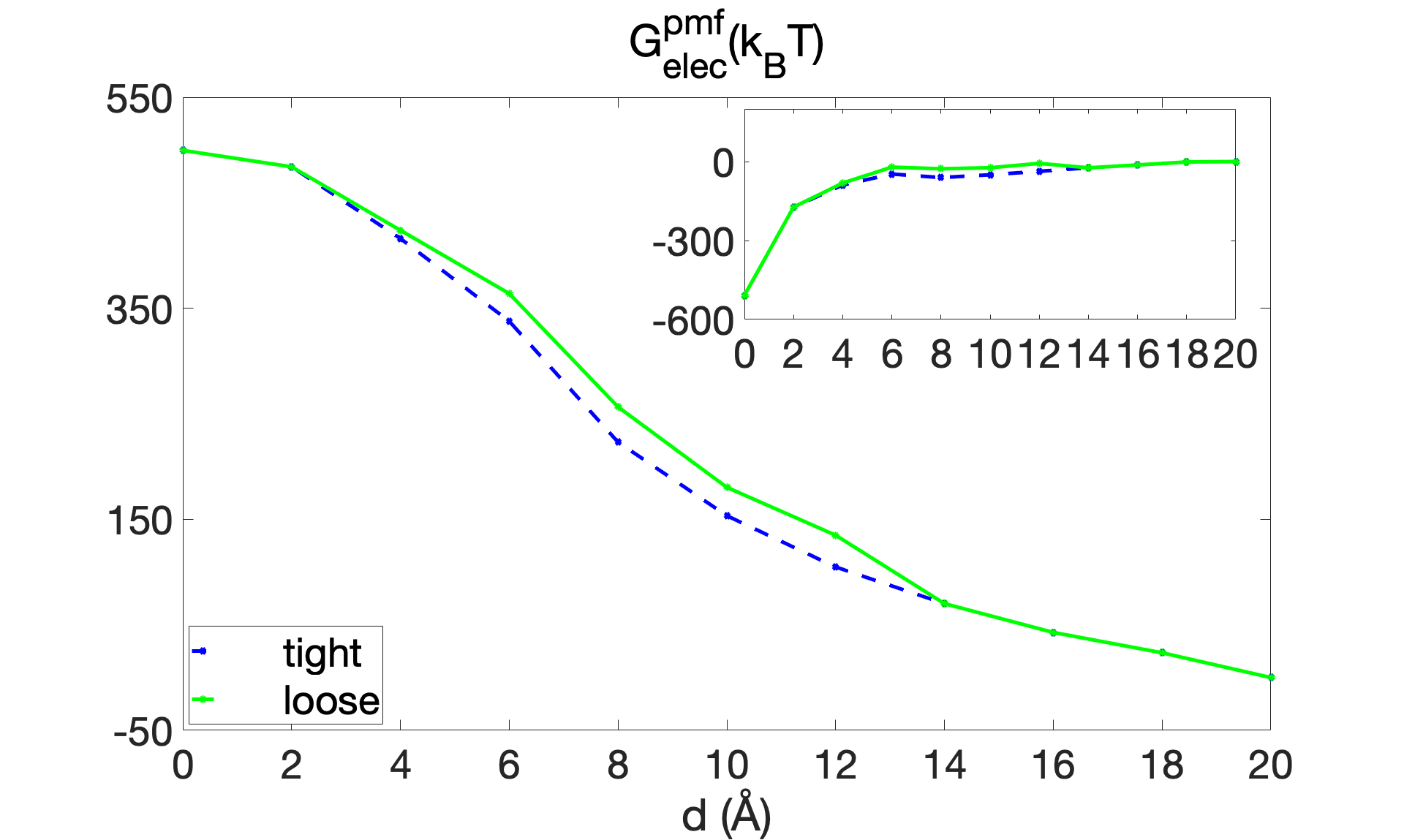}
	\hspace{-8 mm}
	\includegraphics[width=3.3 in, height=2 in]{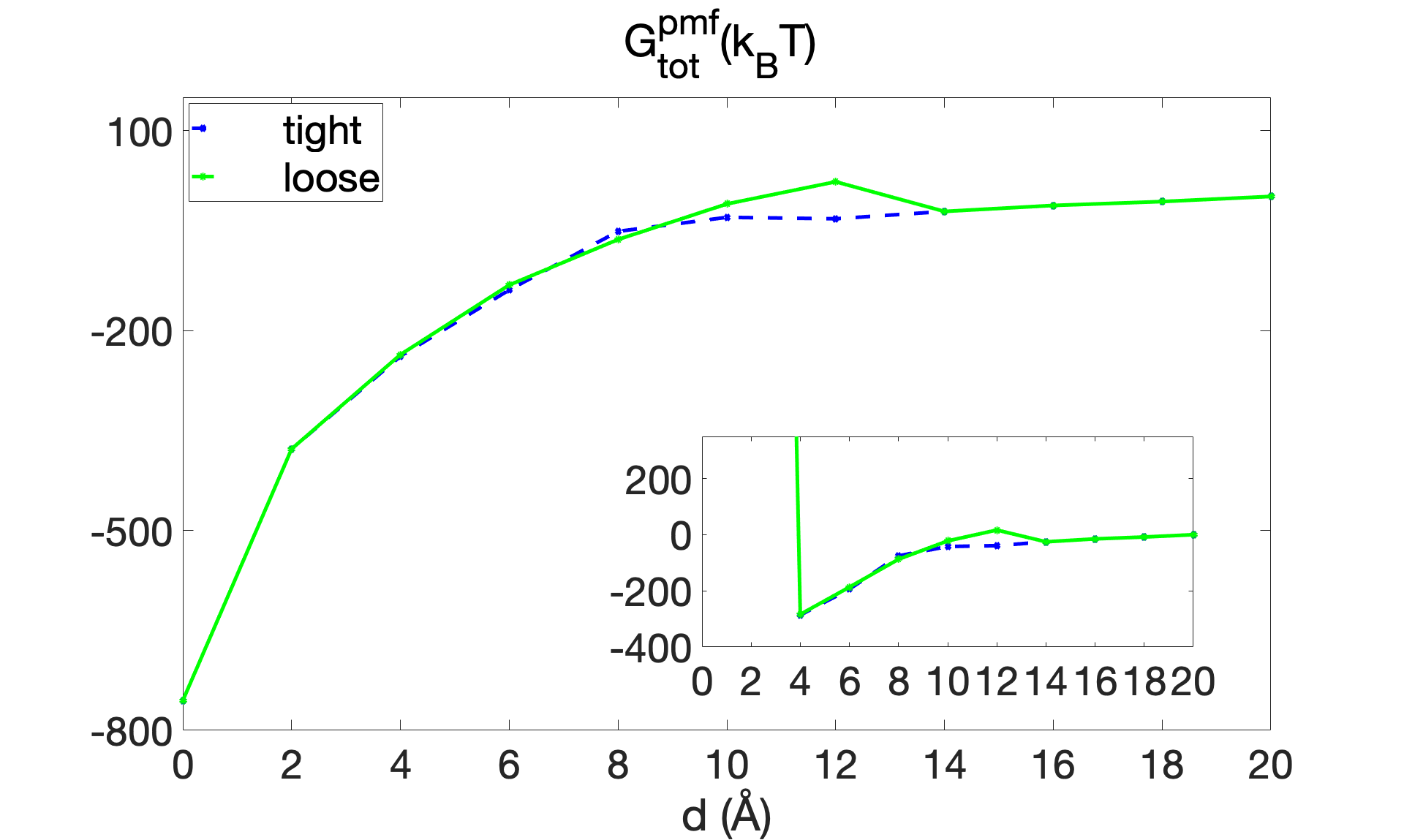}
		\caption{Different parts of the PMF of BphC with respect to the domain separations, with loose and tight initial surfaces. (a) The geometrical part $G_{\rm geom}^{\rm pmf}$. (b) The vdW part $G_{\rm vdW}^{\rm pmf}$. The solute-solute vdW interactions are excluded in the curves in the main frame but included in those in the inset. (c) The electrostatic part $G_{\rm elec}^{\rm pmf}$. 
The solute  charge-charge  interactions are excluded in the curves in the main frame but included in those in the inset. (d) The total PMF $G_{\rm tot}^{\rm pmf}$. The solute-solute vdW interactions are excluded in the curves in the main frame but included in those in the inset.}
	\label{fig:BphC_energy}
\end{figure}

In Figure \ref{fig:BphC_energy}, different parts of the PMF of BphC with respect to the separation of two domains, from $d=0$ to $d=20 \mathring{A}$ obtained by our binary level set calculations using tight and loose initial surfaces are displayed.  These PMFs exclude the solute-solute vdW interactions. We observe the bimodal hydration behavior: the branches of different parts of the PMF of BphC between $4\, \mathring{A}$ and $14 \, \mathring{A}$,   indicating that initial interfaces can strongly affect the PMF of BphC.

In Table \ref{table:BphC}, we show a comparison of the calculation speed and different parts of the free energy  with  the binary level-set  method  between GPU single precision code and CPU double precision code of the BphC. Three different grid sizes are shown. We observe that the  results from CPU double precision code and GPU single precision code are nicely consistent. With the grid size of $50^3$,  $100^3$, and $200^3$, the time cost of CPU code is around 15 times, 78 times, and  216 times  correspondingly to the time cost of the GPU code.

\begin{table}[h]
	\caption{Comparison of GPU (single precision) and CPU (double precision) for free energy ($k_BT$) and its components of BphC with the native configuration in the crystal structure. The unit of time is second. }
    \label{table:BphC}
	\centering
    \resizebox{\columnwidth}{!}{%
		\begin{tabular}{|c|c|c|c|c|c|c|c|c|c|c| }
			\hline
			Grid 
			&  \multicolumn{2}{c|}{Total Energy} &\multicolumn{2}{c|}{Surface Energy} &\multicolumn{2}{c|}{vdW Energy} & \multicolumn{2}{c|}{CFA} & \multicolumn{2}{c|}{Total Time}\\
			\cline{2-11} 
			Points			& GPU& CPU & GPU& CPU& GPU& CPU & GPU& CPU & GPU& CPU  \\
			\hline
			$50^3$&52133.7&52134.2&1588.2&1588.2&53116.8&53117.4&-2571.2&-2571.3&4.1&60.6\\
			\hline
			$100^3$&52251.8&52252.3&1658.3&1658.3&53130.4&53131.0&-2537.0&-2537.0&4.9&381.1\\
			\hline
			$200^3$&52303.7&52304.2&1654.3&1654.3&53146.7&53147.3&-2497.3&-2497.3&13.5&2911.0\\
			\hline
		\end{tabular}
	}
\end{table}

\subsection{ Explicit-solute implicit-sovent free-energy minimization }
\label{ss:VESIS}

In this section, we conduct numerical experiments on two molecuales that are treated as nonpolar
systems (i.e., no charges) to demonstrate the efficiency of our free-energy  minimization algorithm.

\medskip

\noindent
{\bf A two-atom molecule.}
We consider an artificial molecular system of two atoms.  
The solute-water LJ parameters are $\sigma=3.57 \,  \mathring{A}$ and $\varepsilon=0.431 \, k_BT$.  We additionally assume that the two atoms are bonded, with the spring constant in the bond stretching energy $A=800\, k_BT/ \mathring{A}^{2}$. 
We set the computational box to be $(-8,8)^3 \, \mathring{A}^{3}$.  

We design two sets of experiments on the  optimization process and equilibria with different initial configurations. In the first set of experiments, Experiment 1.1.a and Experiment 1.1.b, we set 
the equilibrium bond length $r_0=3\,  \mathring{A}$. In the second set of experiments, Experiment 1.2.a and Experiment 1.2.b, we set the equilibrium bond length $r_0=8\, \mathring{A}$. In each set of experiments, we test two types of initial configurations. In Experiment 1.1.a and Experiment 1.2.a, 
we place initially the two solute atoms far away from each other so that their distance is much larger than the equilibrium bond length. We place the two solute atoms at positions $(7,0,0)$ and $(-7,0,0)$, respectively.  In Experiment 1.1.b and Experiment 1.2.b, we place initially the two solute atoms very close to each other so that their distance is smaller than the equilibrium distance. Specifically, we place the two solute atoms initially at positions $(0.5,0,0)$ and $(-0.5,0,0)$, respectively.

In Figure \ref{fig:two_atoms_Ex1}, the minimization processes of Experiment 1.1.a and Experiment 1.1.b are displayed. The red dots represent two atoms, and the blue segment represents the bond. In the top row of Figure \ref{fig:two_atoms_Ex1}, we can see that initially surface consists of two disconnected spheres, then two atoms get closer, and spheres merge, until the system reaches an equilibrium state. In the bottom row of  Figure \ref{fig:two_atoms_Ex1}, initially, the two atoms are very close to each other, then the atoms are pushed apart due to the force from strong bonding energy, the interface is moved  accordingly, and then the system reaches to an equilibrium state. 

We observe from Table~\ref{table:two_atoms_a} that 
that atoms have the exact same positions and  free energy in the equilibrium
from the two experiments 1.1.a and 1.1.b, indicating that the molecular system in the two simulations 
reached the same equilibrium.


\begin{center}
\begin{tabular}{cccc}
\hspace{-15mm}
\vspace{-2.2mm}
\includegraphics[width=2 in, height=1.5 in]{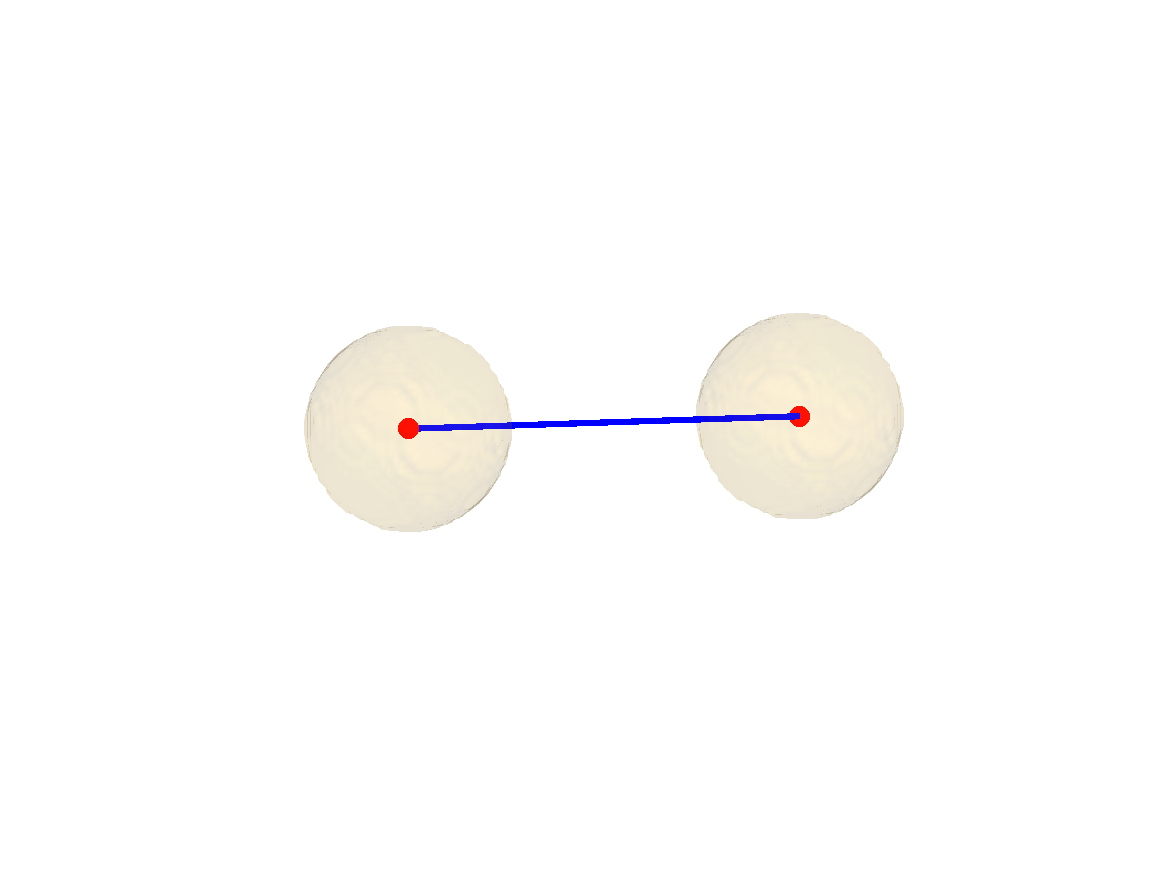} &
\hspace{-16mm}
\vspace{-2.2mm}
\includegraphics[width=2 in, height=1.5 in]{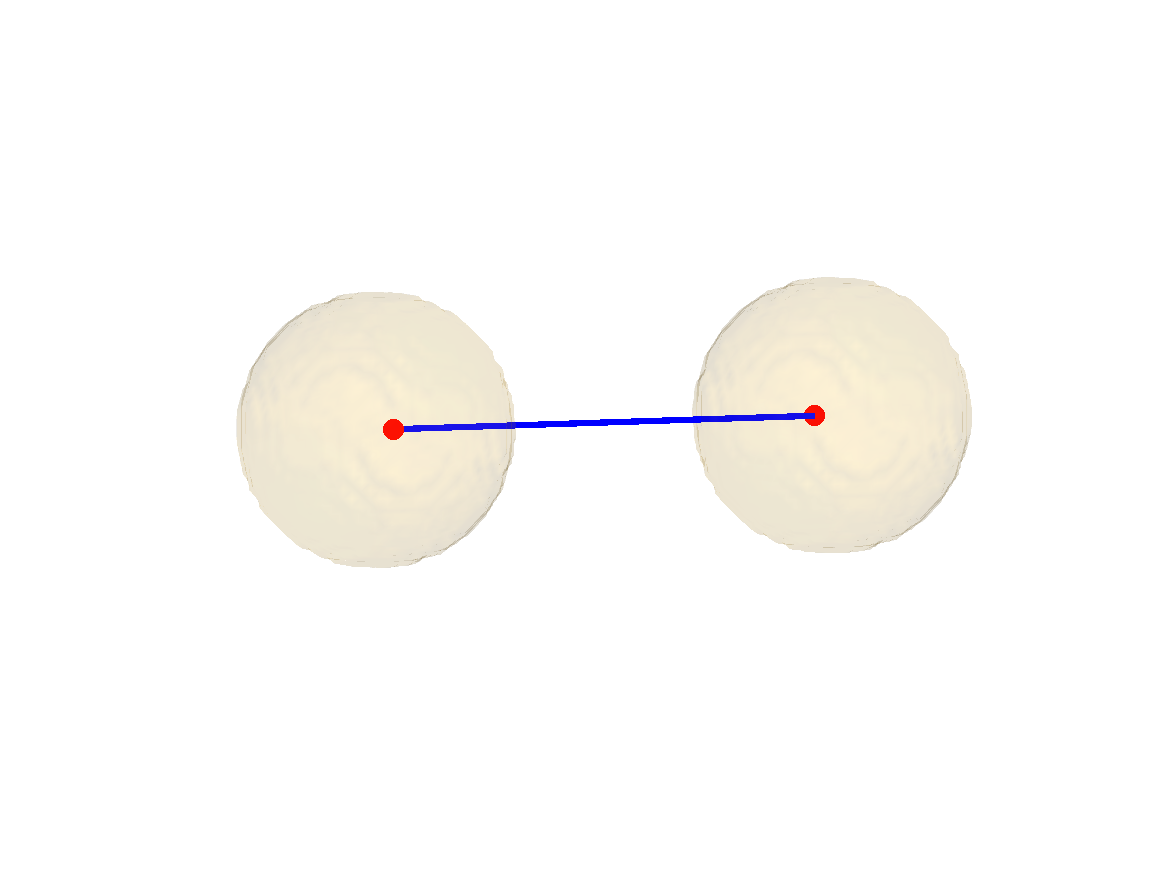}
&
\hspace{-14 mm}
\vspace{-2.2mm}
\includegraphics[width=2 in, height=1.5 in]{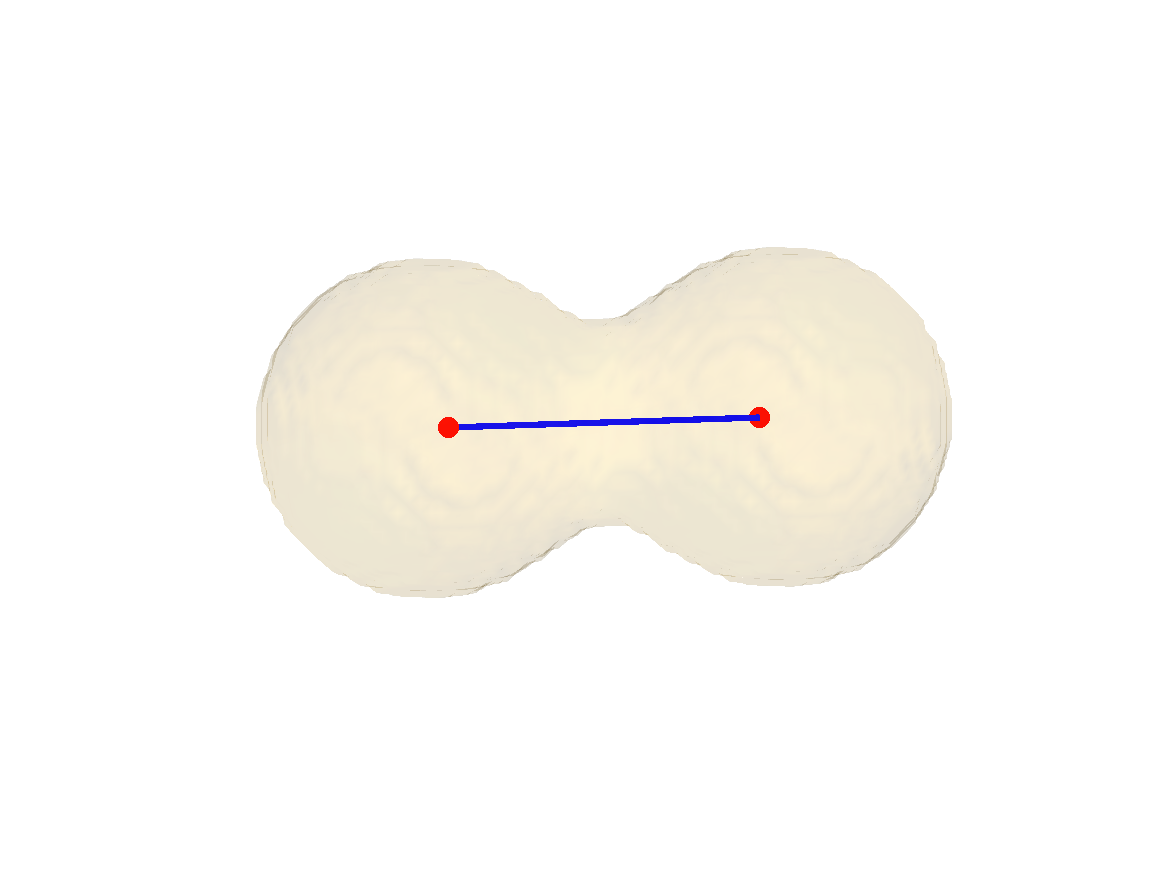}
&
\hspace{-15 mm}
\vspace{-2.2mm}
\includegraphics[width=2 in, height=1.5 in]{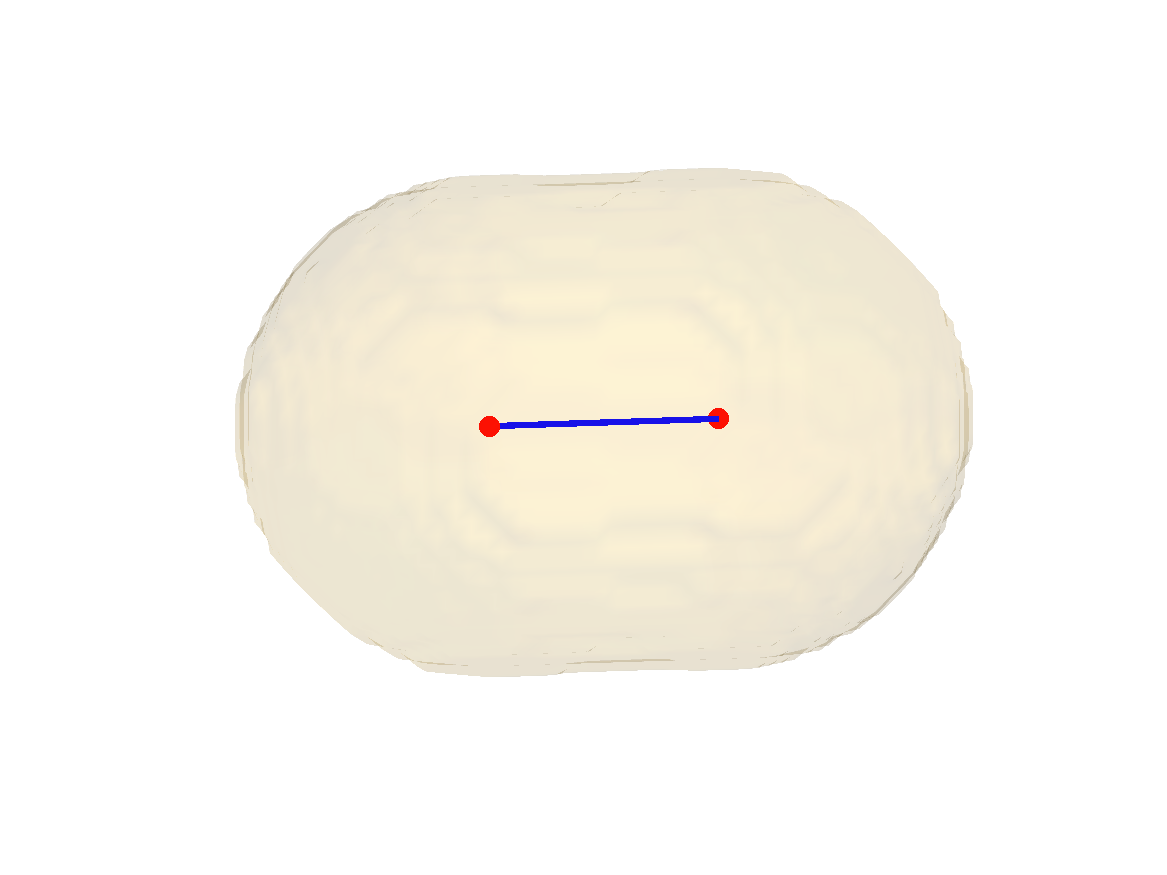}
\\
\hspace{-15mm}(a)   &  \hspace{-16 mm} (b)  &  \hspace{-15 mm} (c)  &	 \hspace{-15mm} (d) \\
\hspace{-15mm}
\vspace{-2.2mm}
\includegraphics[width=2 in, height=1.5 in]{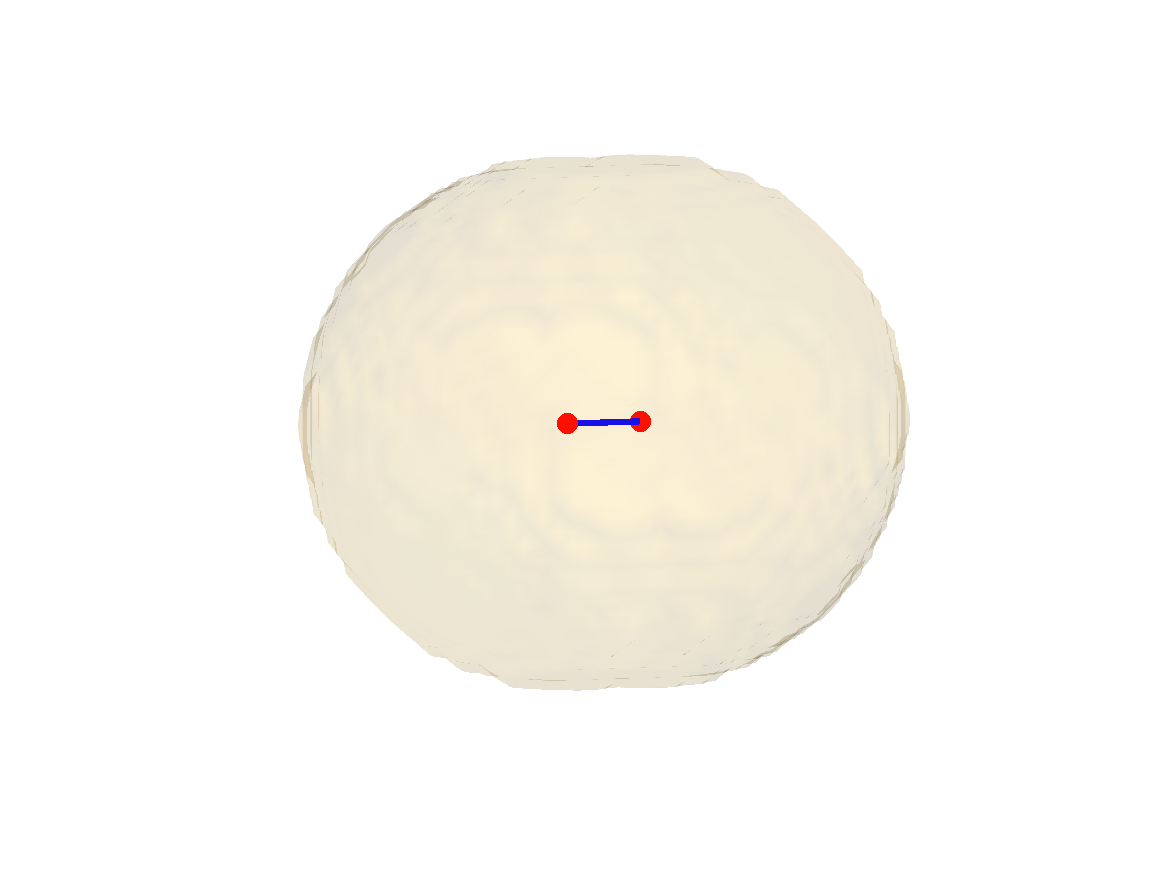} &
\vspace{-2.2mm}
\hspace{-15 mm}
\includegraphics[width=2 in, height=1.5 in]{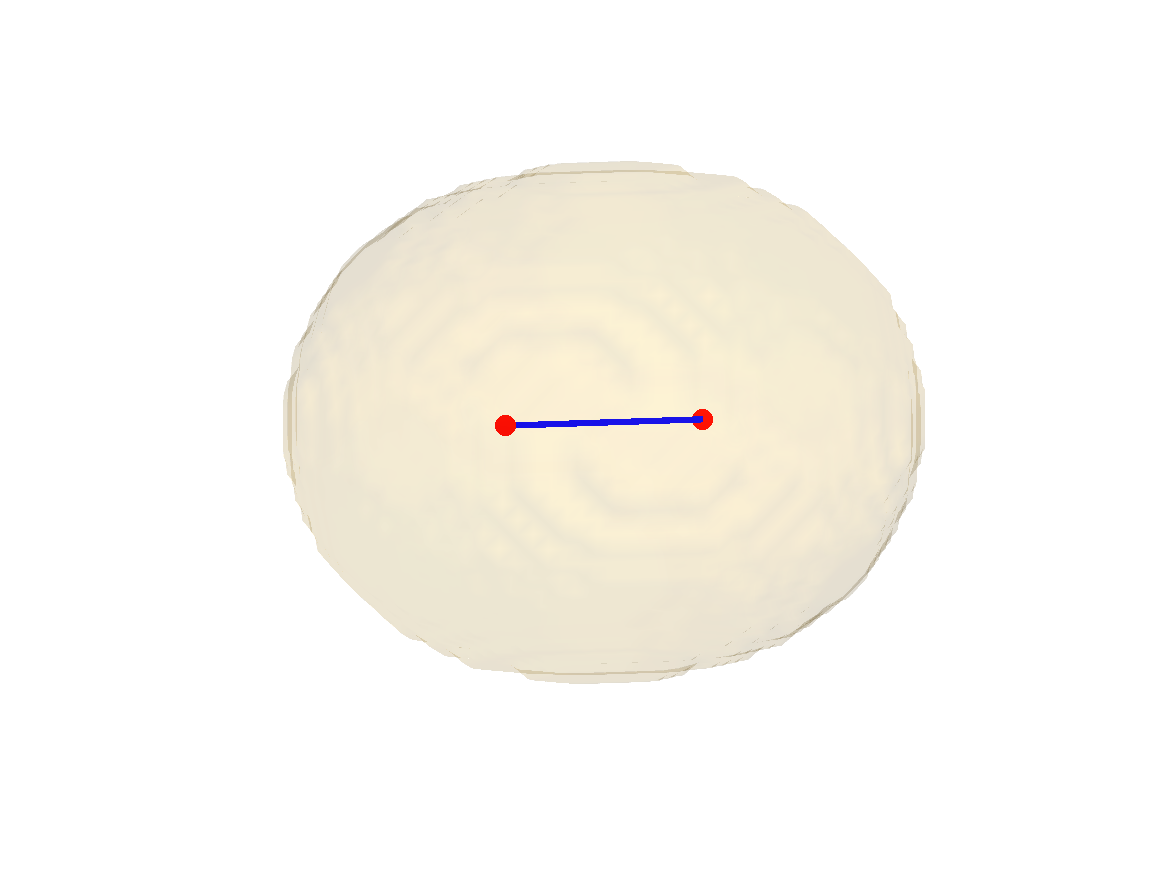} 
&
\vspace{-2.2mm}
\hspace{-16 mm}
\includegraphics[width=2 in, height=1.5 in]{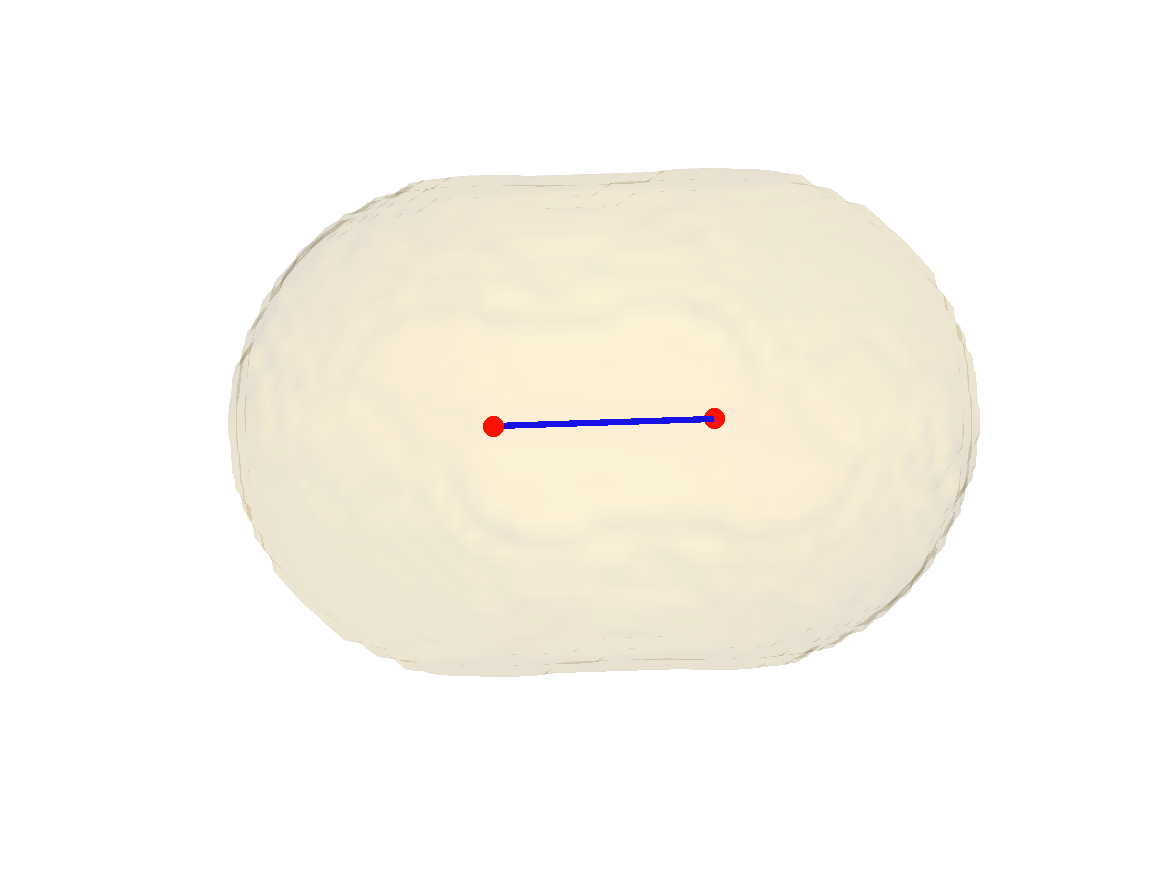} 
&
\vspace{-2.2mm}
\hspace{-14 mm}
\includegraphics[width=2 in, height=1.5 in]{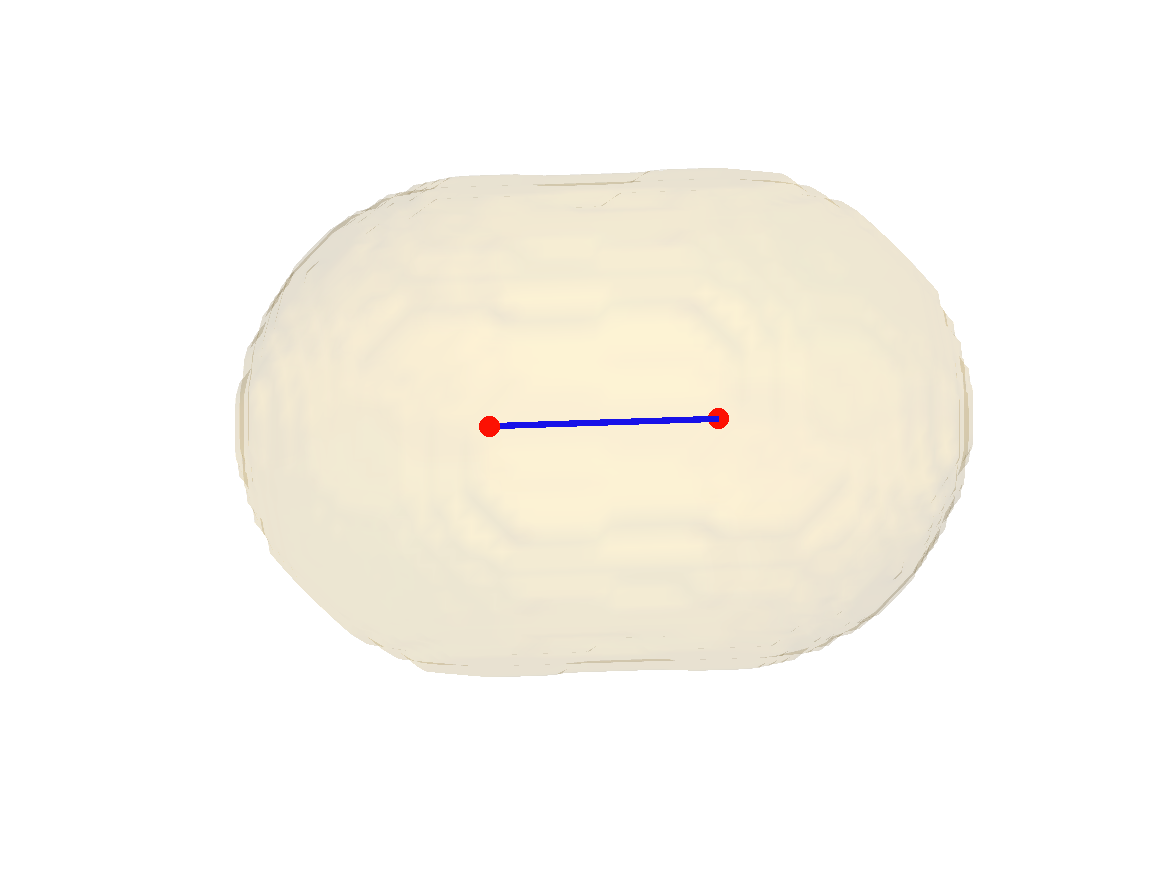} 
\\
 \hspace{-15mm}(e) & \hspace{-16 mm} (f)  &  \hspace{-15 mm} (g) &  \hspace{-15 mm} (h)  \\
\end{tabular}
\captionof{figure}{The free-energy minimization for a two-atom system. Top: Experiment 1.1.a. The 
snapshots  are taken at  (a) initial stage step 0, (b) step 5, (c) step 10,  
and (d) step 63, reaching nearly the steay state. Bottom: Experiment 1.1.b. The 
snapshots are taken at (e) initial stage step 0, (f) step 2, (g) step 4, and (h) step 53, reaching nearly the steay state.}
\label{fig:two_atoms_Ex1}
\end{center}

\vspace{-2 mm}
\begin{table}[h]
	\begin{center}
		\caption{Comparison of experiments 1.1.a and 1.1.b of a two-atom system.}
		\label{table:two_atoms_a}
		\scalebox{0.95}{
			\begin{tabular}{ |c|c|c|c|c|c|c|}
				\hline
				Experiment &  \multicolumn{2}{c|}{ Initial position} &  \multicolumn{2}{c|} {Final position} & Bond length & Free Energy \\
				\hline
				1.1.a &(7,0,0)  & (-7,0,0)&(1.50,0,0)& (-1.498,0,0)&3.00 & 21.85  \\
				\hline
				1.1.b &(0.5,0,0)  & (-0.5,0,0)&(1.50,0,0)& (-1.498,0,0)&3.00& 21.85  \\
				\hline 
			\end{tabular}
		}
	\end{center}
\end{table}

\vspace{-15 mm}
\begin{center}
	\begin{tabular}{cccc}
		\hspace{-15mm}
		\vspace{-3mm}
		\includegraphics[width=2 in, height=1.5 in]{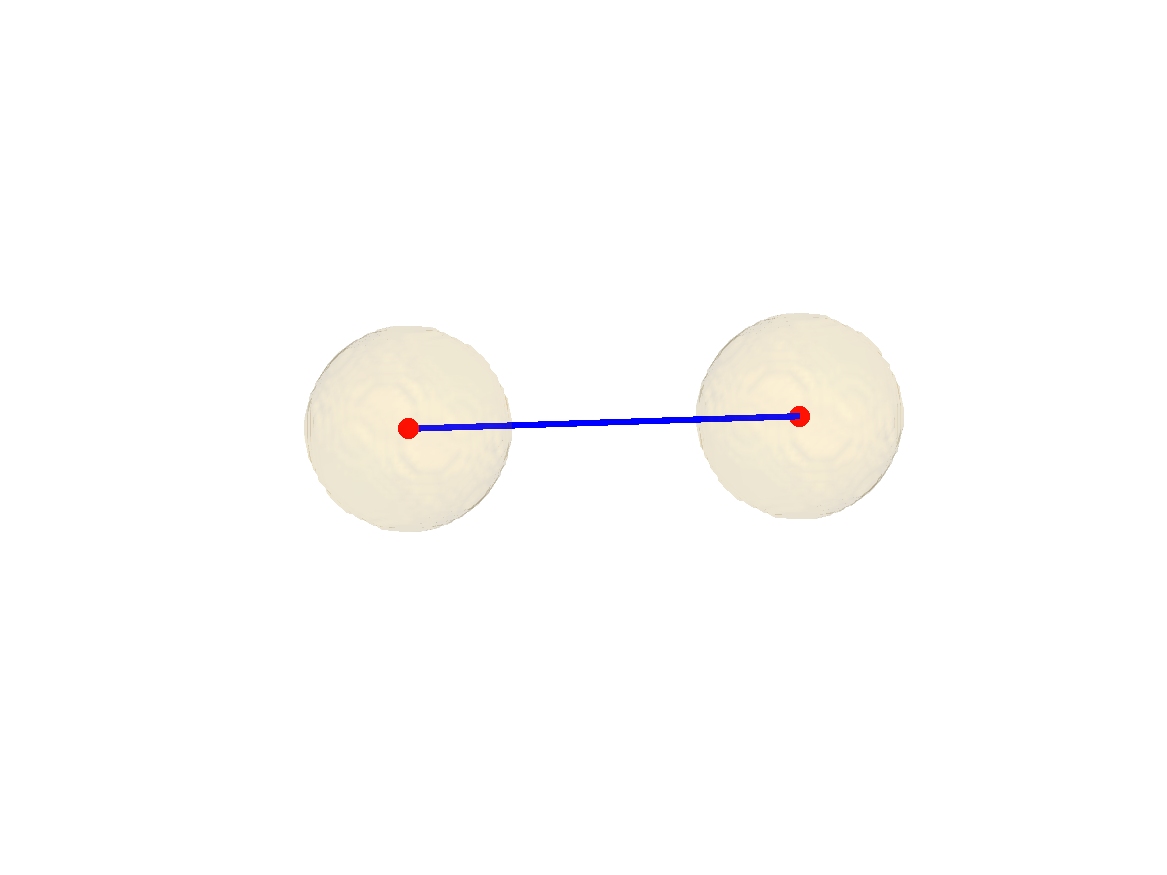} &
		\hspace{-16mm}
		\vspace{-3mm}
		\includegraphics[width=2 in, height=1.5 in]{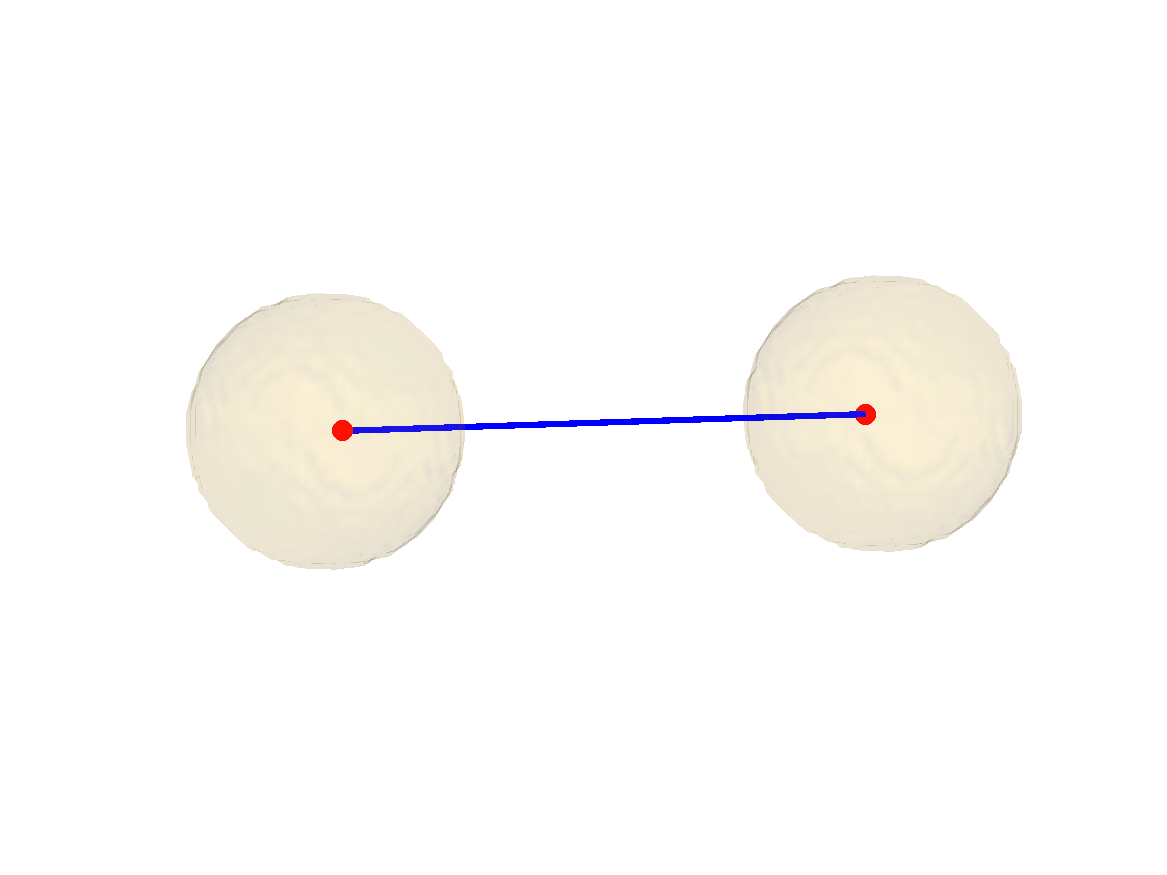}
		&
		\hspace{-11.5 mm}
		\vspace{-3mm}
		\includegraphics[width=2 in, height=1.5 in]{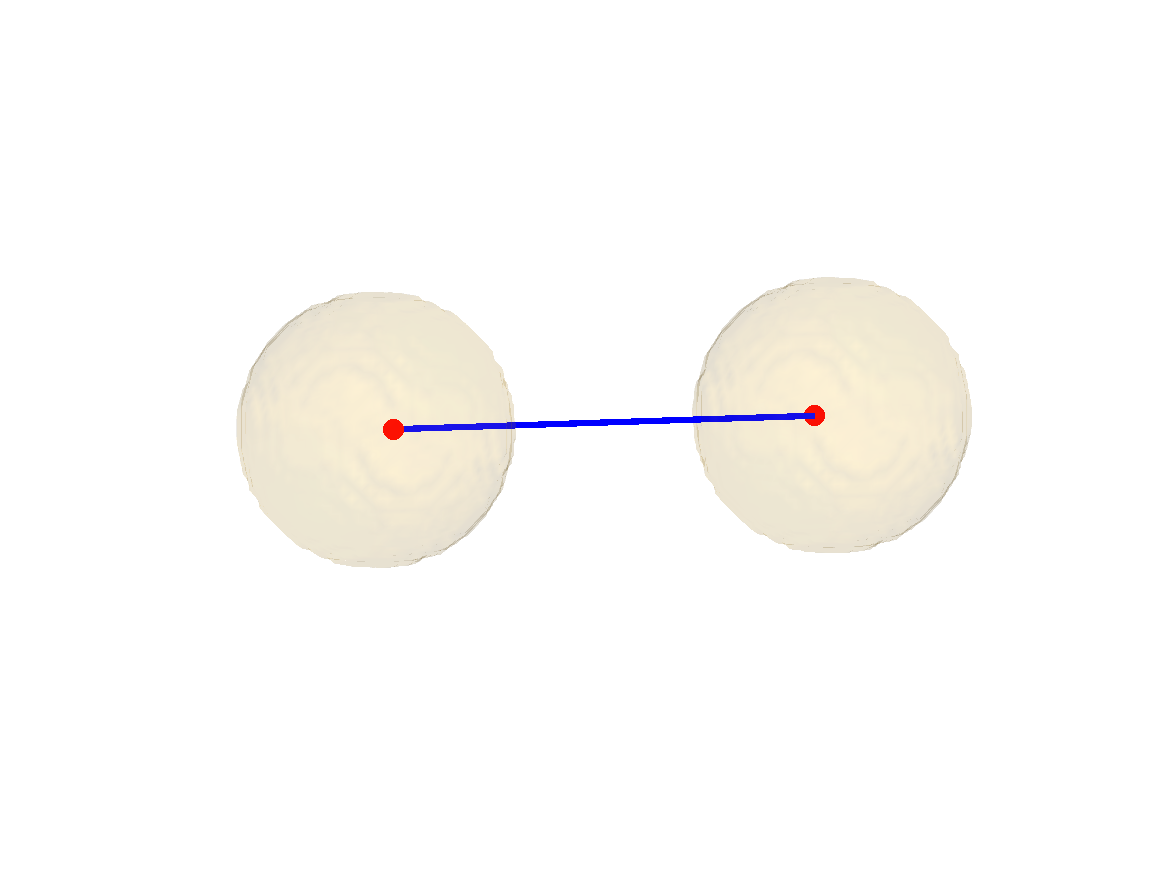}
		&
		\hspace{-14 mm}
		\vspace{-3mm}
		\includegraphics[width=2 in, height=1.5 in]{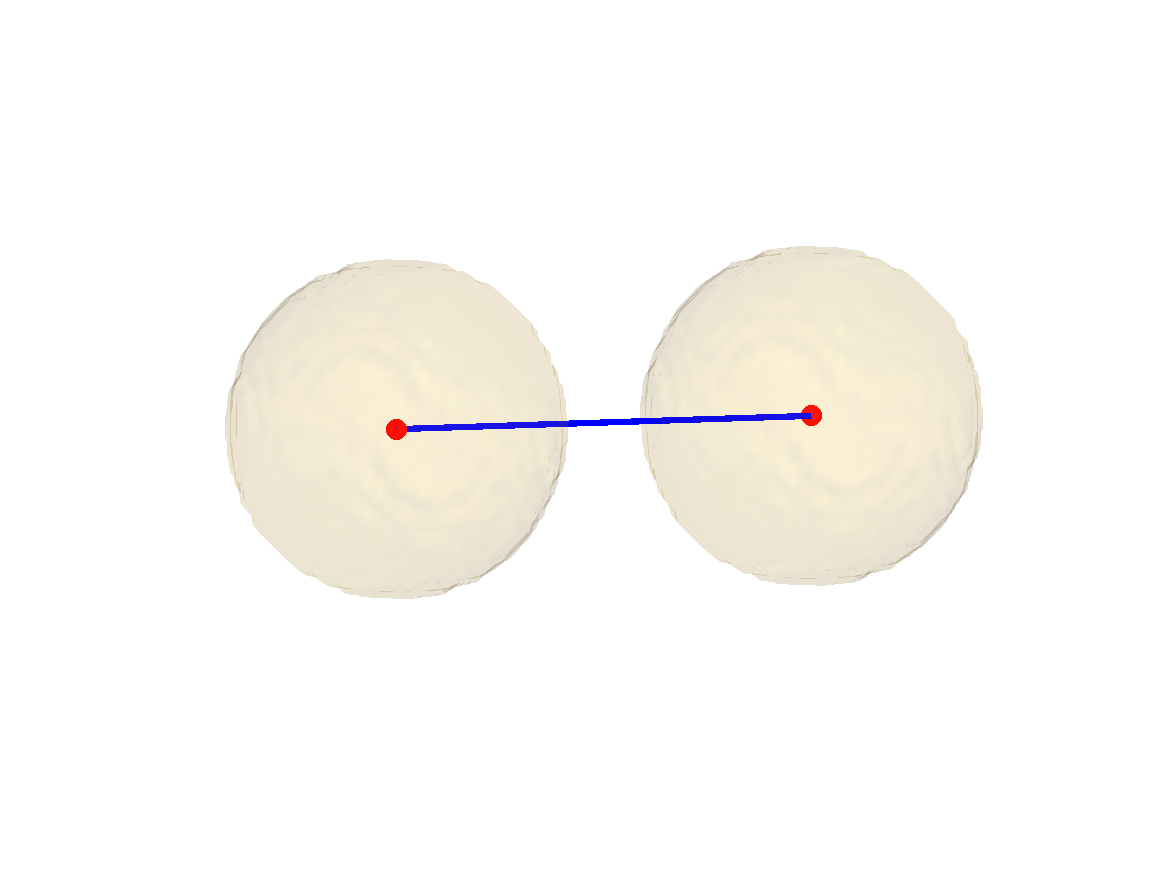}
		\\
		\hspace{-15mm}(a)   &  \hspace{-16 mm} (b)  &  \hspace{-11.5 mm} (c)  &	 \hspace{-15mm} (d) \\
		\hspace{-15mm}
		\vspace{-2.4mm}
		\includegraphics[width=2 in, height=1.5 in]{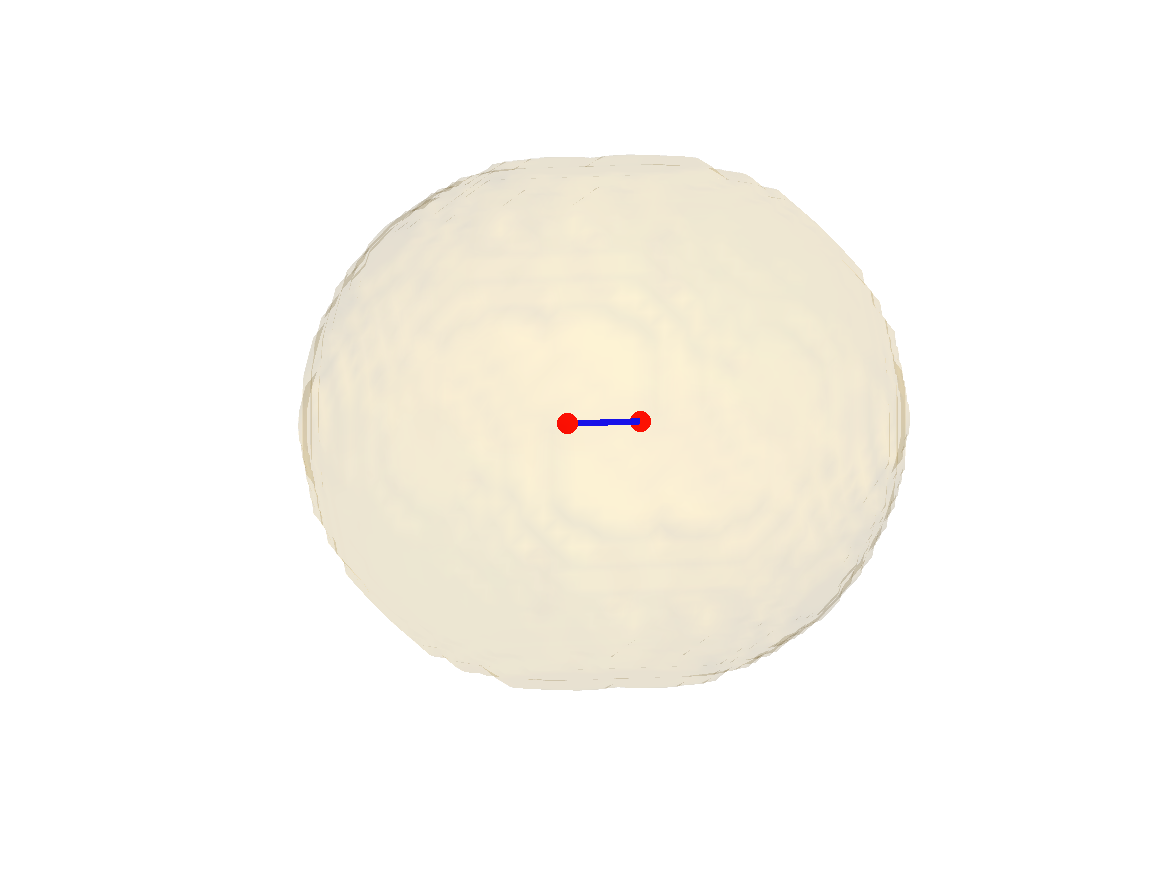} &
		\vspace{-2.4mm}
		\hspace{-14 mm}
		\includegraphics[width=2 in, height=1.5 in]{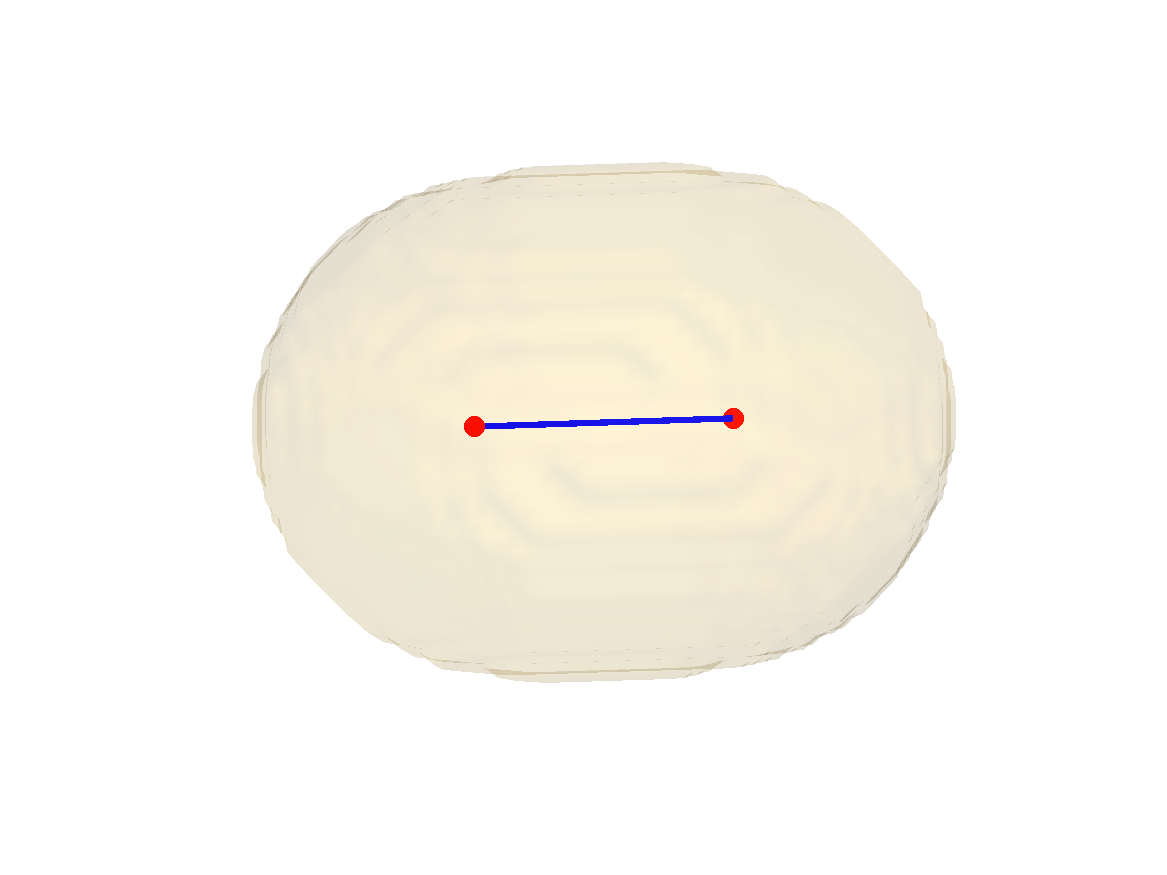} 
		&
		\vspace{-2.5mm}
		\hspace{-11.5 mm}
		\includegraphics[width=2 in, height=1.5 in]{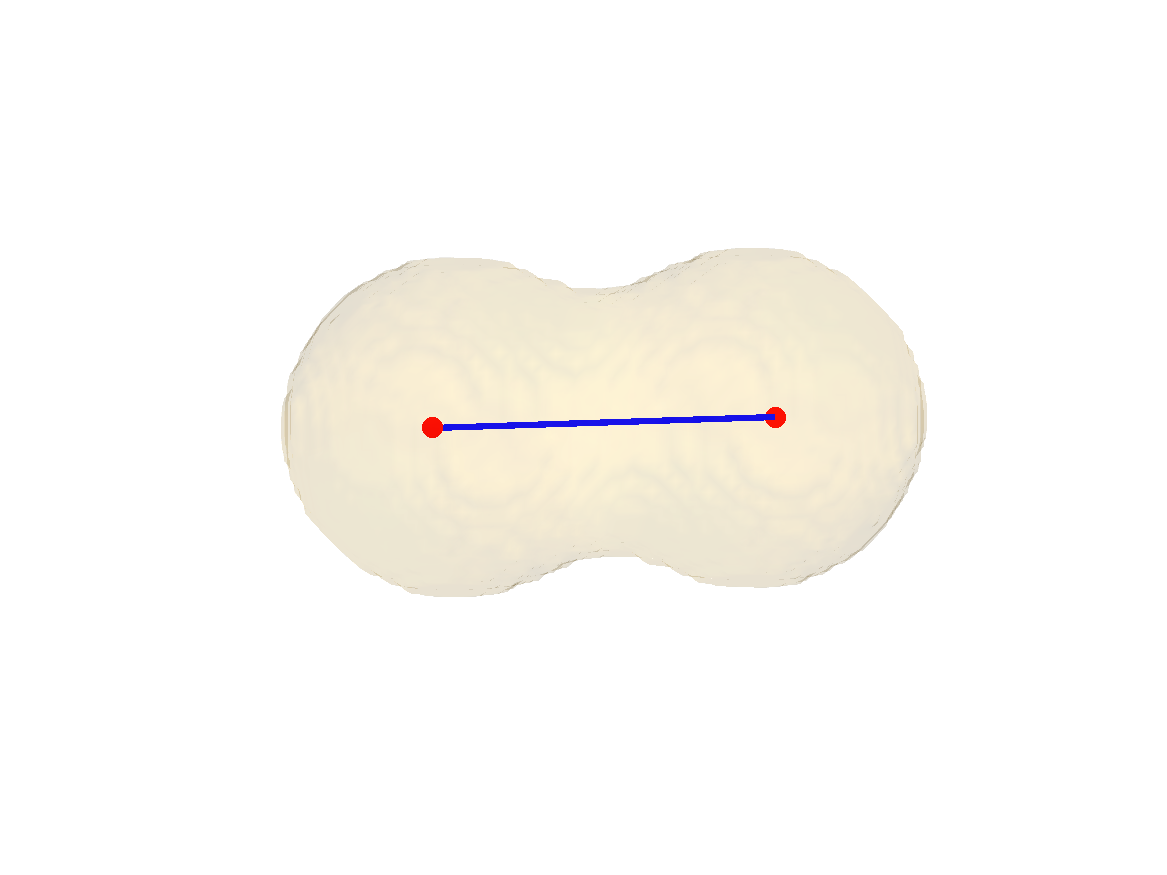}
		&
		\vspace{-2.5mm}
		\hspace{-14 mm}
		\includegraphics[width=2 in, height=1.5 in]{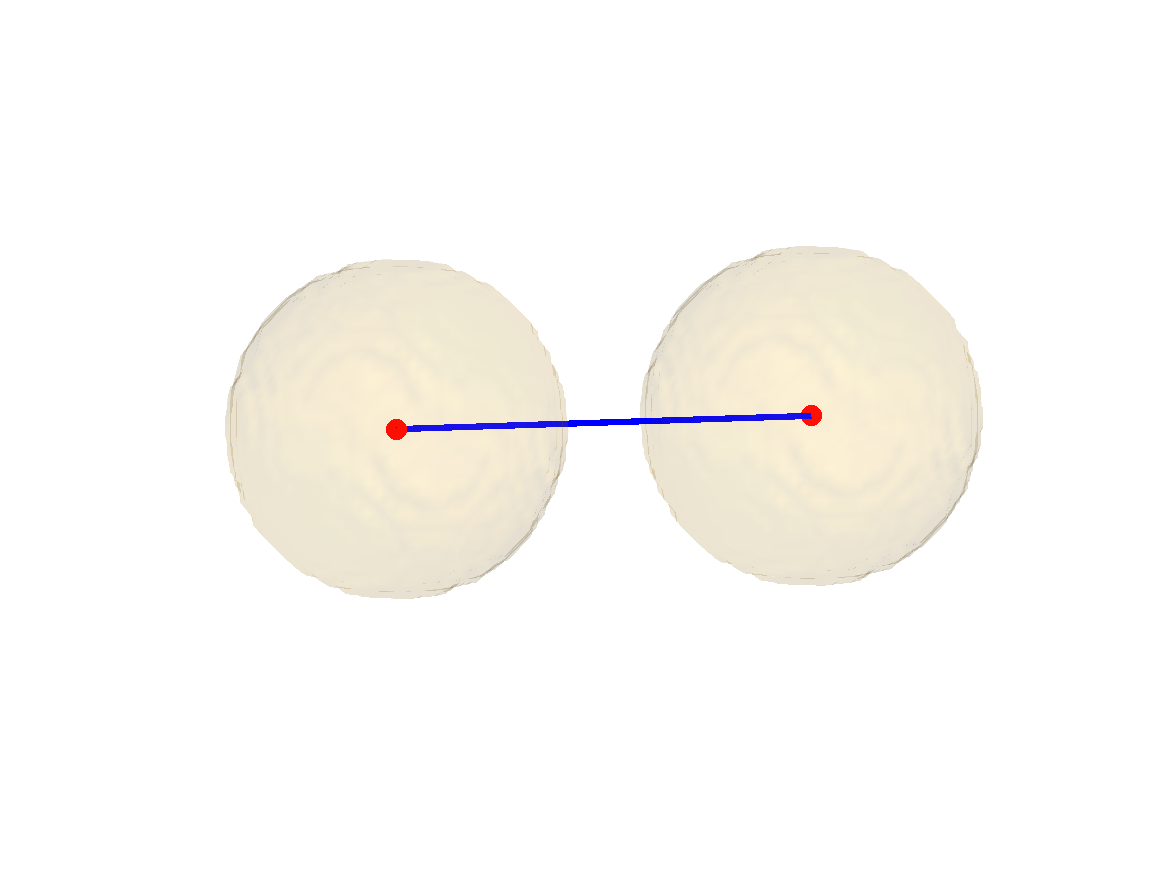}
		\\
		\hspace{-15mm}(e) & \hspace{-14 mm} (f)  &  \hspace{-11.5 mm} (g) &  \hspace{-14 mm} (h)  \\
	\end{tabular}
	\captionof{figure}{The free-energy functional minimization algorithm for a two-atom system. Top: Experiment 1.2.a. The screenshots are taken at (a) initial stage step 0, (b) step 2, (c) step5,  and (d) the steay state step 62.  Bottom: Experiment 1.2.b. The screenshots are taken at (e) initial stage step 0, (f) step 3, (g) step 7, and (h) the steay state step 64.} 
	\label{fig:two_atoms_Ex2}
\end{center}

Figure \ref{fig:two_atoms_Ex2} shows the minimization processes of Experiment 1.2.a and Experiment 1.2.b.  In the top row of Figure \ref{fig:two_atoms_Ex2}, we can see that initially surface consists of two disconnected spheres, then two atoms get closer, until the system reaches an equilibrium state. Comparing with the equilibrium of Experiment 1.1.a, the spheres are not merged due to a larger bond length $8\mathring{A}$. In the bottom row of  Figure \ref{fig:two_atoms_Ex2}, initially, the two atoms are very close to each other, then the atoms are pushed apart due to the force from strong bonding energy, the interface moves  and then splits apart,  then the system reaches to an equilibrium state with which the interface consists of two separate spheres. Table \ref{Table:two_atoms_b} shows that the two experiments 1.2.a and 1.2.b reach the same 
equilibrium. 
\vspace{-2 mm}
\begin{table}[H]
	\begin{center}
		\caption{Comparison of Experiments 1.2.a and 1.2.b of a two-atom system.}
		\label{Table:two_atoms_b}
		\scalebox{0.95}{
			\begin{tabular}{ |c|c|c|c|c|c|c|}
				\hline
				Experiment &  \multicolumn{2}{c|}{ Initial position} &  \multicolumn{2}{c|} {Final position} & Bond length & Free Energy  \\
				\hline
				1.2.a&(7,0,0)  & (-7,0,0)&(4.00,0,0)& (-4.00,0,0)&8.00& 33.93  \\
				\hline
				 1.2.b &(0.5,0,0)  & (-0.5,0,0)&(4.00,0,0)& (-4.00,0,0)&8.00& 33.93  \\
				\hline 
			\end{tabular}
		}
	\end{center}
\end{table}

\noindent
{\bf An ethane molecule.}
We consider an ethane molecule $C_2H_6$ in water and take from \cite{jorgensen1996development, halgren1996merckI, halgren1996merckII} the solute atomic positions and force field parameters. Other parameters are as follows: 
the carbon-water LJ parameters  $\sigma=3.4767  \, \mathring{A}$ and $\varepsilon=0.2311 \, k_BT$,  the hydrogen-water LJ parameters  $\sigma=3.1017\,  \mathring{A}$ and $\varepsilon=0.0989 \, k_BT$, the carbon-carbon LJ parameters  $\sigma=3.4\,  \mathring{A}$  and $\varepsilon=0.344 \, k_BT$, the carbon-hydrogen LJ parameters  $\sigma=3.025\,  \mathring{A}$  and $\varepsilon=\, 0.147 k_BT$,  and the hydrogen-hydrogen LJ parameters  $\sigma=2.650 \, \mathring{A}$  and $\varepsilon=0.063 \, k_BT$. 

In the ethane molecule,  each atom is connected to other atoms through bonding, bending, and torsion structure.   The effect of vdW interaction energy among unbonded pairs of solute atoms is  relatively less important when compared with molecular mechanical interactions, so in our simulation experiment of the ethane molecule, we neglect the vdW  interaction energy among  unbonded pairs of solute atoms. 

We design three different initial configurations of the ethane molecule from its equilibrium:
\begin{compactenum}
	\item[$\bullet$]
  In Experiment 2.1.a,  we stretch all hydrogen-carbon bonds to be $2\mathring{A}$.
	\item[$\bullet$]
	 In Experiment 2.1.b,  we stretch or shrink all hydrogen-carbon bonds such that hydrogen-carbon bonds have the length of $1\, \mathring{A}$,  $1.25\, \mathring{A}$,  $1.5\, \mathring{A}$,  $1.75\, \mathring{A}$,  $2\, \mathring{A}$, and  $2.25\, \mathring{A}$. 
	\item[$\bullet$]
	In Experiment 2.2,  we introduce  a small fluctuation, then rotate one set of three hydrogen-carbon bonds 50 degrees with respect to the carbon-carbon bond. 
\end{compactenum}

In Table \ref{table:ethane}, the bond lengths of hydrogen-carbon bonds and the carbon-carbon bond in their equilibrium of Experiment 2.1.a, Experiment 2.1.b, and Experiment 2.2 are compared. It is clear that in the equilibrium, all hydrogen-carbon bonds in three experiments  have the same length, which is consistent with the reference length of hydrogen-carbon bond  $1.093 \, \mathring{A}$. The carbon-carbon bond in each of the three experiments is the same as the reference length $1.508 \, \mathring{A}$.  This verifies the accuracy of our free-energy minimization algorithm. 

\begin{table}[H]
		\setlength{\abovecaptionskip}{-0.2cm}
	\caption{Comparison of  bond lengths ($ \mathring{A}$) of ethane in equilibrium from different initial configurations. Remark: H3, H4, and H5 are hydrogen atoms bonded with the carbon atom C1,  H6, H7, and H8 are hydrogen atoms bonded with the carbon atom C2.}
	\label{table:ethane}
	\begin{center}
	 \resizebox{\columnwidth}{!}{%
			\begin{tabular}{|c|c|c|c|c| c|c|}
				\hline
				Experiments&\multicolumn{2}{c|}{Experiment 2.1.a}&\multicolumn{2}{c|}{Experiment 2.1.b } & \multicolumn{2}{c|}{Experiment 2.2 }\\ 
				\hline 
				Bond list & Initial  Length  & Final  Length  & Initial Length  & Final Length  &Initial Length  & Final Length \\
				\hline
				\hline
				C1-H3& 2 & 1.093&  1 & 1.093& 1.090 & 1.093\\
				\hline
				C1-H4& 2 & 1.093&  1.25 & 1.093 & 1.090 & 1.093\\
				\hline
				C1-H5& 2 & 1.093&  1.5 & 1.093& 1.090 & 1.093\\
				\hline
				C2-H6& 2 & 1.093&  1.75 & 1.093& 1.090& 1.093\\
				\hline 
				C2-H7& 2 & 1.093&  2 & 1.093 & 1.090& 1.093\\
				\hline 
				C2-H8& 2 & 1.093&  2.25 & 1.093& 1.090 & 1.093\\
				\hline
				C1-C2& 1.77& 1.508 &  1.77 &1.508 &    1.540& 1.508   \\
				\hline
			\end{tabular}
		}
	\end{center}
\end{table}

\begin{figure}[htp]
\centering
\includegraphics[width=1.8in, height=1.8in]{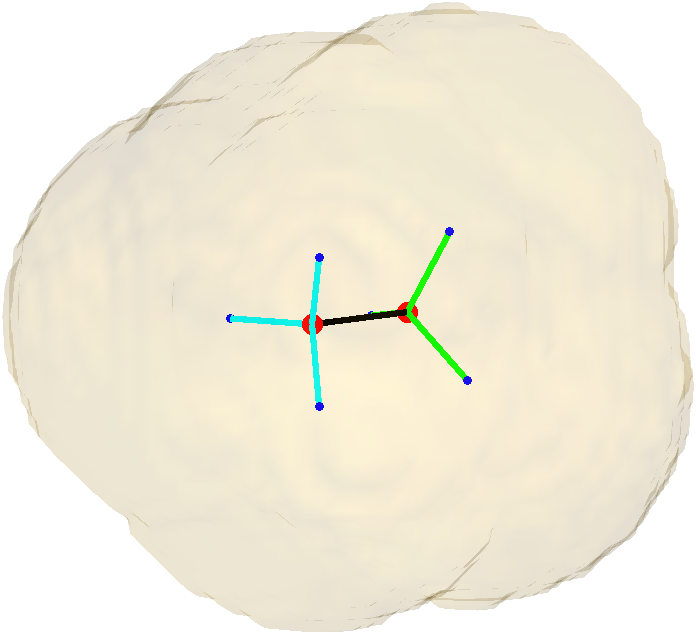}
\hspace{6 mm}
\includegraphics[width=1.8in, height=1.8in]{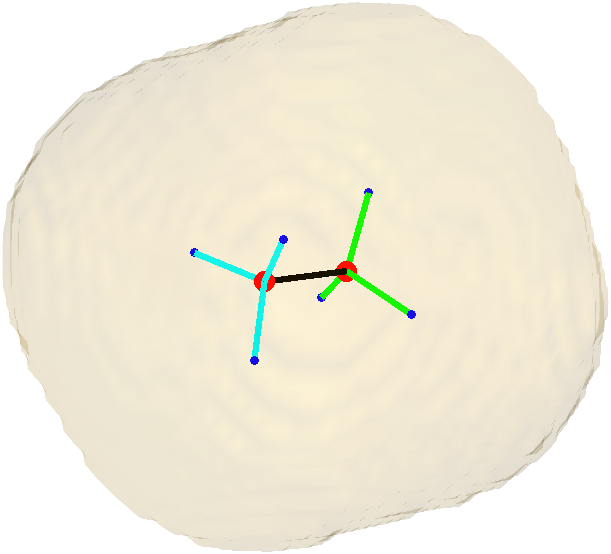}
\hspace{6 mm}
\includegraphics[width=1.8in, height=1.8in]{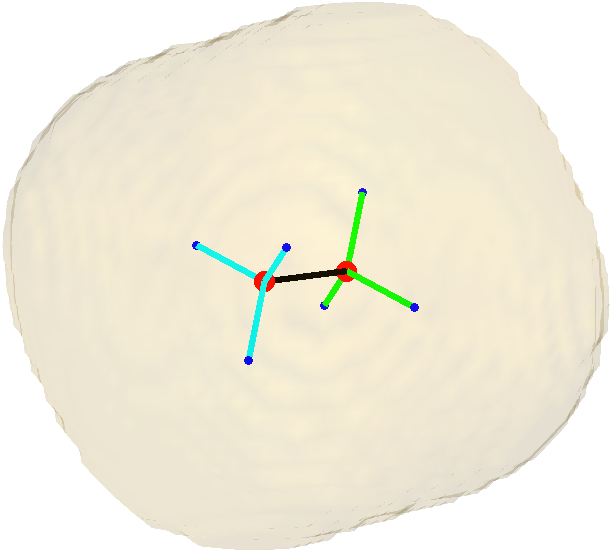}
\caption{The free-energy minimization for ethane from Experiment 2.2. 
Snapshot taken at step 0 (Left),  step 200 (Middle), and step 1500, reaching to a steady state (Right). }
	\label{fig:ethane}
\end{figure}

Figure \ref{fig:ethane}  displays the snapshots of minimization process of Experiment 2.2. The red dots represent carbon atoms, the blue dots represent the hydrogen atoms, the light blue segments and green segments represent the hydrogen-carbon bonds, and the black segment represents the carbon-carbon bond.   It is captured that during the relaxation, the set of three hydrogen-carbon bonds rotated back to their equilibrium, and all hydrogen-carbon bonds have the same length. Free energy of steady state in Experiment 2.1.a is 3.692 $k_BT$, the free energy of steady state in Experiment 2.1.b is 3.685 $k_BT$, and the free energy of steady state in Experiment 2.2 is 3.687 $k_BT$. Thus, the three experiments get to the same equilibria. We remark that the free energy here does not include the vdW interaction energy among unbonded pairs of solute atoms. 

\begin{figure}[!tp]
	\centering 
\hspace{2 mm}
\includegraphics[width=3.1in, height=2in]{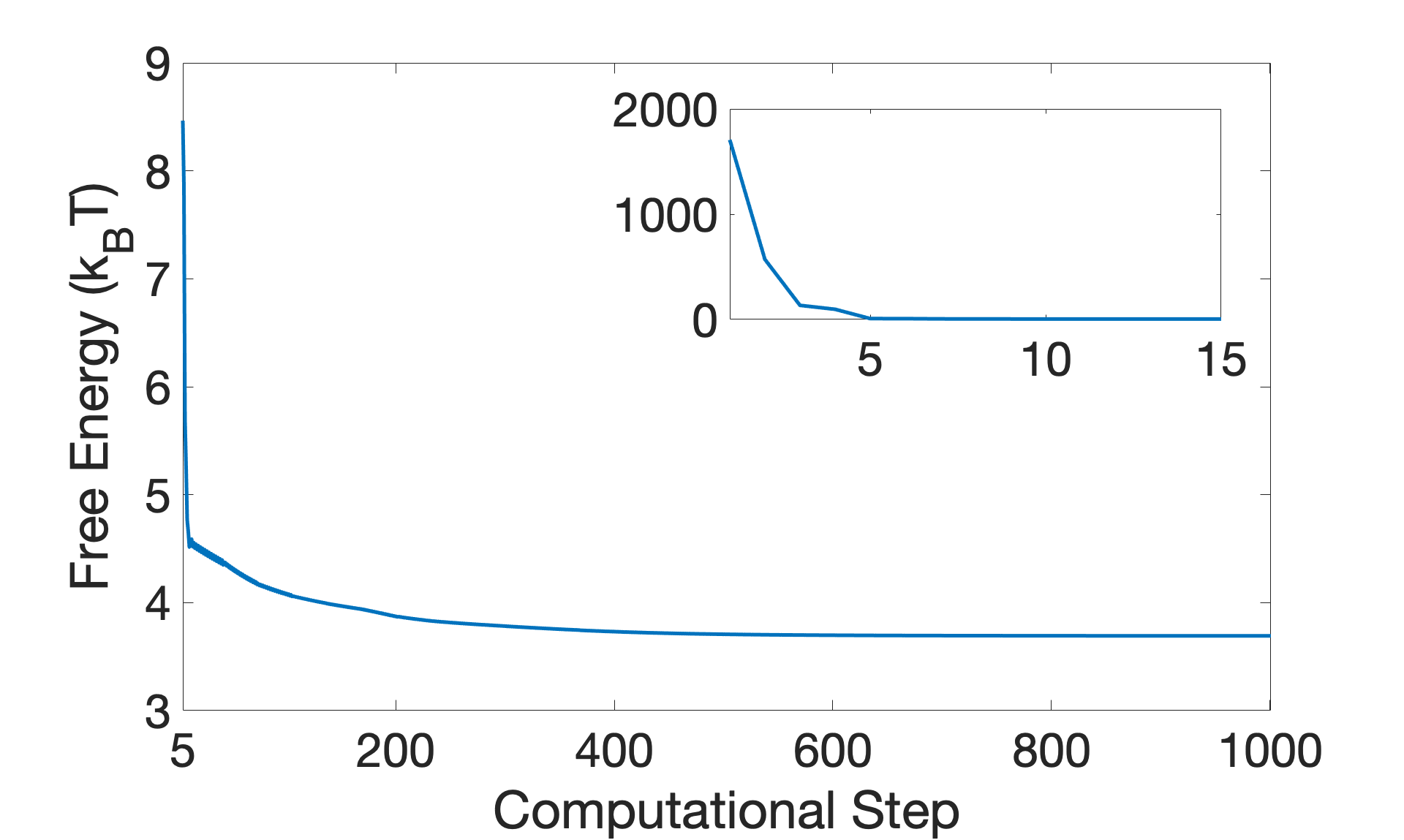}
\hspace{2 mm}
\includegraphics[width=3.1in, height=2in]{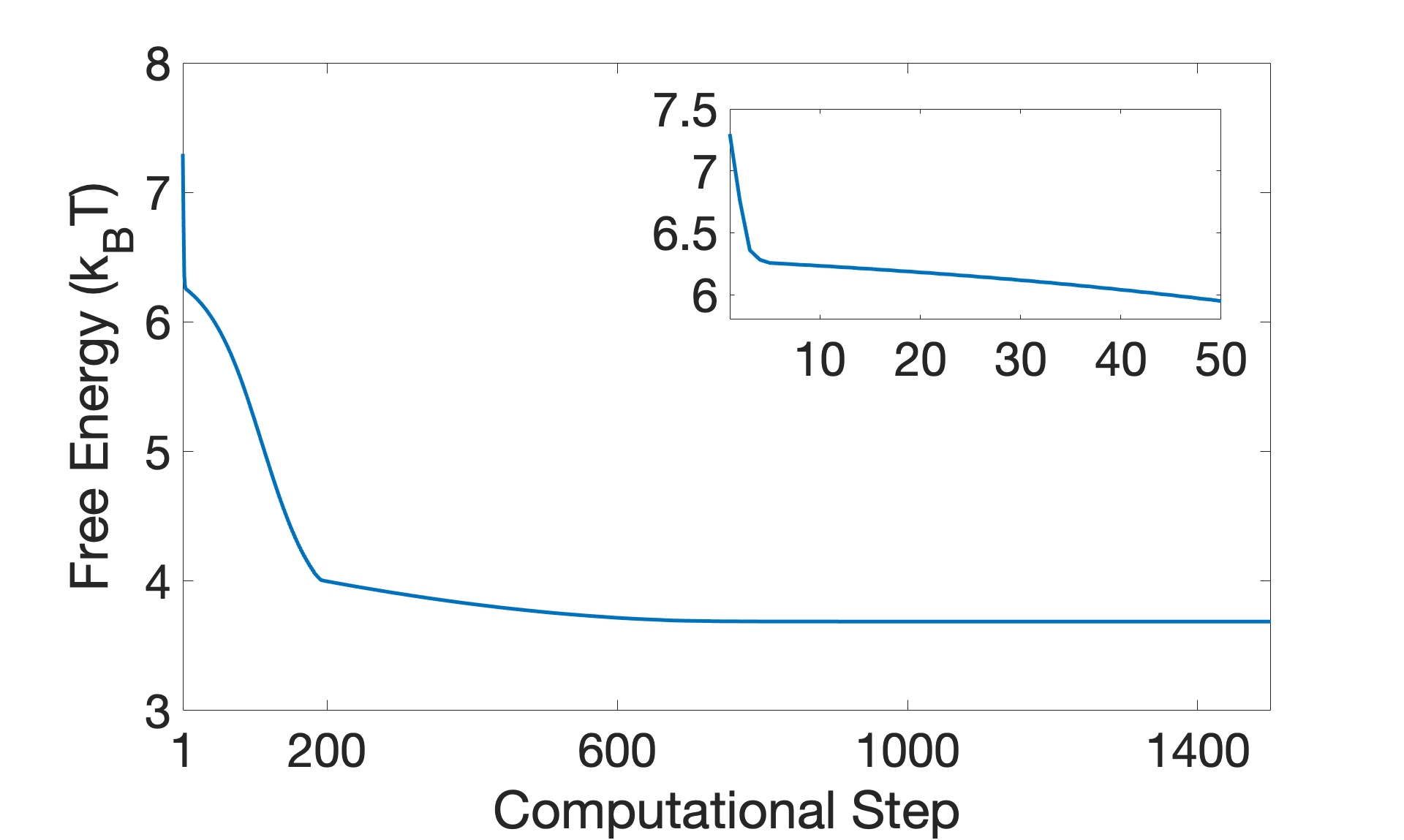}
\caption{The free energy $(k_BT)$ 
vs.\ the computational step in the free-energy minimization for an ethane molecule in Experiment 2.1.a and Experiment 2.2. First 15 (50) computational steps are specified in the inset of each subfigure.}
	\label{fig:ethane_energy}
\end{figure}

In Figure \ref{fig:ethane_energy}, we plot the free energy vs.\ iteration steps in numerical computations for Experiment 2.1.a and Experiment 2.2. In Experiment 2.1.a, initially, the free energy is very large, that is because the stretch or shrink of the initial bond length causes a large value of the bonding energy. 
We can see that  the free energy decays very fast in the first few steps, which is caused by the dominant force from the bonding energy.  It takes around 1000 steps to adjust positions of solute atoms in ethane molecule to reach the equilibrium. 
In contract, the initial free energy of Experiment~2.2 in Figure~\ref{fig:ethane_energy} is a relatively small value, as we only introduced a small fluctuation to the initial atomic positions and the rotation of a set of three hydrogen-carbon bonds did not cause large free energy change.  We observe that the rate  of free-energy change at around the 3rd step and 200th computational step becomes slower and slower, which indicates that our free-energy minimization algorithm was adjusting the suitable mobility factor $M$ during the minimization process. Although the initial free energy is small, it takes more than 1400 steps to rotate the hydrogen-carbon bonds back to the right position and reach the equilibrium. 

\subsection{Simulation of protein-protein interactions}
\label{ss:p53MDM2}

In this section, we choose a biologically important and realistic system, the p53/MDM2 protein complex, and investigate the binding behavior using our free-energy minimization model and algorithm. 
To make the calculations of molecular movement easier, the receptor protein MDM2 here is fixed.

We use the CHARMM36 force field  \cite{huang2017charmm36,best2012optimization,mackerell2004improved,mackerell1998all} for our VISM simulations for the binding of p53 and MDM2.

Table \ref{table:p53_MDM2_rigid_body}
shows the solvation free energy and its components obtained by our VISM simulations
for the bound complex p53/MDM2, and the computational times of the simulations with 
the GPU single precision and the CPU double precision, respectively.

During those simulations, we 
relax the relative difference stop criterion and absolute difference stop criteria to 
be $1$e-$3$, just for the efficiency comparison of the GPU code and the CPU code. 
We set the  initial  configuration of p53/MDM2 to be a tight initial interface for atomic position with small fluctuations of p53/MDM2 in the bound complex, where  the bound complex is taken from the Protein Data Bank (PDB code: 1ycr.pdb).   It can be observed  in Table~\ref{table:p53_MDM2_rigid_body} that the difference between GPU code with single 
precision data type and CPU with double precision data type is less than 1\%, but the cost time of CPU is more than  78 times, and 307 times slower than the cost time of GPU with number of grid points  $50^3$, and $100^3$, correspondingly.

  \begin{table}[h]
  	\caption{Comparison of GPU (single precision) and CPU (double precision) for free energy ($k_BT$) and its components of p53/MDM2 in the steady state. The unit of time is minute. }
  	\label{table:p53_MDM2_rigid_body}
  	\centering
  	\resizebox{\columnwidth}{!}{%
  		\begin{tabular}{|c|c|c|c|c|c|c|c|c|c|c| }
  			\hline
  			Grid 
  			&  \multicolumn{2}{c|}{Total Energy} &\multicolumn{2}{c|}{Surface Energy} &\multicolumn{2}{c|}{vdW Energy} & \multicolumn{2}{c|}{CFA} & \multicolumn{2}{c|}{Total Time}\\
  			\cline{2-11} 
  			Points			& GPU& CPU & GPU& CPU& GPU& CPU & GPU& CPU & GPU& CPU  \\
  			\hline
  			$50^3$&-214.0&-212.0&800.7&802.7&-453.1&-453.0&-561.6&-561.7&10.0&784.7\\
  			\hline
  			$100^3$&-157.3&-159.1&833.8&834.0&-441.9&-440.6&-549.2&-552.4&16.7&5128.6\\
  			\hline
  			$200^3$&-140.9&-&836.5&-&-434.4&-&-543.0&-&62.6&-\\
  			\hline
  		\end{tabular}
  	}
  
  \end{table}

 We further investigate the binding behavior of p53 and MDM2 with our free-energy minimization method and algorithm. Note that, here we use a grid size $h=0.5 \, \mathring{A}$, the relative difference stop criterion and absolute difference stop criterion are $1$e-$5$. We construct the initial configuration with a tight initial interface by pulling p53 away from the MDM2 pocket in the bound complex along the line passing through the geometrical centers.

\begin{figure}[H]
\centering
\includegraphics[width=1.4 in, height=1.4 in]{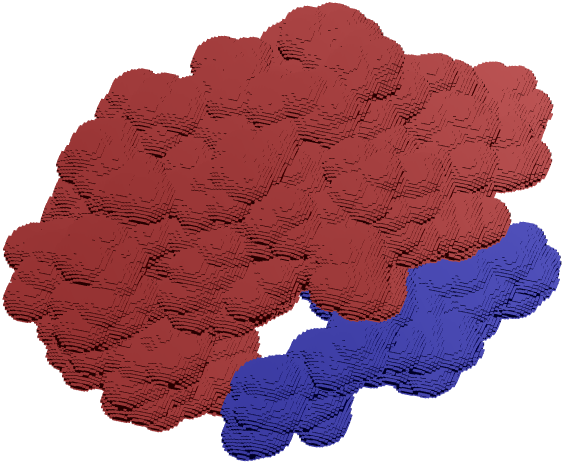}
\hspace{3 mm}
\includegraphics[width=1.4 in, height=1.4 in]{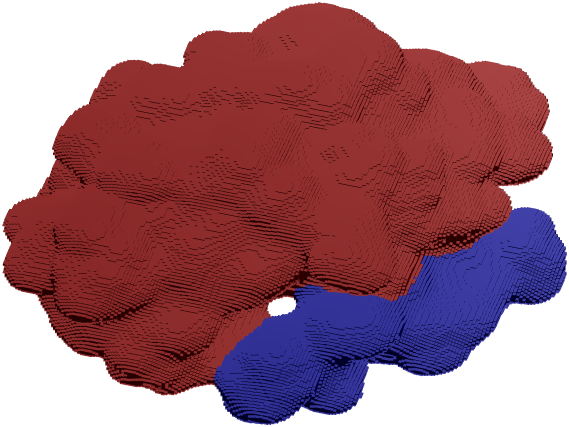}
\hspace{3 mm}
\includegraphics[width=1.4 in, height=1.4 in]{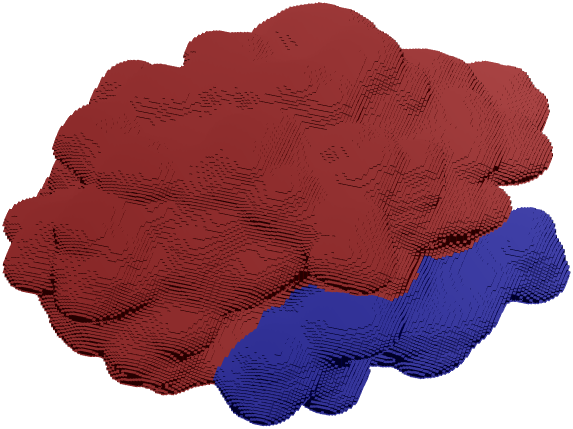}
\hspace{3 mm}
\includegraphics[width=1.4 in, height=1.4 in]{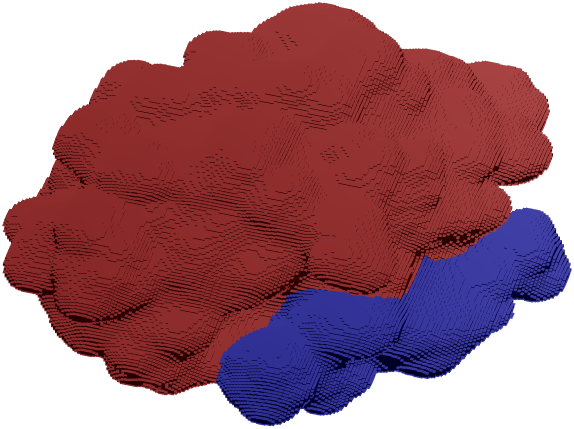}
\label{fig:12b}
	\caption{The free-energy minimization for p53/MDM2.  The snapshots are taken at  (a) initial stage, i.e.,  step 0, (b) step 100, (c) step 200, and (d) step 377, reaching nearly the steady state. Note that, in order to show the relative positions of MDM2 and p53, we color each piece of surface according to whether its closest solute atom comes from MDM2 (red) or p53 (blue).  }
	\label{fig:p53_MDM2_rigid_body}
\end{figure}

 In Figure \ref{fig:p53_MDM2_rigid_body}, a few snapshots of numerical results of p53/MDM2 are displayed, showing  the minimization process. The position of the protein MDM2 is fixed, positions of p53 atoms are adjusted by the free-energy minimization process. We color each piece of surface according to whether its closet solute atom comes from MDM2 (red) or p53 (blue) to show the relative positions of MDM2 and p53.  In the initial configuration, the protein  p53 and the receptor protein MDM2 are separated as we can observe a hole between them; that is a small region filled with water.  In the process, the relative positions of p53 and MDM2 are adjusted, p53 and MDM2 become closer and closer. In the equilibrium, the hole disappears, and the two proteins are combined together.  

\section{Conclusions}
\label{s:Conclusions}

This work presents the development of a  GPU parallel free-energy minimization method and 
algorithm with the fast binary level-set method and an adaptive-mobility gradient descent method for the variational explicit-solute implicit-solvent (VESIS) molecular simulations. 
Minimization of the free-energy functional determines an equilibrium interface and an equilibrium molecular structure.

Our free-energy minimization is an iterative process with two stages.  
In the first stage, we fix solute atoms, then minimize the solvation free energy to obtain an optimal solute-solvent interface. In the second stage, with a fixed interface, we relax the solute atoms using a gradient descent type method. 

The proposed minimization algorithm is implemented in parallel on GPUs  with single precision.

We have presented a series of numerical experiments and have demonstrated the  accuracy and  efficiency of our numerical methods and algorithm for the free-energy minimization. In particular, our numerical experiments of potentials of mean force for two charged systems, two charged parallel plates, and the protein BphC, have shown that VISM with the binary level-set method can capture well the sensitive response of capillary evaporation to the charge in hydrophobic confinement and the polymodal hydration behavior. Moreover, our numerical experiments for small molecular systems, the two-atom system and an ethane molecule, have demonstrated that our algorithm can capture topological changes of the solute-solvent interfaces as well as describe the equilibrium molecular structure. A key application of our algorithm  is for large biomolecular simulations. We have applied our free-energy minimization method and algorithm to a realistic system, the  p53/MDM2 protein complex. Our model and method describes the relaxation process of the binding of these two molecules. 

To verify the performance of our GPU parallel implementation of our algorithm, 
we have compared for different molecular systems our computational results 
and computational times with both of the CPU with double precision and the GPU with single precision. 
We observe that the GPU implementation is much more efficient than the CPU implementation. 
The GPU with single precision combining the pairwise summation 
can efficiently limit the grow of round-off errors.  
For small molecular systems such as the two charged parallel plates the computational time with 
the CPU is around 5 times of that with the GPU.  
For a relatively large molecular system, such as p53/MDM2, and fine finite-difference grids, 
our GPU implementation works especially well, reaching a speed about 100 times faster than that
of the CPU implementation.  
In the meantime, both implementations lead to the same minimum free energies and even their
individual components.

To speed up the computations further, our immediate next step is to construct a hybrid CPU-GPU architecture to combine CPU parallel computing and GPU parallel computing together. For the CPU parallel computing, we can use  a standard domain decomposition approach. The communication between sub-domains is based on the message-passing interface (MPI). 

With our fast algorithm and GPU code, we can now carry out flexible VESIS-Monte Carlo simulations for the binding of two proteins in which both the solute-solvent interface and the set 
of solute atomic positions change in each step of the Monte Carlo move.

\section*{Appendix}
\label{s:Appendix}
\renewcommand{\thesection}{A}
\setcounter{equation}{0}

\noindent
\paragraph{Gradient of $G[\Gamma,\bR].$} 
Fix $n$ with $1 \le n \le N.$ We have 
\begin{align}
\label{GradientG}
	\nabla_{\mathbf{r}_n}G[\Gamma,\mathbf{R}] &=\frac{1}{8\pi^2\varepsilon_0} 
\left (\frac{1}{\varepsilon_{\rm w}}-\frac{1}{\varepsilon_{\rm m}}\right) \int_{\Omega_{\rm w}} 
\left ( \sum_{i=1}^{N}\frac{Q_i(\mathbf{r}-\mathbf{r}_i)}{|\mathbf{r}-\mathbf{r}_i|^3} \right ) 
\left (\frac{Q_n}{|\mathbf{r}-\mathbf{r}_n|^3} \right) dV_{\mathbf{r}}
\nonumber \\ 
&\qquad +\rho_0\int_{\Omega_{\rm w}}U'_{sw}(|\mathbf{r}_n-\mathbf{r}|)
\frac{\mathbf{r}_n-\mathbf{r}}{|\mathbf{r}_n-\mathbf{r}|}dV_{\mathbf{r}}
	+ \sum_{(i,j)'}\delta_{ni}G_{\rm elec}^{'\rm ss}(|\mathbf{r}_i-\mathbf{r}_j|)
\frac{\mathbf{r}_i-\mathbf{r}_j}{|\mathbf{r}_i-\mathbf{r}_j|} 
\nonumber \\
&\qquad +\sum_{(i,j)'}\delta_{ni}U'_{ss}(|\mathbf{r}_i-\mathbf{r}_j|)
\frac{\mathbf{r}_i-\mathbf{r}_j}{|\mathbf{r}_i-\mathbf{r}_j|}
+\sum_{(i,j)}\delta_{ni}A_{ij}(|r_{ij}-r_{0ij}|)
\frac{\mathbf{r}_i-\mathbf{r}_j}{|\mathbf{r}_i-\mathbf{r}_j|}
\nonumber \\
	&\qquad +\sum_{(i,j,k)}\nabla_{\mathbf{r}_n}W_{\rm bend}(\mathbf{r}_i,\mathbf{r}_j,\mathbf{r}_k)
+\sum_{(i,j,k,l)}\nabla_{\mathbf{r}_n}
W_{\rm torsion}(\mathbf{r}_i,\mathbf{r}_j,\mathbf{r}_k,\mathbf{r}_l),  
\end{align}
where $\delta_{ni} = 1$ if $n=i$ and $0$ otherwise. 

 \paragraph{Force calculations of molecular mechanical interactions.}
 For fixed $\mathbf{r}_i,\mathbf{r}_j$, and $\mathbf{r}_k$, denote  the vector from $\mathbf{r}_j$ to $\mathbf{r}_i$  by  $\mathbf{q}_{ji}=\mathbf{r}_i-\mathbf{r}_j$ for any $i$ and $j$ and the length of $\mathbf{q}_{ji}$ by $q_{ji}=|\mathbf{q}_{ji}|$.
We have 
 \begin{equation*}
 	\nabla_{\mathbf{r}_n}W_{\rm bend}(\mathbf{r}_i,\mathbf{r}_j,\mathbf{r}_k)=B_{ijk}(\theta_{ijk}-\theta_{0ijk})\nabla_{\mathbf{r}_n} \theta_{ijk}, \quad n=i,j,k, 
 \end{equation*}
where 
  \begin{align*}
 \nabla_{\mathbf{r}_i}\theta_{ijk}&=\frac{1}{\sin \theta_{ijk}}\left ( \frac{\mathbf{q}_{ji} \cdot \mathbf{q}_{jk}}{q_{ji}^3q_{jk}}\mathbf{q}_{ji}-\frac{1}{q_{ji}q_{jk}}\mathbf{q}_{jk}   \right),\\
 \nabla_{\mathbf{r}_k}\theta_{ijk}&=\frac{1}{\sin \theta_{ijk}}\left ( \frac{\mathbf{q}_{ji} \cdot \mathbf{q}_{jk}}{q_{jk}^3q_{ji}}\mathbf{q}_{ji}-\frac{1}{q_{ji}q_{jk}}\mathbf{q}_{ji}   \right),\\
 \nabla_{\mathbf{r}_j}\theta_{ijk}&=\frac{1}{\sin \theta_{ijk}}\left [\left ( \frac{1}{q_{ji}q_{jk}}-\frac{\cos \theta_{ijk}}{q_{ji}^2} \right) \mathbf{r}_{ji}+\left ( \frac{1}{q_{ji}q_{jk}}-\frac{\cos \theta_{ijk}}{q_{jk}^2} \right) \mathbf{r}_{jk}\right]. 
  \end{align*}
Recall for for fixed $\mathbf{r}_i,\mathbf{r}_j, \mathbf{r}_k$, and $\mathbf{r}_l$ that  
  \begin{equation*}
 	W_{\rm torsion} (\mathbf{r}_i,\mathbf{r}_j, \mathbf{r}_k, \mathbf{r}_l)=\sum_{n=0}^{6}C_n[1+\cos(n\tau-\psi_n)],
 \end{equation*}
where $\psi_n$ is the phase factor, which is introduced to shift the zero of the torsion potential. The phase angles $\psi_n$ are usually chosen so that terms with positive $C_{n}$ has minima at $180^{\circ}$ (i.e., for odd $n$, $\psi_n=0^{\circ}$ and for even $n$, $\psi_n=180^{\circ}$ ).
We denote  
\begin{align*}
	&\mathbf{q}_1=	\mathbf{q}_{ij}, \quad  \mathbf{q}_2=	\mathbf{q}_{jk}, \quad \mathbf{q}_3=	\mathbf{q}_{kl}, \\
	&\mathbf{u}=\mathbf{q}_1 \times \mathbf{q}_2, \quad \mathbf{v}=\mathbf{q}_2 \times \mathbf{q}_3,\\
	&\tau=\tau_{ijkl}, \quad \Lambda=\Lambda_{ijkl}=\cos \tau=\frac{\mathbf{u} \cdot \mathbf{v}}{|\mathbf{u}||\mathbf{v}|}, \\
	&C_{n}=C_{n,{ijkl}}, \quad n=1,2,3,4,5,6.
\end{align*}
 Due to the fact that $\psi_n=0^{\circ}$ or $180^{\circ}$, we derive
 \begin{align*}
 & \nabla_{\mathbf{r}_n}W_{\rm torsion}(\mathbf{r}_i,\mathbf{r}_j,\mathbf{r}_k,\mathbf{r}_l)
\\
& \quad 
=\nabla_{\mathbf{r}_n} \Lambda [(C_1\cos(\psi_1)-3C_3\cos(\psi_3)+5C_5\cos(\psi_5) ) 
\\
 &\qquad +\Lambda(4C_2\cos(\psi_2)-16C_4\cos(\psi_4)+36C_6\cos(\psi_6))  +\Lambda^2(12C_3\cos(\psi_3)-60C_5\cos(\psi_5))  \\
 &\qquad +\Lambda^3(32C_4\cos(\psi_4)-192C_6\cos(\psi_6)) +\Lambda^4(80C_5\cos(\psi_5)) +\Lambda^5(192C_6\cos(\psi_6))], 
 \end{align*}
 where
 \begin{align*}
 	&\nabla_{\mathbf{r}_i}\Lambda=-	\nabla_{\mathbf{q}_1}\Lambda, \\
 	&\nabla_{\mathbf{r}_j}\Lambda=\nabla_{\mathbf{q}_1}\Lambda-\nabla_{\mathbf{q}_2}\Lambda, \\
 	&\nabla_{\mathbf{r}_k}\Lambda=\nabla_{\mathbf{q}_2}\Lambda-\nabla_{\mathbf{q}_3}\Lambda, \\
 	&\nabla_{\mathbf{r}_l}\Lambda=\nabla_{\mathbf{q}_3}\Lambda, 
 \end{align*}
 and 
  \begin{align*}
 	&\nabla_{\mathbf{q}_1}\Lambda=\frac{(\mathbf{q}_2\times \mathbf{v})|\mathbf{u}|^2-(\mathbf{u}\cdot \mathbf{v})(\mathbf{q}_2\times \mathbf{u})}{|\mathbf{u}|^3 |\mathbf{v}|},\\
 	&\nabla_{\mathbf{q}_2}\Lambda=\frac{(-\mathbf{q}_1\times \mathbf{v})|\mathbf{u}|^2+(\mathbf{u}\cdot \mathbf{v})(\mathbf{q}_1\times \mathbf{u})}{|\mathbf{u}|^3 |\mathbf{v}|}+\frac{(\mathbf{q}_3\times \mathbf{u})|\mathbf{v}|^2-(\mathbf{u}\cdot \mathbf{v})(\mathbf{q}_3\times \mathbf{v})}{|\mathbf{u}| |\mathbf{v}|^3},\\
 	&\nabla_{\mathbf{q}_3}\Lambda=\frac{(-\mathbf{q}_2\times \mathbf{u})|\mathbf{v}|^2+(\mathbf{u}\cdot \mathbf{v})(\mathbf{q}_2\times \mathbf{v})}{|\mathbf{u}| |\mathbf{v}|^3}.
 \end{align*}

\paragraph{Potentials of Mean Force.}
 The potential of mean force (PMF) is a general term for the effective interaction between solutes that stems from direct solute-solute interactions and that is mediated by the solvent.
 It is usually defined with respect to a reaction coordinate as the difference between the free energy of solvated state at a given coordinate  $d$ and that at a fixed, reference coordinate $d_{ref}$.  
Here, we recall the definition of PMF for our VISM \cite{WangEtal_VISMCFA_JCTC12}.

For a solute-solvent interface $\Gamma$, we denote by $G_{\rm geom}[\Gamma]$, $G_{\rm vdW}[\Gamma]$, and 
$G_{\rm elec}[\Gamma]$ the first, second, and last term in $G_{\rm VISM}[\Gamma]$  
\reff{G_VISM}, respectively. 
 Fix now a finite coordinate $d$. Denote by $\Gamma_d$ a corresponding VISM optimal surface, i.e., a stable equilibrium solute-solvent interface minimizing locally the VISM solvation free-energy functional. We define the (total) PMF to be the sum of its separate contributions
 \begin{equation*}
 	G_{tot}^{\rm pmf}(d)=G_{\rm geom}^{\rm pmf}(d) + G_{\rm vdW}^{\rm pmf}(d) + G_{\rm elec}^{\rm pmf}(d), 
 \end{equation*}
where 
 \begin{align*}
&G_{\rm geom}^{\rm pmf}(d)=G_{\rm geom}[\Gamma_d] - G_{\rm geom}[\Gamma_{\infty}],\\
&G_{\rm vdW}^{\rm pmf}(d)=G_{\rm vdW}[\Gamma_d]-G_{\rm vdW}[\Gamma_{\infty}]+\sum_{i=1}^{M}\sum_{j=M+1}^{N} U_{i,j}(|\mathbf{x}_i-\mathbf{x}_j|),\\
&G_{\rm elec}^{\rm pmf}(d)=G_{\rm elec}[\Gamma_d]-G_{\rm elec}[\Gamma_{\infty}]+\frac{1}{4\pi\varepsilon_m\varepsilon_0} \sum_{i=1}^{M}\sum_{j=M+1}^{N}\frac{Q_iQ_j}{|\mathbf{x_i}-\mathbf{x_j}|}. 
 \end{align*}
 Here a quantity at $\infty$ is understand as the limit of that quantity at a coordinate $d'$ as $d' \rightarrow \infty$. 
The double-sum terms above are the solute-solute vdW and charge-charge interactions.

 As $d$ becomes large, the VISM optimal solute solvent interface $\Gamma_d$ becomes the union of two separate VISM optimal solute-solvent interface $\Gamma_I$ and $\Gamma_{II}$, both independent of $d$. 
 They are obtained by minimizing the VISM free energy functional for the corresponding groups of fixed, solute atoms. If we denote by $G_{\Gamma_I}$ and $G_{\Gamma_{II}}$ the corresponding minimum VISM free energies for these individual groups of atoms, then 
 $
 G_{\Gamma_{\infty}}=G_{\Gamma_{I}}+G_{\Gamma_{II}}
 $
 Similarly, each component of the VISM free energy is the sum of
 that for the two groups of solute atoms, i.e., $G$ in the above equation can be replaced by $G_{\rm geom}$, or $G_{\rm vdW}$ or $G_{\rm elec}$.

\bigskip

\noindent
\textbf{\large Acknowledgment.} This work was supported in part by 
an AMS Simons Travel Grant (SL), 
the US National Science Foundation through grant DMS-1913144 (LTC $\&$ BL), and 
the US National Institutes of Health through grant R01GM132106 (BL). 
The authors thank Dr.\ Clarisse G.\ Ricci and Professor Shenggao Zhou for helpful discussions.

\bibliographystyle{plain}
\bibliography{SLbibfile}

\begin{thebibliography}{10}

\bibitem{BatesEtal_JMB09}
P.~W. Bates, Z.~Chen, Y.~H. Sun, G.~W. Wei, and S.~Zhao.
\newblock Geometric and potential driving formation and evolution of
  biomolecular surfaces.
\newblock {\em J. Math. Biol.}, 59:193--231, 2009.

\bibitem{best2012optimization}
R.~B. Best, X.~Zhu, J.~Shim, P.~E.~M. Lopes, M.~Jeetain, M.~Feig, , and A.~D.
  {MacKerell Jr.}
\newblock Optimization of the additive {CHARMM} all-atom protein force field
  targeting improved sampling of the backbone $\phi$, $\psi$ and side-chain
  $\chi$1 and $\chi$2 dihedral angles.
\newblock {\em J. Chem. Theory Comput.}, 8(9):3257--3273, 2012.

\bibitem{CDML_JCP07}
L.-T. Cheng, J.~Dzubiella, J.~A. McCammon, and B.~Li.
\newblock Application of the level-set method to the implicit solvation of
  nonpolar molecules.
\newblock {\em J. Chem. Phys.}, 127:084503, 2007.

\bibitem{CXDMCL_JCTC09}
L.-T. Cheng, Y.~Xie, J.~Dzubiella, J.~A. McCammon, J.~Che, and B.~Li.
\newblock Coupling the level-set method with molecular mechanics for
  variational implicit solvation of nonpolar molecules.
\newblock {\em J. Chem. Theory Comput.}, 5:257--266, 2009.

\bibitem{cheng2010level}
Li-Tien Cheng, Bo~Li, and Zhongming Wang.
\newblock Level-set minimization of potential controlled hadwiger valuations
  for molecular solvation.
\newblock {\em Journal of computational physics}, 229(22):8497--8510, 2010.

\bibitem{DSM06a}
J.~Dzubiella, J.~M.~J. Swanson, and J.~A. McCammon.
\newblock Coupling hydrophobicity, dispersion, and electrostatics in continuum
  solvent models.
\newblock {\em Phys. Rev. Lett.}, 96:087802, 2006.

\bibitem{DSM06b}
J.~Dzubiella, J.~M.~J. Swanson, and J.~A. McCammon.
\newblock Coupling nonpolar and polar solvation free energies in implicit
  solvent models.
\newblock {\em J. Chem. Phys.}, 124:084905, 2006.

\bibitem{Esdoglu_JCP2017}
S.~Esdo$\bar{g}$lu, M.~Jacobs, and P.~Zhang.
\newblock Kernels with prescribed surface tension and mobility for threshold
  dynamics schemes.
\newblock {\em J. Comput. Phys.}, 337:62--83, 2017.

\bibitem{gibou2005fast}
F.~Gibou and R.~Fedkiw.
\newblock A fast hybrid k-means level set algorithm for segmentation.
\newblock In {\em 4th Annual Hawaii International Conference on Statistics and
  Mathematics}, pages 281--291, 2005.

\bibitem{GNF_Che}
Z.~Guo, B.~Li, J.~Dzubiella, L.-T. Cheng, J.~A. McCammon, and J.~Che.
\newblock Evaluation of hydration free energy by the level-set variational
  implicit-solvent model with the coulomb-field approximation.
\newblock {\em J. Chem. Theory Comput.}, 9:1778--1787, 2013.

\bibitem{GNF_Che2014}
Z.~Guo, B.~Li, J.~Dzubiella, L.-T. Cheng, J.~A. McCammon, and J.~Che.
\newblock Heterogeneous hydration of p53/mdm2 complex.
\newblock {\em J. Chem. Theory Comput.}, 10:1302--1313, 2014.

\bibitem{halgren1996merckI}
T.~A. Halgren.
\newblock Merck molecular force field. {I.} {B}asis, form, scope,
  parameterization, and performance of {MMFF94}.
\newblock {\em J. Comput. Chem.}, 17:490--519, 1996.

\bibitem{halgren1996merckII}
T.~A. Halgren.
\newblock Merck molecular force field. {II.} {MMFF94} van der {W}aals and
  electrostatic parameters for intermolecular interactions.
\newblock {\em J. Comput. Chem.}, 17:520--552, 1996.

\bibitem{hopfinger1985computer}
AJ~Hopfinger.
\newblock Computer-assisted drug design.
\newblock {\em Journal of medicinal chemistry}, 28(9):1133--1139, 1985.

\bibitem{huang2017charmm36}
J.~Huang, S.~Rauscher, G.~Nawrocki, T.~Ran, M.~Feig, B.~L. {de Groot},
  H.~Grubm{\"u}ller, and A.~D. MacKerell.
\newblock {CHARMM36:} {A}n improved force field for folded and intrinsically
  disordered proteins.
\newblock In {\em 61st Annual Meeting of the Biophysical Society}, pages
  175a--176a, 2017.

\bibitem{jorgensen1996development}
W.~L. Jorgensen, D.~S. Maxwell, and J.~Tirado-Rives.
\newblock Development and testing of the {OPLS} all-atom force field on
  conformational energetics and properties of organic liquids.
\newblock {\em J. Amer. Chem. Soc.}, 118(45):11225--11236, 1996.

\bibitem{Tai_IEEE2006}
J.~Lie, M.~Lysaker, and X.-C. Tai.
\newblock A binary level set model and some applications to {M}umford--{S}hah
  image segmentation.
\newblock {\em IEEE Trans. Image Proc.}, 15:1171--1181, 2006.

\bibitem{Tai_MathComp2006}
J.~Lie, M.~Lysaker, and X.-C. Tai.
\newblock A variant of the level set method and applications to image
  segmentation.
\newblock {\em Math. Comput}, 75(255):1155--1174, 2006.

\bibitem{LCW99}
K.~Lum, D.~Chandler, and J.~D. Weeks.
\newblock Hydrophobicity at small and large length scales.
\newblock {\em J. Phys. Chem. B}, 103:4570--4577, 1999.

\bibitem{mackerell1998all}
A.~D. {MacKerell Jr.}, D.~Bashford, M.~L. D.~R. Bellott, R.~L. {Dunbrack Jr},
  J.~D. Evanseck, M.~J. Field, S.~Fischer, J.~Gao, H.~Guo, S.~Ha, et~al.
\newblock All-atom empirical potential for molecular modeling and dynamics
  studies of proteins.
\newblock {\em The J. Phys. Chem. B}, 102(18):3586--3616, 1998.

\bibitem{mackerell2004improved}
A.~D. {MacKerell Jr.}, M.~Feig, and Charles~L. Brooks.
\newblock Improved treatment of the protein backbone in empirical force fields.
\newblock {\em J. Amer. Chem. Soc.}, 126(3):698--699, 2004.

\bibitem{MBO_1992}
B.~Merriman, J.~Bence, and S.~Osher.
\newblock Diffsion generated motion by mean curvature.
\newblock In J.~Taylor, editor, {\em Computational Crystal Growers Workshop},
  pages 73--83. Amer. Math. Soc., 1992.

\bibitem{Borgis_JPCB2005}
R.~Ramirez and D.~Borgis.
\newblock Density functional theory of solvation and its relation to implicit
  solvent models.
\newblock {\em J. Phys. Chem. B}, 109:6754--6763, 2005.

\bibitem{Ricci_Rev2018}
C.~G. Ricci, B.~Li, L.-T. Cheng, J.~Dzubiella, and J.~A. McCammon.
\newblock Tailoring the variational implicit solvent method for new challenges:
  {B}iomolecular recognition and assembly.
\newblock {\em Front. Mol. Biosci.}, 5(13), 2018.

\bibitem{Ruuth_DiffuGen_JCP1998}
S.~J. Ruuth.
\newblock Efficient algorithms for diffusion-generated motion by mean
  curvature.
\newblock {\em J. Comput. Phys.}, 144:603--625, 1998.

\bibitem{RuuthMerriman_Rev_JCP2001}
S.~J. Ruuth and B.~Merriman.
\newblock Convolution-thresholding methods for interface motion.
\newblock {\em J. Comput. Phys.}, 169:678--707, 2001.

\bibitem{Ruuth_Convol_JCP1999}
S.~J. Ruuth, B.~Merriman, and S.~Osher.
\newblock Convolution generated motion as a link between cellular automata and
  continuum pattern dynamics.
\newblock {\em J. Comput. Phys.}, 151:836--861, 1999.

\bibitem{wang_efficient_2017}
D.~Wang, H.~Li, X.~Wei, and X.-P. Wang.
\newblock An efficient iterative thresholding method for image segmentation.
\newblock {\em J. Comput. Phys.}, 350:657--667, 2017.

\bibitem{WangEtal_VISMCFA_JCTC12}
Z.~Wang, J.~Che, L.-T. Cheng, J.~Dzubiella, B.~Li, and J.~A. McCammon.
\newblock Level-set variational implicit solvation with the {C}oulomb-field
  approximation.
\newblock {\em J. Chem. Theory Comput.}, 8:386--397, 2012.

\bibitem{watkins2004fundamentals}
D.~S. Watkins.
\newblock {\em Fundamentals of matrix computations}, volume~64.
\newblock John Wiley \& Sons, 2004.

\bibitem{zhang2021coupling}
Z.~Zhang, C.~G. Ricci, C.~Fan, L.-T. Cheng, B.~Li, and J.~A. McCammon.
\newblock Coupling {M}onte {C}arlo, variational implicit solvation, and binary
  level-set for simulations of biomolecular binding.
\newblock {\em J. Chem. Theory Comput.}, 17:2465--2478, 2021.

\bibitem{Zhou_VISMPB_JCTC2014}
S.~Zhou, L.-T. Cheng, J.~Dzubiella, B.~Li, and J.~A. McCammon.
\newblock Variational implicit solvation with {P}oisson--{B}oltzmann theory.
\newblock {\em J. Chem. Theory Comput.}, 10(4):1454--1467, 2014.

\bibitem{zhou_ls-vism_2015}
S.~Zhou, L.-T. Cheng, H.~Sun, J.~Che, J.~Dzubiella, B.~Li, and J.~A. McCammon.
\newblock {LS}-{VISM}: {A} software package for analysis of biomolecular
  solvation.
\newblock {\em J. Comput. Chem.}, 36:1047--1059, 2015.

\bibitem{Zhou_StochLSM_JCP2015}
S.~Zhou, H.~Sun, L.-T. Cheng, J.~Dzubiella, B.~Li, , and J.~A. McCammon.
\newblock Stochastic level-set variational implicit-solvent approach to
  solute-solvent interfacial fluctuations.
\newblock {\em J. Chem. Phys.}, 145:054114, 2016.

\end{thebibliography}
\end{document}